\documentclass[useAMS,usenatbib]{mn2e}

\usepackage[pdftex]{color}
\usepackage[pdftex]{graphicx}
\usepackage[fleqn]{amsmath}     
\usepackage{amssymb}

\newcommand{\swd}{Schwarzschild }

\newcommand{\sfg}{S$^4$G }

\begin{document}

\title[Orbital decomposition]{Orbital decomposition of CALIFA spiral galaxies}
\author[Zhu., et al]{Ling Zhu$^1$\thanks{E-mail: lzhu@mpia.de}, Remco~van~den~Bosch$^1$, 
    	Glenn~van~de~Ven$^{1,8}$, Mariya Lyubenova$^{2,8}$,  \and Jes\'us Falc\'on-Barroso$^{3,4}$,  Sharon E. Meidt$^1$, Marie Martig$^1$, Juntai Shen$^5$, \and  Zhao-Yu Li$^5$,  Akin Yildirim$^1$, C. Jakob Walcher$^6$, Sebastian F. Sanchez$^7$\\
    	$^1$ Max Planck Institute for Astronomy, K\"onigstuhl 17, 69117 Heidelberg, Germany \\
    	$^2$ Kapteyn Astronomical Institute, P.O. box 800 9700 AV Groningen, the Netherlands\\
	$^3$ Instituto de Astrof\'isica de Canarias (IAC), E-38205 La Laguna, Tenerife, Spain\\
	$^4$ Universidad de La Laguna, Dpto. Astrof\'isica, E-38206 La Laguna, Tenerife, Spain\\
	$^5$ Key Laboratory for Research in Galaxies and Cosmology, Shanghai Astronomical Observatory, Chinese Academy of Sciences,\\ 80 Nandan Road, Shanghai 200030, China \\
	$^6$ Leibniz-Institut für Astrophysik Potsdam (AIP), Germany\\
	$^7$ Instituto de Astronom\'ia, Universidad Nacional Auton\'oma de M\'exico, A.P. 70-264, 04510, M\'exico, D.F.\\
	$^8$ European Southern Observatory, Karl-Schwarzschild-Str. 2, 85748 Garching b. M\"unchen, Germany\\
	}

\maketitle
\date{}

\label{firstpage}
\begin{abstract}
\swd orbit-based dynamical models are widely used to uncover the internal dynamics of early-type galaxies and globular clusters. 
Here we present for the first time the \swd models of late-type galaxies: an SBb galaxy NGC 4210 and an S0 galaxy NGC 6278 from the CALIFA survey. 
The mass profiles within $2\,R_e$ are constrained well with $1\sigma$ statistical error of $\sim 10\%$. The luminous and dark mass can be disentangled with uncertainties of $\sim 20\%$ and $\sim 50\%$ respectively. From $R_e$ to $2\,R_e$, the dark matter fraction increases from $14\pm10\%$ to $18\pm10\%$ for NGC 4210 and from $15\pm10\%$ to $30\pm20\%$ for NGC 6278. 
The velocity anisotropy profiles of both $\sigma_r/\sigma_t$ and $\sigma_z/\sigma_R$ are well constrained. 
The inferred internal orbital distributions reveal clear substructures.
The orbits are naturally separated into three components: a cold component with near circular orbits, a hot component with near radial orbits, and a warm component in between. 
The photometrically-identified exponential disks are predominantly made up of cold orbits only beyond $\sim 1\,R_e$, while they are constructed mainly with the warm orbits inside. Our dynamical hot components are concentrated in the inner regions, similar to the photometrically-identified bulges.  
The reliability of the results, especially the orbit distribution, are verified by applying the model to mock data. 
\end{abstract}


\section{Introduction}
\label{S:intro}
The orbital structure of a galaxy is a fundamental diagnostic of its formation and evolution. 
The stars on dynamically cold near circular orbits were born and lived in quiescent times \citep[e.g.,][]{White1978, Fall1980}, while stars on dynamically hot box or radial orbits could be born from unsettled gas or been heated by violent processes; such as major mergers \citep{Davies1983}, minor mergers \citep{Quinn1993} and internal disk instabilities (\citealt{Minchev2006}; \citealt{Saha2010}).

For decades, morphological type (Hubble sequence) and photometric bulge-disk decomposition have been used as proxies of the orbit distribution of galaxies (\citealt{Freeman1970}; \citealt{Laurikainen2010}; \citealt{Weinzirl2009}). 
However, the morphological type is not always a good indication of the underlying orbital structure \citep{Krajnovic2013}. Simulations have for instance shown cases of galaxies with exponential disks but stars on kinematically hot orbits (\citealt{Teklu2015}; \citealt{Obreja2016}), and cases of galaxies where bulge-disk decomposition based on photometry and kinematics give significantly different results (\citealt{Scannapieco2010}; \citealt{Martig2012}; \citealt{Obreja2016}). 

Integral field spectroscopic surveys, such as \emph{CALIFA} \citep{califa2012}, \emph{SAMI} \citep{Croom2012} and \emph{MaNGA} \citep{Bundy2015}, provide stellar kinematic maps for thousands of nearby galaxies across the Hubble sequence. 
Dynamical models are needed to infer from these observations the orbital structures of galaxies.

A powerful dynamical modelling technique is the Schwarzschild's (1979) orbit-superposition method, which builds galactic models by weighting the orbits generated in a gravitational potential.
The \swd method has been widely applied to model the dynamics of early type galaxies: from the central black hole mass (e.g \citealt{vdB2012}, \citealt{Gebhardt2011}) to the outer dark matter distribution 
(e.g \citealt{Thomas2007}, \citealt{Murphy2011})
to the internal orbital structures of the galaxies \citep{vdV2008}, and globular clusters \citep{vdV2006}.
The orbit distributions obtained by these models have been used to identify different dynamical components (\citealt{vdV2006}; \citealt{vdB2008}; \citealt{Cappellari2007SAURONX}; \citealt{Breddels2014}).

Here, we extend the application of \swd models to late-type galaxies and perform dynamical decomposition.
The paper is organized in the following way: in Section~\ref{S:model}, we describe the modeling steps and the technical details; in Section~\ref{S:califa} we apply it to two \emph{CALIFA} galaxies; in Section~\ref{S:discussion} we compare with the results from photometric decompositions and we conclude in Section~\ref{S:summary}.  In appendix~\ref{S:orbit-cut}, we explain in detail why the \swd models had problems on modelling fast rotating galaxies and present our solution. In appendix~\ref{S:simulation}, we apply the model to mock data from simulated galaxies, and evaluate the model's reliability of recovering the underlying mass profiles and orbit distribution. 

\section{Schwarzschild models of spiral galaxies}
\label{S:model}

The main steps to create a \swd model are, first, to create a suitable model for the underlying gravitational potential, second, to calculate a representative library of orbits with this gravitational potential, and third, to find the combination of orbits that reproduces the observed kinematic maps and luminosity distribution.
We illustrate these steps of creating \swd models for spiral galaxies in this section. The first two steps directly follow \citet{vdB2008}, and we only briefly describe them here to illustrate model parameter choices. 

About half of the nearby spiral galaxies in optical wavelengths show bar features (e.g \citealt{Marinova2007}; \citealt{Menendez2007}). However, bar does not significantly affect galaxy's global kinematics (\citealt{Barrera2014}, \citealt{Seidel2015}), stellar kinematics tend to still follow the gravitational potential of the disc, even for the galaxies with strong bars (see our tests to simulated barred galaxy in Appendix~\ref{S:simulation}). Thus we neglect the non-axisymmetry features regarding to the bar in the model.
\subsection{Gravitational potential}
The gravitational potential is generated by a combination of the stellar and dark matter (DM) distributions. The resolution of \emph{CALIFA} data is well beyond the influence radius of black hole (BH), so that the BH mass does not affect our results significantly, and is fixed by adopting a value following the relation between the BH mass and the stellar velocity dispersion from \citet{vdB2016}.

\subsubsection{Stellar mass distribution}
\label{SS:stellar}
The images of the galaxy trace the stellar light, which can be de-projected to get the intrinsic luminosity.

To make the de-projection and the following calculations of gravitational force mathematically convenient,
we use 2D Multiple Gaussian Expansion (MGE; \citealt{Emsellem1994}; \citealt{Cappellari2002}) to describe the flux on 2D plane (in unit of $L_{\mathrm{sun}}/\mathrm{pc}^2$ converted from surface brightness in unit of $\mathrm{magnitude}/\mathrm{arcsec}^2$):
\begin{equation}
S(x',y') = \sum_{i=1}^N {\frac{L_i}{2\pi\sigma'^2_iq'_i} \times \exp[-\frac{1}{2\sigma'^2_i} (x'^2 + \frac{y'^2}{q'^2_i})]},
\end{equation}
where $(L_i, \sigma'_i,q'_i)$ describes the observed total luminosity, size and flattening of each Gaussian component. 

After assuming the space orientation of the galaxy, described by three viewing angles $(\vartheta, \phi, \psi)$, we can de-project the 2D axisymmetric MGE flux to a 3D triaxial MGE luminosity density:
\begin{equation}
\rho(x,y,z) = \sum_{i=1}^N { \frac{L_i}{(\sigma_i \sqrt{2\pi})^3 q_i p_i} \times \exp[-\frac{1}{2\sigma^2_i} (x^2 + \frac{y^2}{p^2_i} + \frac{z^2}{q^2_i})] }.
\end{equation} 
where $p_i  = B_i/A_i$ and $q_i=C_i/A_i$ with $A_i$,$B_i$,$C_i$ representing the major, medium and minor axis of the 3D triaxial Gaussian component.
The relations between the observed quantities $(\sigma'_i,q'_i)$ and the intrinsic ones $(\sigma_i, p_i, q_i)$ are given by \citet{Cappellari2002} and described in detail for application in \swd models in \citet{vdB2008}. The space orientation $(\vartheta, \phi, \psi)$ and the intrinsic shape $(p_i, q_i,u_i = \sigma'_i/\sigma_i)$ can be converted to each other directly, with the requirement of $q_i\leq p_i \leq 1$, $q_i\leq q'_i$ and $\max (q_i/q'_i, p_i) \leq u_i \leq \min(p_i/q'_i, 1)$.

As argued in \citet{vdB2008}, the intrinsic shape are more natural parameters than the space orientation to use as our model free parameters. 
The flattest Gaussian component, having the minimum flattening $q'_{\mathrm{min}}$, dictates the allowed space orientation for the de-projection. 
In practice, we adopt 
$(p_{\mathrm{min}}, q_{\mathrm{min}},u_{\mathrm{min}})$ as free parameters of our model. 

Assuming a constant stellar mass-to-light ratio $\Upsilon_{\ast}$ as free parameter in our models, we obtain the 3D stellar mass density to generate the contribution of stars to the gravitational potential. The correctness of the assumption of constant $\Upsilon_{\ast}$ depends on how well an image taken at a certain wavelength range tracers the underlying mass distribution. We investigate this in Section~\ref{SS:data}.


\subsubsection{Dark matter distribution}
\label{SS:DM}
For the DM distribution, we adopt a spherical NFW (\citealt{Navarro1996}) halo with the enclosed mass profile:
\begin{equation}
M(<r) = M_{200}  g(c) \Bigg [\ln (1+ c r/r_{200}) - \frac{c r/r_{200}}{1+c r/r_{200}} \Bigg],
\end{equation}
where $c$ is the concentration of the DM halo,  $g(c) = [\ln(1+c) -c/(1+c) ]^{-1}$, the virial mass $M_{200}$
is defined as the mass within the virial radius $r_{200}$, i.e., $M_{200} = \frac{4}{3}\pi 200 \rho_c^0 r_{200}^3 $, with the critical density $\rho_c^0 = 1.37 \times 10^{-7} \mathrm{M}_{\odot} \mathrm{pc}^{-3}$.
There are two free parameters in an NFW halo: the concentration $c$ and the virial mass $M_{200}$.

Kinematic data extending to large radius are required to constrain the concentration $c$ and the virial mass $M_{200}$, separately. 
With \emph{CALIFA} kinematic data extending to $\sim 2 R_e$ of the galaxies, the degeneracy between these two parameters is significant (e.g. \citealt{Zhu2014}). Therefore, we fix the DM concentration $c$ following the relation from \citet{Dutton2014}:
\begin{equation}
\log_{10} c = 0.905 - 0.101\log_{10} (M_{\mathrm{200}}/[10^{12} h^{-1} M_{\odot}]),
\end{equation}
with $h=0.671$ (Planck Collaboration 2013).  This assumption should not significantly affect the enclosed DM profiles within the CALIFA data coverage, thus not affect the stellar mass-to-light ratio. 

Combining stellar mass-to-light ratio $\Upsilon_{\ast}$, the three parameters representing the space orientation $(q_{\mathrm{min}}, p_{\mathrm{min}}, u_{\mathrm{min}})$, and the DM halo mass $M_{200}$, we have five free parameters in total.

\subsection{The orbit library}
In a separable triaxial potential, all orbits are regular and conserve three integrals of motion $E$, $I_2$ and $I_3$ which can be calculated analytically. Four different types of orbits exist: three types of tube orbits (the short axis tubes, outer and inner long axis tubes) and the box orbits. 

In the more general potential as generated with the MGE of luminosity density,
the three type of loop orbits are still supported \citep{Schwarzschild1993}, the box orbits are transformed into boxlets \citep{Miralda1989}.

We sample the initial conditions of the orbits via $E$ and their position on the $(x, z)$ plane \citep{vdB2008}. 
The orbit energy $E$ is sampled implicitly through a logarithmic grid in radius, each energy is linked to a grid radius $r_i$ by calculating the potential at the position $(x, y, z) = (r_i, 0, 0)$. Then for each energy, the starting point $(x,z)$ is selected from a linear open polar grid $(R, \theta)$ in between the location of the thin orbits and the equipotential of this energy.
We refer to \citet{vdB2008} for the details of the orbits sampling. This orbit library includes mostly short axis tubes, long axis tubes and a few box orbits in the inner region. 

The number of points we sampled across the three integrals is $n_E \times n_{\theta} \times n_{R}  =  21 \times 10 \times 7$, where $n_E$, $n_{\theta}$, $n_{R}$ are the number of intervals taken across the energy $E$, the azimuthal angle $\theta$ and radius $R$ on the (x, z) plane.
The $r_i$ representing energy spans the region from $0.5\sigma'_{\mathrm{min}} \, (< 1'')$ to $5 \sigma'_{\mathrm{max}}$, where $\sigma'_{\mathrm{min}}$ and $\sigma'_{\mathrm{max}}$ are the minimum and maximum $\sigma'$ of the Gaussian components from the MGE fit. 

The above sampling may not include enough box orbits for creating a possible triaxial shape, we include additional box orbits dropped from the equipotential surface, using linear steps in the two spherical angles $\theta$ and $\phi$. Combining with the energy $E$, the number of points we sample across the three dimensional set are $n_E \times n_{\theta} \times n_{\phi}  =  21 \times 10 \times 7$. The set of energies $E$ and angles $\theta$ are designed to be identical for the two sets of orbit libraries.

In order to smooth the model, 5 ditherings for each value of integrals are introduced. So for each orbit, we create $5\times 5\times 5$ dithering orbits to form an orbit bundle. 

\subsection{Weighing the orbits}
\label{S:weighing}
Consider we have 2D MGE described flux and kinematic data in hundreds of observational apertures forming kinematic maps (each aperture denoted as $l$), used as the model constraints. 
We are going to reproduce these data simultaneously by a superposition of $\sim 1000$ orbit bundles, with each orbit bundle $k$ weighted by $w_k$. The solution of orbit weights is a linear least $\chi^2$ problem, the $\chi^2$ to be minimized is:
\begin{equation}
\chi^2 = \chi^2_{\mathrm{lum}} + \chi^2_{\mathrm{kin}}.
\end{equation}
In practice, the luminosity distribution is easy to fit and $\chi^2_{\mathrm{lum}}$ is much smaller than $\chi^2_{\mathrm{kin}}$. $\chi^2$ is dominated and highly correlated with $\chi^2_{\mathrm{kin}}$.

Throughout the paper, we keep the subscript $l$ denoting observational apertures and the subscript/superscript $k$ denoting orbit bundles. 

\subsubsection{Fitting luminosity distribution}
Luminosity distribution of the model is constrained by both the observed 2D MGE described flux and deprojected 3D MGE luminosity density. We bin the 2D flux as the same binning scheme as the kinematic data ($S_l$ for each aperture $l$) and divide the 3D luminosity distribution into a 3D grids with 360 bins ($\rho_n, n = 1,2,3...360$). We allow relative errors of $1\%$ for $S_l$ and $2\%$ for $\rho_n$.

Each orbit bundle $k$ contributes linearly $S_l^k$ to flux in each aperture $l$ and $\rho_n^k$ to the intrinsic luminosity density in each bin $n$, thus the fitting to luminosity distribution is just minimizing
\begin{multline}
\label{eqn:SB}
\chi^2_{\mathrm{lum}} = \chi^2_{\mathrm{S}} + \chi^2_{\mathrm{\rho}}
\\
\chi^2_{\mathrm{S}} = \sum_{l} \Bigg[ \frac{S_l - \sum_k w_k S_l^k}{0.01S_l} \Bigg]^2
\\
\chi^2_{\mathrm{\rho}} = \sum_{n} \Bigg[ \frac{\rho_n - \sum_k w_k \rho_n^k }{0.02\rho_n} \Bigg]^2.
\\
\end{multline}

\subsubsection{Fitting kinematic maps}
\label{SS:GH1}
Consider a velocity distribution profile (VP) $f_l$ observed at the aperture $l$, several orbit bundles in the model may contribute partly to this aperture; each orbit bundle $k$ contributes with a VP of $f_l^k$.
Our purpose is to get the best fitting of $f_l$, that is solving the orbit weights $w_k$ to get $\sum_{k} w_k f_l^k$ as close as possible to $f_l$ for all the observational apertures. 

The observed VP $f_l$ itself is usually a complicated profile. 
Gaussian-Hermite (GH) expansion is used to describe the VPs (\citealt{Gerhard1993}, \citealt{vdM&Franx1993}, \citealt{Rix1997}):
\begin{multline}
\mathcal{GH}(v; \gamma, V, \sigma, h_m) = 
\\
\frac{\gamma}{\sqrt{2\pi} \sigma} \exp \Bigg [-\frac{1}{2} (\frac{v-V}{\sigma})^2  \Bigg] \sum_{m=0}^{4}h_m H_m(\frac{v-V}{\sigma}),
\end{multline}
where $H_m$ are the Hermite polynomials and for which we usually truncate at the fourth order or at the second order depending on the data quality. 
We just keep the description of methodology in generality by including $h_3$, $h_4$. If there is no reliable $h_3$, $h_4$, the following description still holds with only up to the second order moments.

For each observational aperture $l$, the best GH fit of $f_l$ yields the parameters $(V_{l}, \sigma_{l}, h_{3,l}, h_{4,l})$, given errors of $(dV_{l}, d\sigma_{l}, dh_{3,l}, dh_{4,l})$, with fixed $h_{0,l} = 1, h_{1,l} = 0, h_{2,l} = 0$. In this way $(V_{l}, \sigma_{l})$ also determines the best Gaussian fit of $f_l$ (\citealt{vdM&Franx1993}). 

We then describe the distribution of each orbit bundle $f_l^k$ by GH expansion around the same 
$(V_{l}, \sigma_{l})$, resulting in the coefficients of $(h_{0,l}^k, h_{1,l}^k, h_{2,l}^k, h_{3,l}^k, h_{4,l}^k)$. 

Adopting option A as described in Appendix~\ref{S:orbit-cut}, $h_{0,l}$ will not be included in the fitting, the parameters we are going to fit are $(h_{1,l} = 0, h_{2,l} = 0, h_{3,l}, h_{4,l})$, with errors of $(dh_{1,l} = dV_{l}/(\sqrt{2}\sigma_l), dh_{2,l} = d\sigma_l/(\sqrt{2}\sigma_l), dh_{3,l}, dh_{4,l})$ (\citealt{vdM&Franx1993}; \citealt{Magorian&Binney1994}).
Each orbit $k$ contributes linearly to these parameters of the VP $f_l$:
\begin{equation}
\label{eqn:kin}
\chi^2_{\mathrm{kin}} = \sum_{l} \sum_{m=1}^{4} \Bigg[ \frac{S_l h_{m,l} - \sum_k w_k S_l^k h_{m,l}^k}{S_l dh_{m,l}} \Bigg]^2.
\end{equation}
The solution of the orbit weights becomes a linear least $\chi^2$ problem and it includes the non-Gaussian components of the VPs.

The algorithm causes problems in modelling fast rotating galaxies \citep{Cretton1999} in the present \swd models.
In appendix~\ref{S:orbit-cut}, we explain in detail the reason of this problem and show two possible solutions, while option A is chosen for the modelling of \emph{CALIFA} galaxies. In appendix~\ref{S:simulation}, we show that our model works reasonably well for recovering the true orbit distribution of a simulated spiral galaxy.  

\section{Application to two CALIFA galaxies}
\label{S:califa}

\subsection{Stellar imaging and kinematics}
\label{SS:data}
The two galaxies we selected are in the \emph{CALIFA} as well as the \sfg survey. The basic information of the galaxies is listed in Table~\ref{tab:info}.  We use the half-light radius $R_e$ measured from  the $r$-band images.  $M_{\mathrm{sun},r} = 4.64$ and $M_{\mathrm{sun}, 3.6} = 3.24$ are used to convert magnitude to solar luminosity. 

\begin{table}
\caption{The basic information of the two galaxies. From left to right, they are the Hubble type of the galaxies, the absolute magnitude in SDSS $r$-band, the absolute magnitude of \sfg 3.6-$\mu$m image,  the distance in Mpc, and the half-light radius of the $r$-band images in arcsec.}
\label{tab:info}
\begin{tabular}{*{6}{l}}
\hline
 -                 & Hubble Type      &  $M_{\mathrm{r}}$  &  $M_{\mathrm{3.6}}$ & D & $R_e$ \\
\hline
 NGC 4210   & Sb(B)    &    -20.57    &  -20.85  &  43.65  & 23\\
 NGC 6278   & S0(AB)  &    -20.98    &  -21.25   &  39.95 & 14\\
  \hline
 \end{tabular}
\end{table}

We use the stellar kinematic data from the \emph{CALIFA} survey (\citealt{califa2012}; \citealt{CALIFAdr1}). The stellar kinematics are extracted from \emph{CALIFA} spectrum which cover the range 3400-4750 $\AA\,$ at a spectrum resolution of $\sim 1650$. The stellar kinematic maps are obtained by using Voronoi binning, getting a signal to noise threshold of $S/N = 20$. Please refer to \citet{Falcon-Barroso2016} for the details of the \emph{CALIFA} stellar kinematics.  

As mentioned in Section~\ref{SS:stellar} and Section~\ref{SS:GH1}, we need an image to construct the gravitational potential and an image to constrain the luminosity distribution of the orbit-superposed model. The former image has to follow the assumption of constant stellar mass-to-light ratio and the latter one has to trace the luminosity of the kinematic tracer. The \swd model does not need to be self-consistent, thus the two images are not necessarily the same.

The \emph{CALIFA} kinematic data are drawn from spectra with wavelength coverage close to the SDSS $r$-band, so we take the SDSS $r$-band image to constrain the luminosity distribution of the orbit-superposed model. 
 
As a simplest option, we also take the same $r$-band image to construct the stellar mass contribution to the gravitational potential, assuming a constant stellar mass-to-light ratio in the $r$-band ($\Upsilon_r$). This is our first set of models and we call it the $r$-band model in the following sections. 

Alternatively, we use the \sfg stellar light map at 3.6-$\mu$m to construct the stellar mass contribution to the gravitational potential.
The stellar light at 3.6-$\mu$m is isolated from the ``contaminating" emission by using independent component analysis (\citealt{Meidt2014}; \citealt{Querejeta2015}). The (corrected) 3.6 micron image traces primarily the light from old stars and is therefore a more direct tracer of stellar mass than the $r$-band image. A constant stellar mass-to-light ratio at 3.6-$\mu$m $\Upsilon_{3.6}$ is supposed to be a more reasonable assumption than a constant $\Upsilon_r$. We create another set of models with this option, which is called the 3.6-$\mu$m model in the following sections. Note that even for the 3.6-$\mu$m models, the $r$-band images are always used to constrain the luminosity distribution of the models, because the stellar kinematics are in the optical and not near-infrared.

\begin{figure*}
\centering\includegraphics[width=7.9cm]{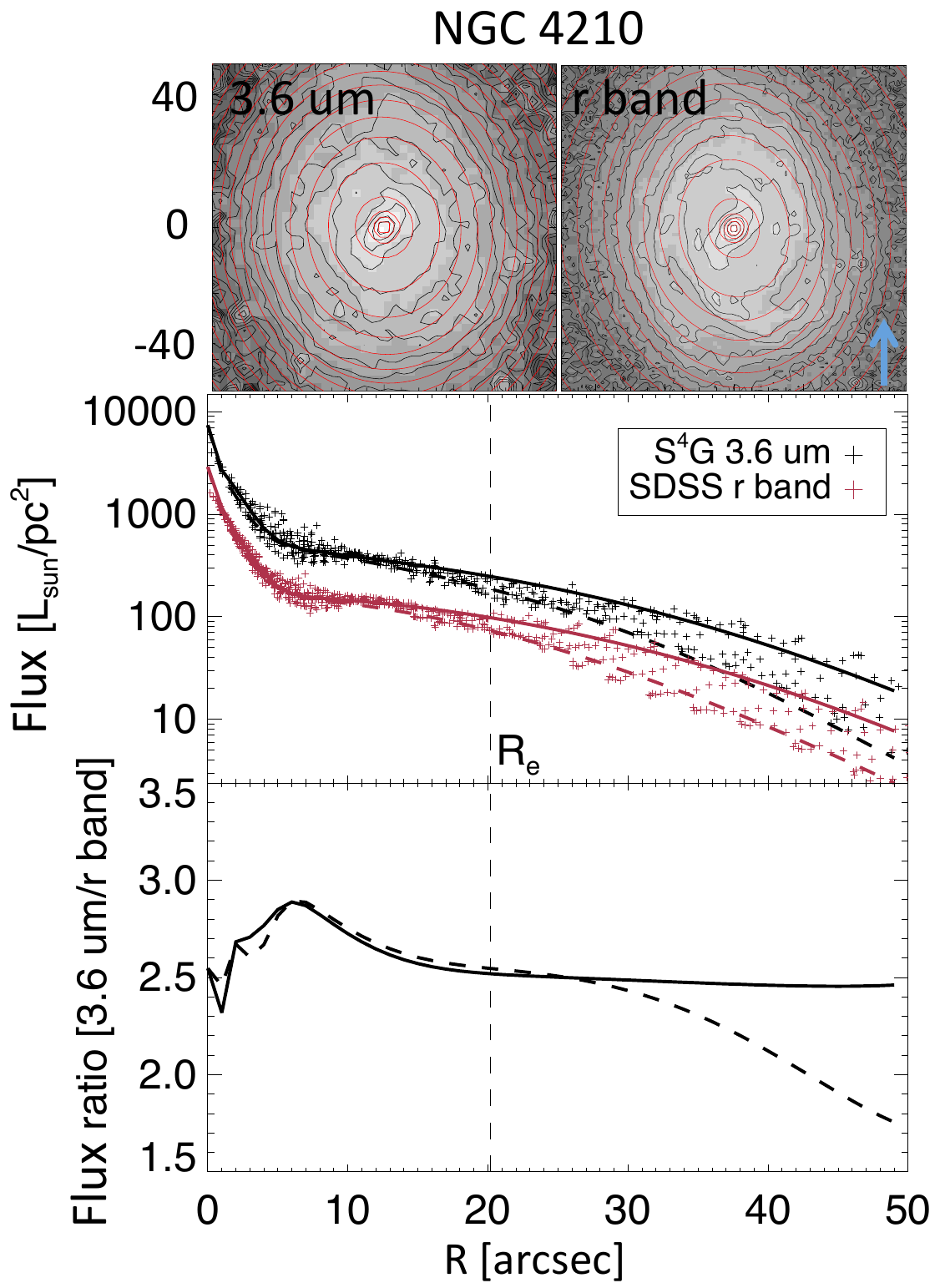}\includegraphics[width=7.4cm]{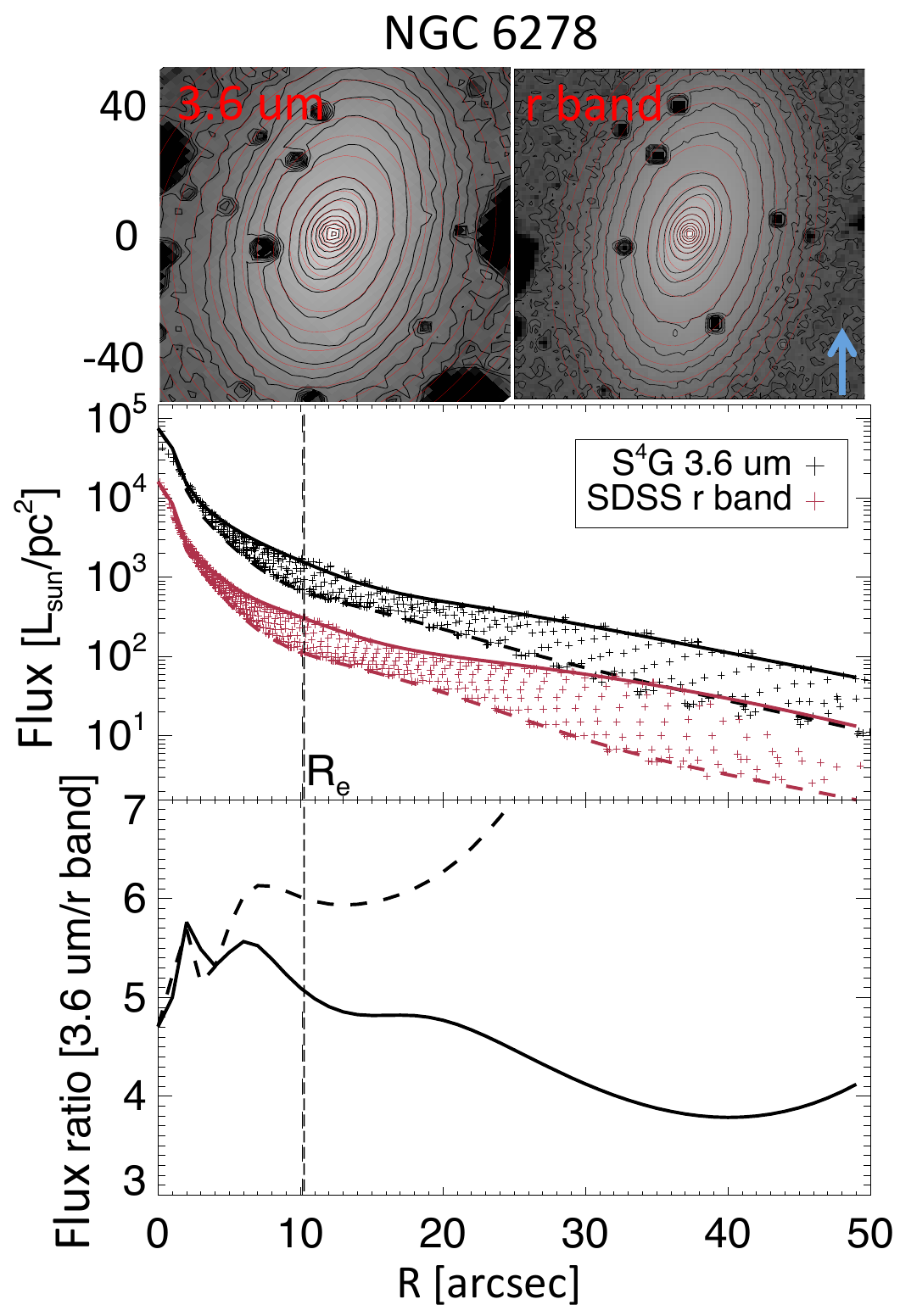}
\caption{The stellar surface brightness at SDSS $r$-band and \sfg 3.6-$\mu$m; NGC 4210 on the left and NGC 6278 on the right. {\bf Top panel}: the black contours represent the original images of 3.6-$\mu$m and $r$-band, the red contours are the two dimensional MGE fits correspondingly. The arrows point the north direction. 
{\bf Middle panel}: the dots represent the flux radial profiles when dividing the original image to several sectors. The solid and dashed curves are the MGE fit along the major and the minor axis. Red is for the $r$-band image and black is for the 3.6-$\mu$m image.
{\bf Bottom panel}: the solid and dashed curves are the flux ratio of \sfg 3.6-$\mu$m to SDSS $r$-band image along the major and the minor axis. The vertical dashed line represent the position of $1\,R_e$}
\label{fig:mge}
\end{figure*}

Figure~\ref{fig:mge} shows the images of the two galaxies and the corresponding MGE fit, with the parameters of MGEs are shown in a table in appendix. The top panels are the ($r$-band and 3.6-$\mu$m) images (the black contours), with the 2D MGE fit overplotted (the red ellipses). The middle panels are the radial flux with the MGE fit along the major (solid curves) and minor axis (dashed curves). The bottom panels are the flux ratio of 3.6-$\mu$m to $r$-band along the major (solid curves) and minor axis (dashed curves). 

NGC 4210 has a bar in the inner $0.5 \, R_e$. The bar is elongated and has a different orientation than the major axis of the galaxy defined as the photometric shape in the outer regions.
The non-axisymmetry of the bar is neglected in the MGE fits.
NGC 4210 is brighter at 3.6-$\mu$m in the inner $0.5 \, R_e$ where the bulge/bar dominates.
The flux ratio of \sfg 3.6-$\mu$m to SDSS $r$-band, calculated by their MGE fits, along the major axis are generally consistent with that along the minor axis within $2\,R_e$. The broader bar in 3.6-$\mu$m image is not reflected in the MGE fits. 

NGC 6278 has a prominent bulge and possibly also an embedded bar revealed from the slightly twisted surface brightness contour in the center, however the 2D axisymmetric MGEs fits the surface brightness reasonably well.  
The two images have different flattening, which may be caused by a significant contribution of bulge component extending to large radius, which is round and older than the disk, thus causing the 3.6-$\mu$m image to be rounder.  

As described in Section~\ref{SS:stellar}, when de-projecting the image, the flattest Gaussian component with 
${q'}_{\mathrm{min}}$ dictates the lower limit of inclination angle. During the MGE fit, we choose ${q'}_{\mathrm{min}}$ as large as possible to allow for a wide region of inclination angles.

\subsection{Best-fitting models}
\label{SS:califa_fit}
\begin{figure*}
\centering\includegraphics[width=14cm]{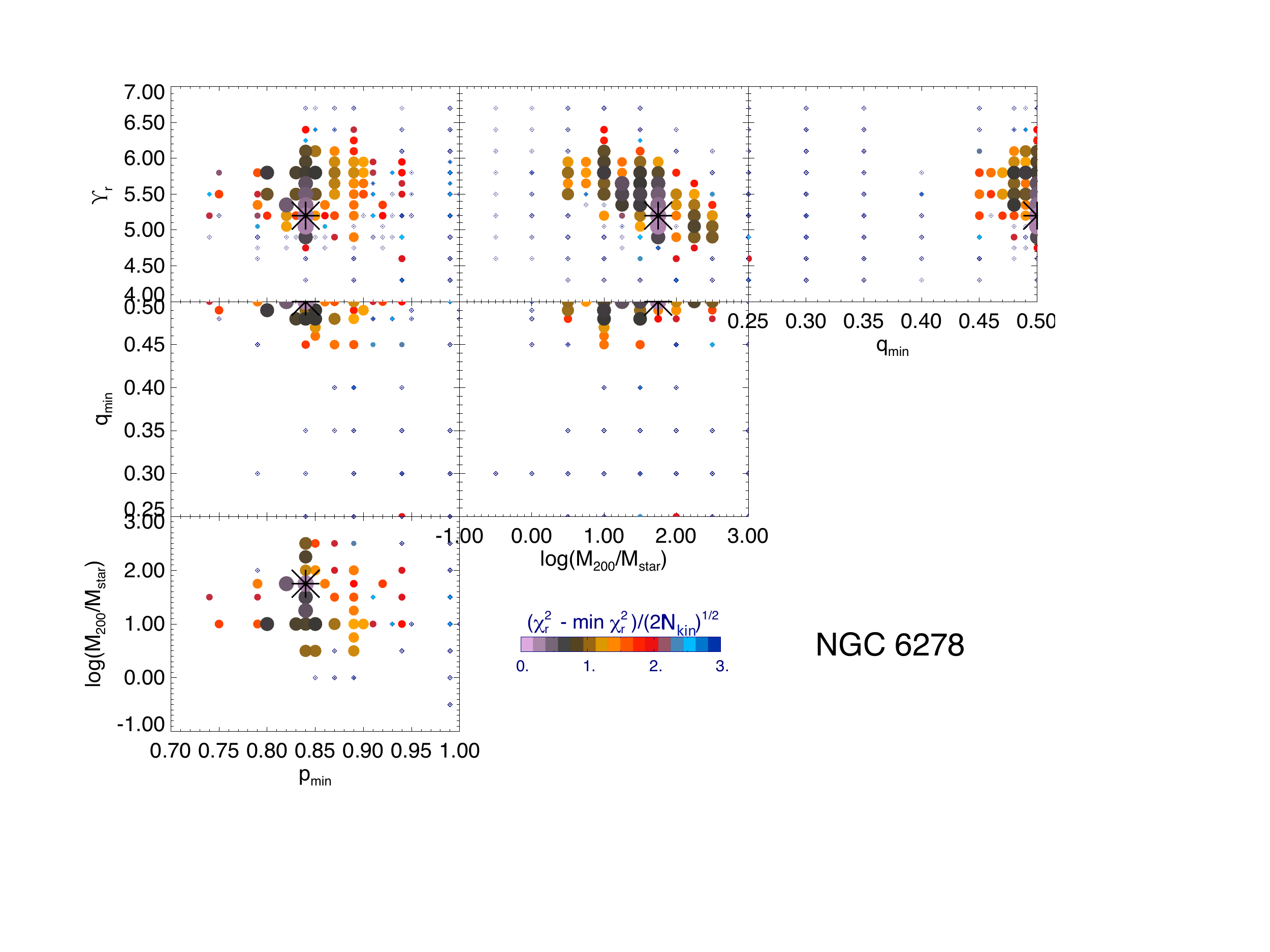}
\caption{Illustration of the parameters grid we span with NGC 6278. The black asterisk indicates the best-fitting model. The colored large dots represent the models within $3\sigma$ confidence, as the color bar indicated, where $\chi_r^2$ are renormalised with $\min (\chi_r^2)/N_{\mathrm{kin}} = 1$. The small dark dots are the models outside $3\sigma$ confidence.}
\label{fig:grid}
\end{figure*}

\begin{figure*}
\centering\includegraphics[width=7.8cm]{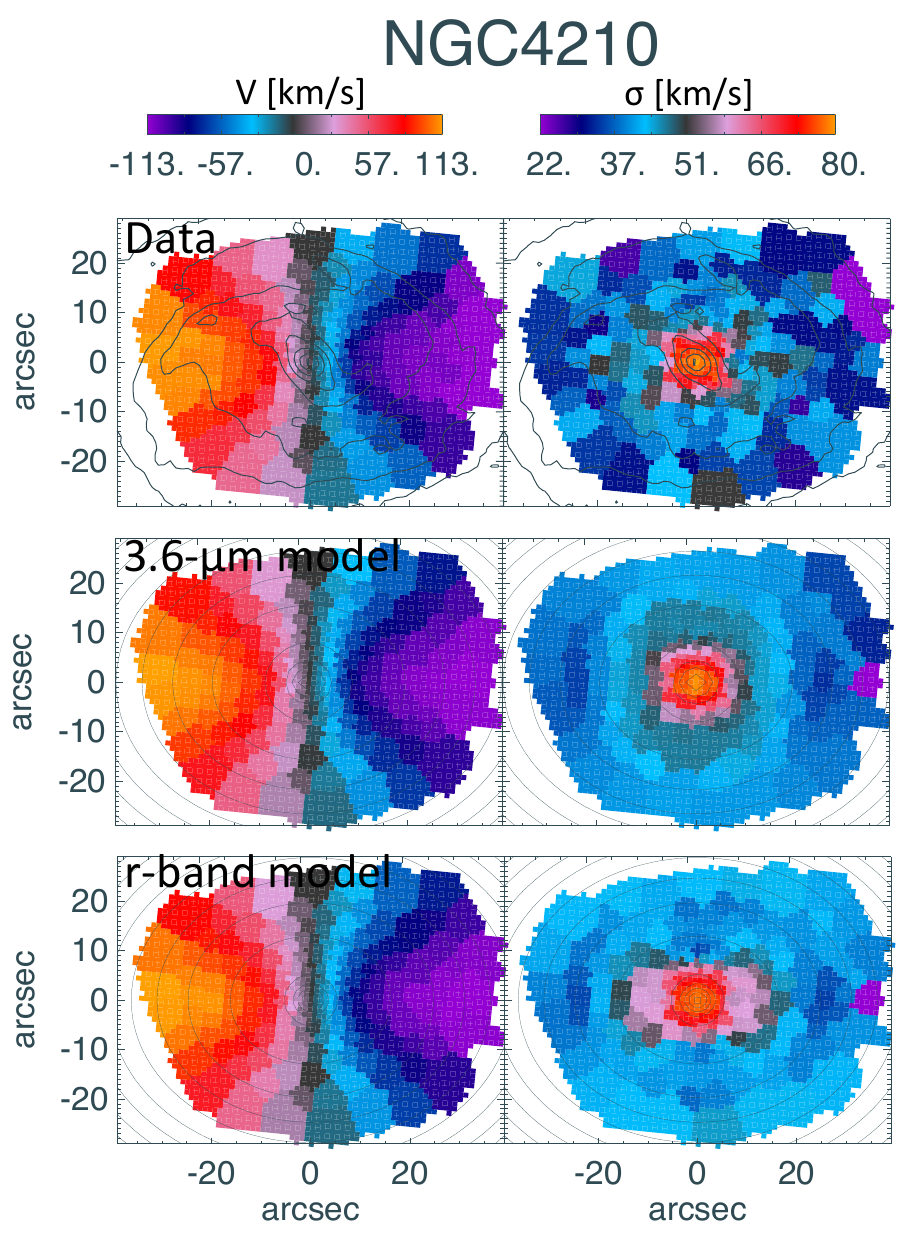}\includegraphics[width=7.8cm]{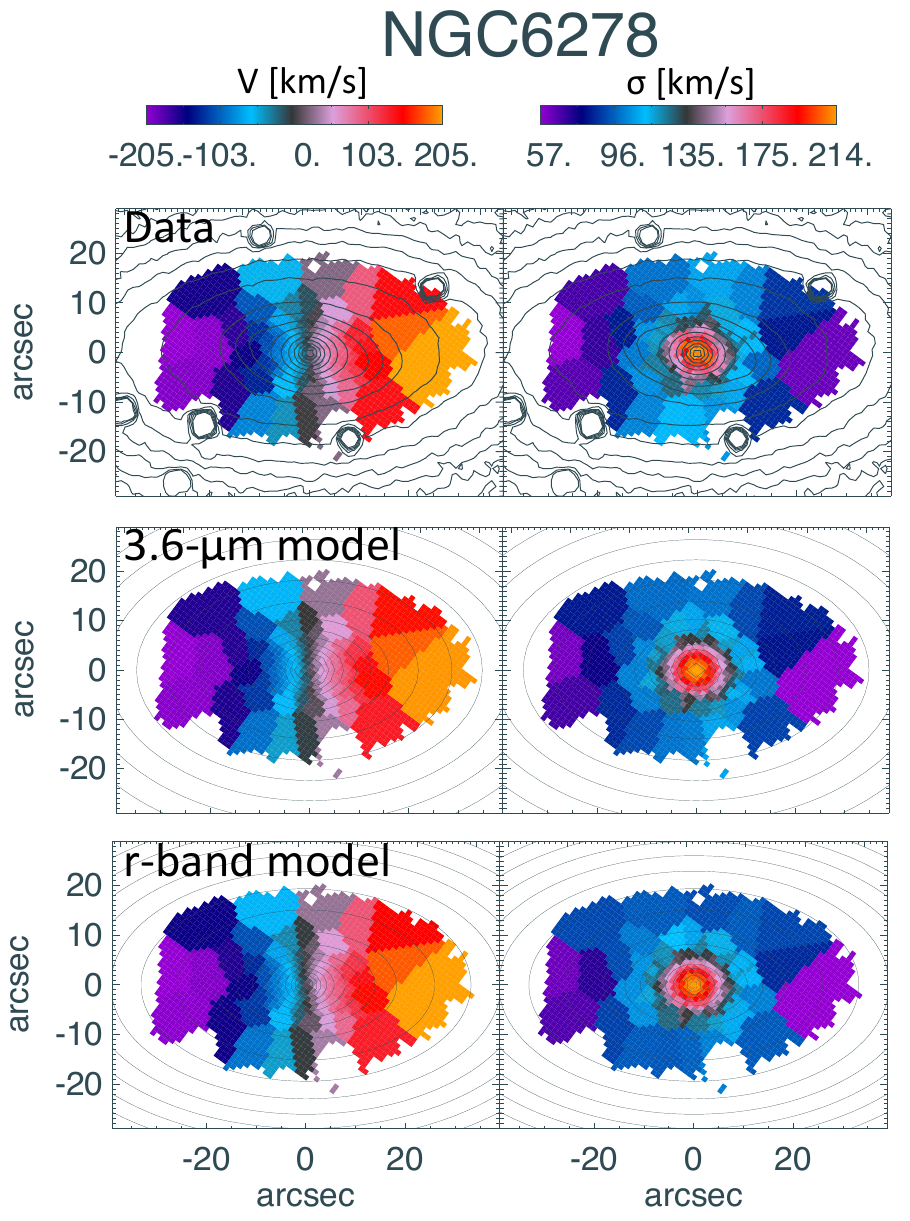}
\caption{The kinematic maps for NGC 4210 on the left and NGC 6278 on the right. For each galaxy, the top panels are the point-symmetrized observed mean velocity (left) and velocity dispersion (right), with the contours of $r$-band image overplotted. The middle panels are the best-fitting 3.6-$\mu$m model, with the corresponding MGE fit of the surface brightness overplotted. The bottom panels are the best-fitting $r$-band model, with the corresponding MGE fit overplotted. The observed and modelled kinematic maps are scaled in the same values as indicated in the colour bars.}
\label{fig:sdss_kin}
\end{figure*}

We construct the \swd models as described in Section~\ref{S:model}. We have five free parameters in the model, the stellar mass-to-light ratio $\Upsilon_{\ast}$, the intrinsic shape of the flattest Gaussian component ($p_{\mathrm{min}}, q_{\mathrm{min}}, u_{\mathrm{min}}$), which represents the space orientation $(\vartheta, \phi, \psi)$, and the DM halo mass $M_{\mathrm{200}}$. 
A central BH is included with the mass fixed as described in Section~\ref{S:model}. We fix $u_{\mathrm{min}} = 0.9999$ to reduce the degeneracy thus only four free parameters left. No specific prior constraints have been applied to $p_{\mathrm{min}}$ and $q_{\mathrm{min}}$. $\Upsilon_{\ast}$ and $\log M_{\mathrm{200}} / M_{\ast}$ are allowed in a wide range: $\Upsilon_{\ast}$ from 0.1 to 10 and $\log M_{\mathrm{200}} /M_{\ast}$ from -3 to 3. 

We adopt a parameter grid with intervals of 0.1, 0.05, 0.05 and 0.5 in $\Upsilon_{\ast}$, $q_{\mathrm{min}}$, $p_{\mathrm{min}}$ and $\log (M_{\mathrm{200}} / M_{\ast})$, and perform iterative process searching for the best-fitting models.
The modelling is started with an initial, then iterating starts after the first models finished. We select the best-fitting models after each iteration by using $\chi^2 - \min (\chi^2) < \chi^2_s$ with $\chi^2_s = 2$, then create new models around, by walking two steps in every direction of the parameter grid from each of the selected models.
In this way, the searching process goes in the direction of smaller $\chi^2$ on the parameter grid, and it stops until the minimum $\chi^2$ model is found. 
Then we continue the iteration by using larger $\chi^2_s$, ensures all the models within $1\sigma$ confidence are calculated before the iteration finishes. The values of $\chi^2_s$ are chosen 
empirically: it is neither too small that preventing the models stucking at local minimum, nor too large that keeping the searching process efficient enough. Finally, we reduce the parameter intervals by half and find the more precise position for the best-fitting parameters. 

As mentioned in Section~\ref{SS:GH1}, The luminosity distribution can almost always be reproduced up to numerical precision (\citealt{vdB2008}; \citealt{vanderMarel1998}; \citealt{Poon2002} and is thus not
relevant for finding the best-fitting solution. We only use $\chi^2_{\mathrm{kin}}$ to estimate the statistical uncertainties.  

As shown in Table~\ref{tab:ML}, $\min(\chi^2_{\mathrm{kin}}) / N_{\mathrm{kin}}$ of $ \sim 0.7$ \footnote{Ignoring the
orbital weights, we only have four free parameters regarding to gravitational potential and orientation of the galaxies in our model. which is much smaller than $N_{\mathrm{kin}}$, thus $N_{\mathrm{kin}}$ is used as the number of freedom for each model.} 
is obtained for the best-fitting models. $N_{\mathrm{kin}}$ is the total number of Voronoi-binning kinematic data (the number of apertures times the number of GH moments). 
The kinematic data are point-symmetrized before the modelling; the data in different bins are not independent. 
We introduce a normalised $\chi_r^2 = \chi^2_{\mathrm{kin}} N_{\mathrm{kin}}/\min(\chi^2_{\mathrm{kin}})$ to ensure $\min (\chi_r^2) / N_{\mathrm{kin}} = 1$, as expected for models constrained by independent data points.



$\chi^2_r$ ($\chi^2_{\mathrm{kin}}$) fluctuate significantly in \swd models, with a standard deviation of $\sim \sqrt{2N_{\mathrm{kin}}}$, which enlarges the model confidence level ({\citealt{Thomas2005}, \citealt{Morganti2013}). We use $\Delta\chi^2_r = \sqrt{2N_{\mathrm{kin}}}$ as our model's $1\sigma$ confidence level, and qualify its statistical meaning in Appendix~\ref{S:simulation} when applying the model to mock data sets.

The parameters space we span and the best-fitting modelling for NGC 6278 are illustrated in Figure~\ref{fig:grid}. The colored dotted represents all the models within $3\sigma$ confidence interval, while the black dots represent that outside. $q_{min}$ of the best models hit the boundary set by the flattest Gaussian component with $q'_{min}$, which just indicates that NGC 6278 prefers the models near edge-on. 
In the end, we typically get a few hundreds of models for each set, with 10-50 models within the $1\sigma$ confidence intervals.

The parameters of the best-fitting models of these two galaxies are summarized in Table~\ref{tab:ML}, the inclination angle $\vartheta$ of the system has been inferred from the intrinsic shape of the flattest Gaussian $(q_{\mathrm{min}},p_{\mathrm{min}})$, while $\overline{q}$, $\overline{p}$ represents the intrinsic shape of our model measured at $2R_e$. The total DM mass $M_{\mathrm{200}}$ is hard to constrain due to the limited data coverage and the DM vs. luminous matter degeneracy, so that we show the constraint on DM mass within $R_e$ and $2\,R_e$, $M_{\mathrm{dm}}(<R_e)$ and $M_{\mathrm{dm}}(<2R_e)$, instead. 

Most of the analysis below is based on the best-fitting models, while errors of the parameters are calculated using the models within $1\sigma$ confidence intervals.

The best-fitting models provide good fits to the surface brightness and kinematic maps for the two galaxies as shown in Figure~\ref{fig:sdss_kin}. The top panels are the observed mean velocities and velocity dispersions, the middle panels are the best-fitting 3.6-$\mu$m models, and the bottom panels are the best-fitting $r$-band models. 
The two sets of models fit the date equally well; $\chi_{\mathrm{kin}}^2/N_{\mathrm{kin}}$ of the best-fitting $r$-band model and the 3.6-$\mu$m model are similar. 
 
\begin{table*}
\caption{The best-fitting parameters for NGC 4210 and NGC 6278 using images in $r$-band and $3.6-\mu$m, respectively. The second column gives the stellar mass-to-light ratio $\Upsilon_{\ast, \mathrm{Chab}}$ from stellar population synthesis assuming Chabrier IMF from \citet{Meidt2014} and \citet{Walcher2014}, the remaining columns show the parameters obtained from the best-fitting \swd models: the stellar mass-to-light ratio $\Upsilon_{\ast}$, inclination angle $\vartheta$, the intrinsic shape of the model $\overline{q}$ and $\overline{p}$ measured at $2\,R_e$, $\overline{u}$ fixed at 0.9999, the DM mass $M_{dm}(<R_e)\,[10^{10} \,M_{sun}]$,  $M_{dm}(<2R_e)\,[10^{10} \,M_{sun}]$ and $\min(\chi_{\mathrm{kin}}^2)/N_{\mathrm{kin}}$ from our best-fitting models. }
\label{tab:ML}
\begin{tabular}{llllllllll}
\hline
 -              &  $\Upsilon_{\ast,\mathrm{Chab}}$    &  $\Upsilon_{\ast}$   &   $\vartheta$   & $\overline{q} (2\,R_e)$ & $\overline{p} (2\,R_e)$ & $\overline{u}$ &  $M_{dm}(<R_e)$  &  $M_{dm}(<2R_e)$  & $\min(\chi^2_{\mathrm{kin}})/N_{\mathrm{kin}}$ \\
\hline
 N4210 &&&&& \\
r-band      &  1.1 & $3.8\pm0.5$    &  $(42\pm1)^o$  &    $0.26_{-0.14}^{+0.1}$   &$1.0_{-0.02}^{0.0}$ & 0.9999   & $0.6_{-0.3}^{0.1}$    &  $1.6_{-0.8}^{+0.6}$    &   0.74    \\
3.6-$\mu$m   & 0.6  &    $1.6\pm0.2$     &  $(42\pm1)^o$   &    $0.26_{-0.14}^{+0.1}$ & $1.0_{-0.02}^{0.0}$ & 0.9999   & $0.5_{-0.2}^{0.3}$     &   $1.0_{-0.3}^{+1.3}$   &   0.76    \\
 \\
  N6278 &&&&& \\
r-band    &  2.9  &  $5.5\pm0.3$    &  $(83\pm10)^o$    &    $0.53_{-0.02}^{+0.00}$ &$0.85\pm0.05$ & 0.9999   &  $0.9_{-0.6}^{+0.5}    $ &$3.4_{-2.0}^{+2.0}$     & 0.60   \\
3.6-$\mu$m  & 0.6  &     $1.0\pm0.1$    &  $(79\pm10)^o$    &    $0.50_{-0.02}^{+0.01}$& $0.95\pm0.05$ & 0.9999  & $0.5_{-0.2}^{0.2}       $ &  $2.0_{-1.3}^{+1.0}$   & 0.70   \\

  \hline
 \end{tabular}
\end{table*}

\subsection{The cumulative mass profiles}
\label{SS:califa_mass}
Figure~\ref{fig:mass} shows the cumulative mass profiles for the two galaxies. The mass profiles obtained by the two sets of models, $r$-band model and 3.6-$\mu$m model, are shown in solid and dashed lines respectively. The black, red and blue curves represent the enclosed total mass, stellar mass and dark matter mass profiles, respectively. 

The total mass profiles are well constrained with statistic $1\sigma$ uncertainties of $\sim 10\%$ within $2\,R_e$. The total mass obtained by the two sets of models are consistent with each other within $1\sigma$ errors for both galaxies. 

There is significant degeneracy between the contribution of stellar mass and dark matter mass, which causes $\sim 20\%$ uncertainties on the stellar mass, and $\sim 50\%$ uncertainties on the dark matter mass within $2\,R_e$. 
The $1\sigma$ errorbars of the stellar mass (or DM mass) obtained from the two sets of models can overlap, but are not so large as to overlap with the median value of the other model; an increase in the error bar by a factor of $\sim 2$ would lead to greater overlap and thus statistical similarity of stellar mass (or DM mass) obtained by $r$-band model and 
3.6-$\mu$m model.

Considering the statistical errorbars and the difference of $r$-band model and 
3.6-$\mu$m model, from $R_e$ to $2\,R_e$, the DM fraction increases from $14 \pm 10 \%$ to $18 \pm 10\%$ for NGC 4210 and from $15 \pm 10 \%$ to $30 \pm 20 \%$ for NGC 6278. 

The flux ratio of 3.6 $\mu$m to $r$-band image is higher in the inner 10 arcsec for these two galaxies (see Figure~\ref{fig:mge}). If the stellar mass-to-light ratio is constant in 3.6 $\mu$m, we thus expect the stellar mass-to-light ratio in the $r$-band to be $\sim10\%$ higher in the inner regions than that in the outer regions.
In our model, we assumed constant $\Upsilon_r$ as well as constant $\Upsilon_{3.6}$. 
The possibly higher $\Upsilon_r$ in the inner regions is compensated by including different combinations of DM mass and luminous mass in the $r$-band models.
 
The assumption of constant $\Upsilon_r$ affects the estimate of DM and stellar matter separately, 
thus we expect our imperfect stellar mass model to lead to systematic errors in both the DM and stellar masses. As seen in Figure \ref{fig:mass}, the formal errors on the DM mass and the stellar mass could be larger than their pure statistical errors by a factor of $\sim 2$. 
However, constant $\Upsilon_r$ does not systematically affect the total mass profile within the data coverage, thus does not affect our estimates of the internal dynamics as we show in Section~\ref{SS:califa_anisotropy} and Section~\ref{SS:orbital}.  

\begin{figure*}
\centering\includegraphics[width=7.5cm]{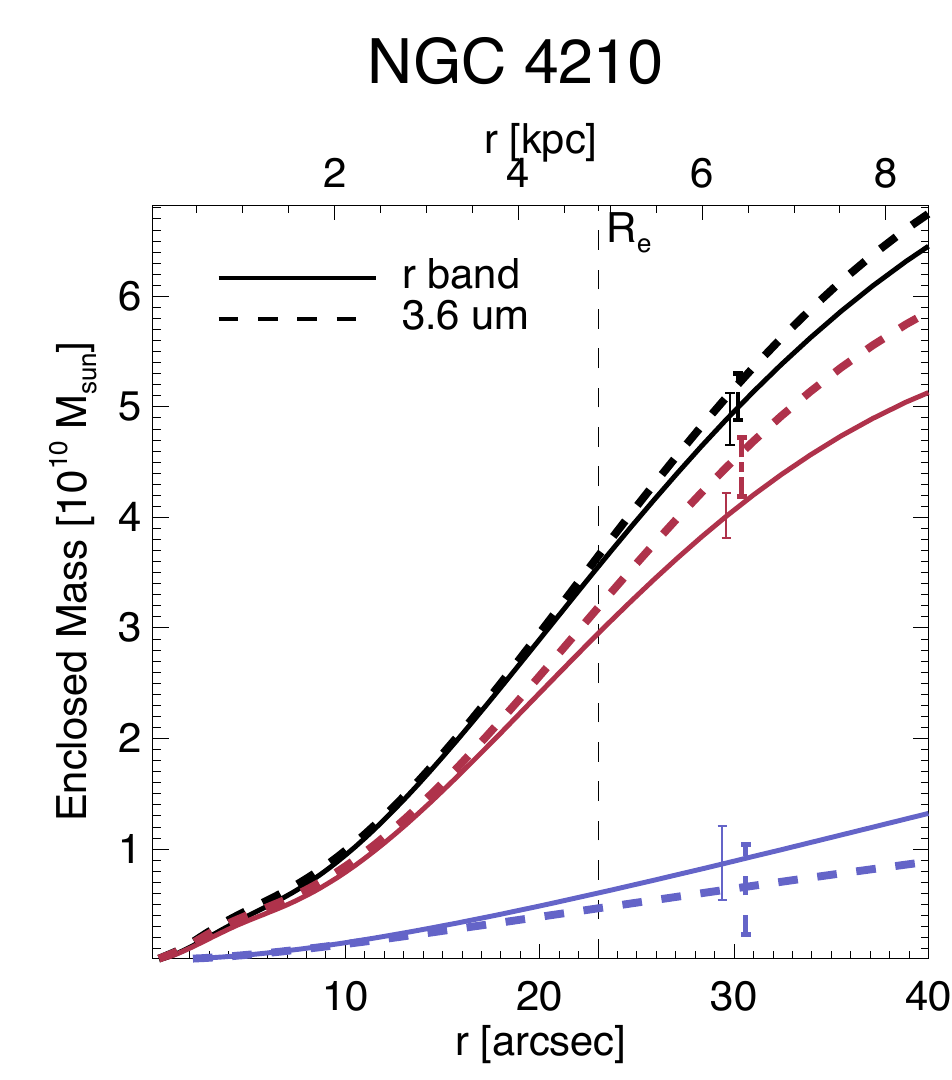}\includegraphics[width=7.5cm]{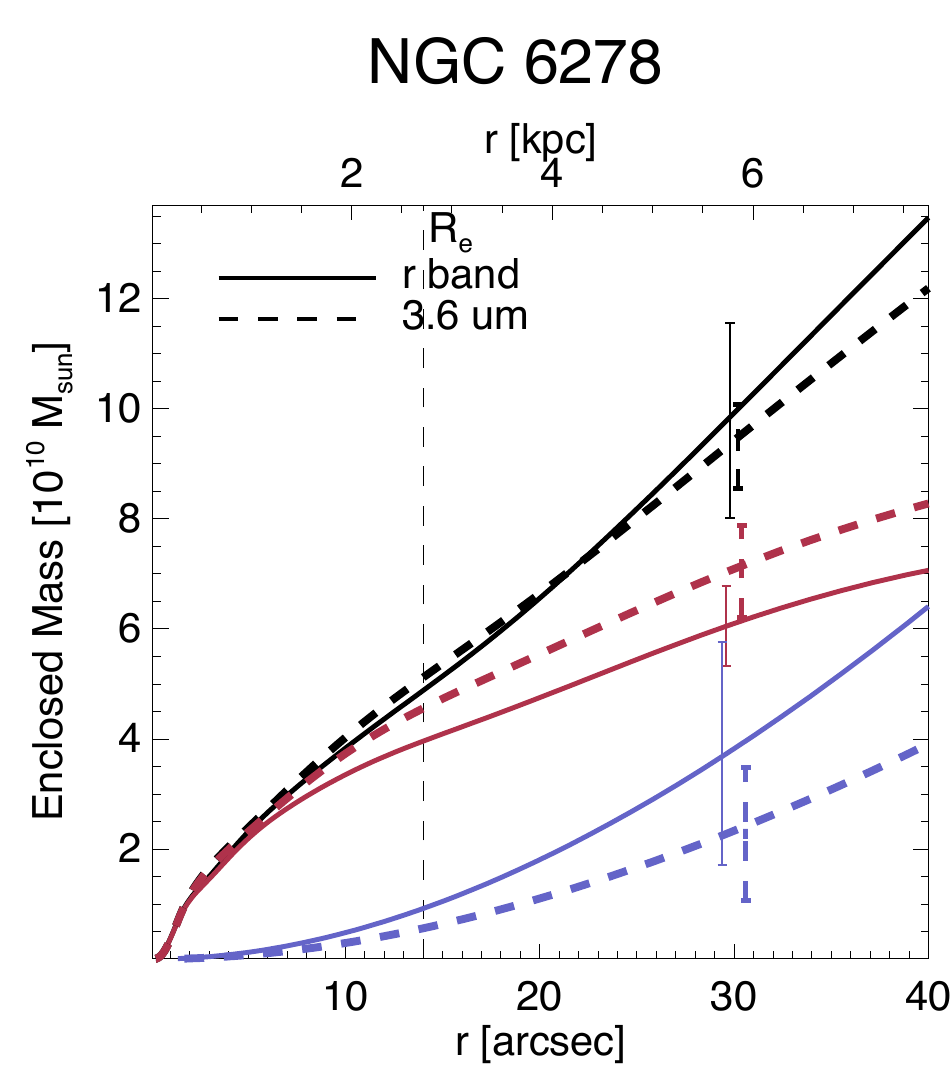}
\caption{The cumulative mass profiles of the two galaxies; NGC 4210 on the left and NGC 6278 on the right. 
The black, red and blue curves represent the mass profiles of total mass, stellar mass and dark matter, respectively (solid curves: $r$-band models -- dashed curves: 3.6-$\mu$m models).
Error bars at $r=30''$ indicate the $1\,\sigma$ error of the mass profiles at that point.  The vertical dashed line represent the position of $1\,R_e$.}
\label{fig:mass}
\end{figure*}

The $\Upsilon_r$ and $\Upsilon_{3.6}$ we obtained for the best-fitting models are listed in Table~\ref{tab:ML}.  
By assuming Chabrier initial mass function (IMF), the stellar population synthesis gives $\Upsilon_{r, \mathrm{Chab}} = 1.1$ for NGC 4210 and $\Upsilon_{r, \mathrm{Chab}} =2.9 $ for NGC 6278 \citep{Walcher2014}, and average $\Upsilon_{3.6, \mathrm{Chab}} = 0.6$ \citep{Meidt2014} for all galaxies regardless of their age and metallicity. 
The dynamical stellar mass-to-light ratio is a few times higher ($\sim3.0$ for NGC 4210, and $\sim1.7$ for NGC 6278) from stellar population systhesis with Chabrier IMF.

The dynamical stellar mass-to-light ratio $\Upsilon_{\ast}$ is reliable regarding to its independence of the stellar age, metallicity, star formation history and IMF, while all these factors affect the estimate of stellar mass-to-light ratio from stellar population synthesis. For the S0 galaxy NGC 6278, a higher stellar mass-to-light ratio could be inferred from a more bottom-heavy IMF \citep{Cappellari2013}.
While for the SBb galaxy NGC 4210, the discrepancy between $\Upsilon_{\ast}$ and $\Upsilon_{\ast, \mathrm{Chab}}$ could be partly caused by the content of gas in the disk plane \citep{Huang2012} and partly the old stellar population of this galaxy (\citealt{Sanchez2014},  \citealt{Meidt2014}). 



\subsection{The velocity anisotropy}
\label{SS:califa_anisotropy}
\begin{figure*}
\centering\includegraphics[width=8cm]{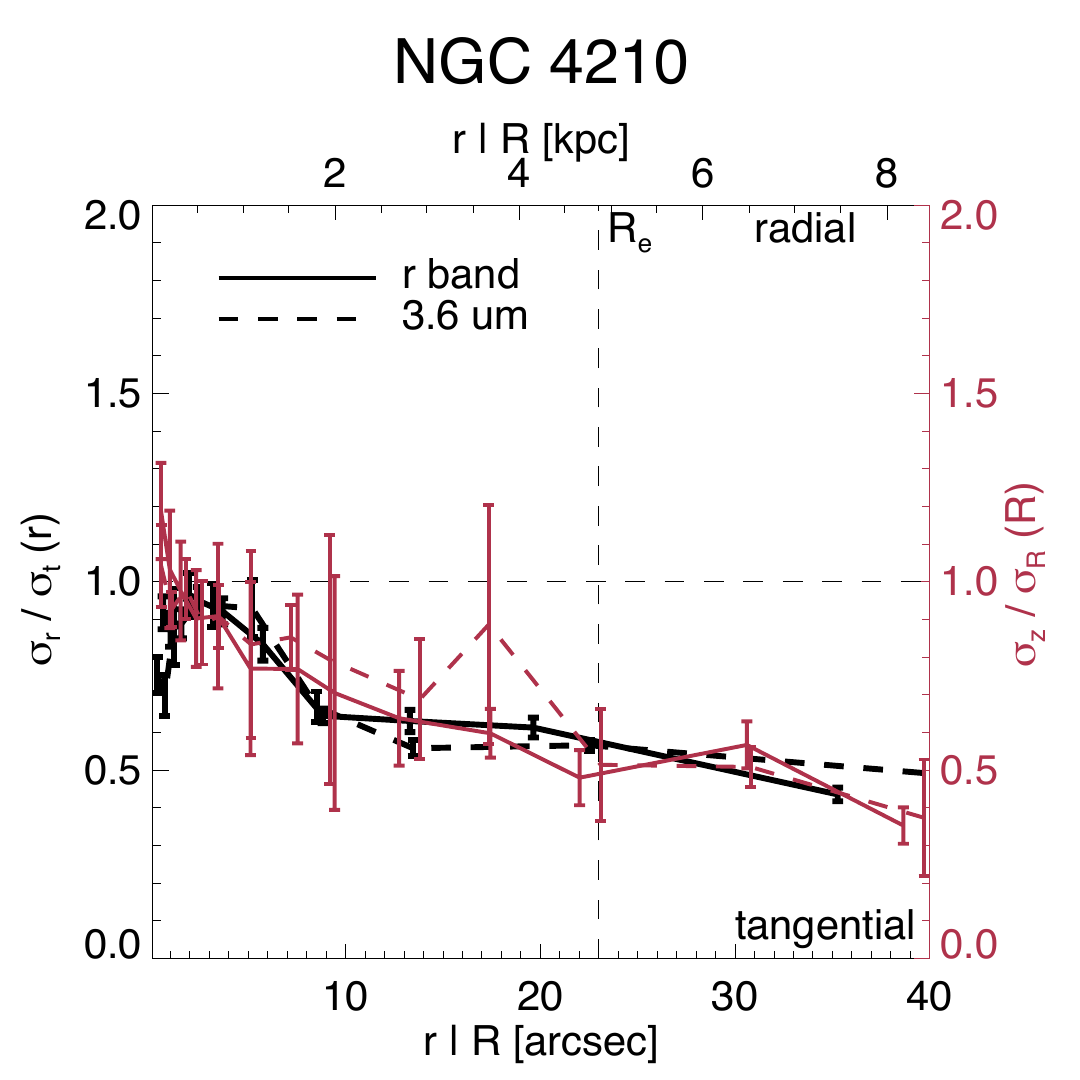}\includegraphics[width=8cm]{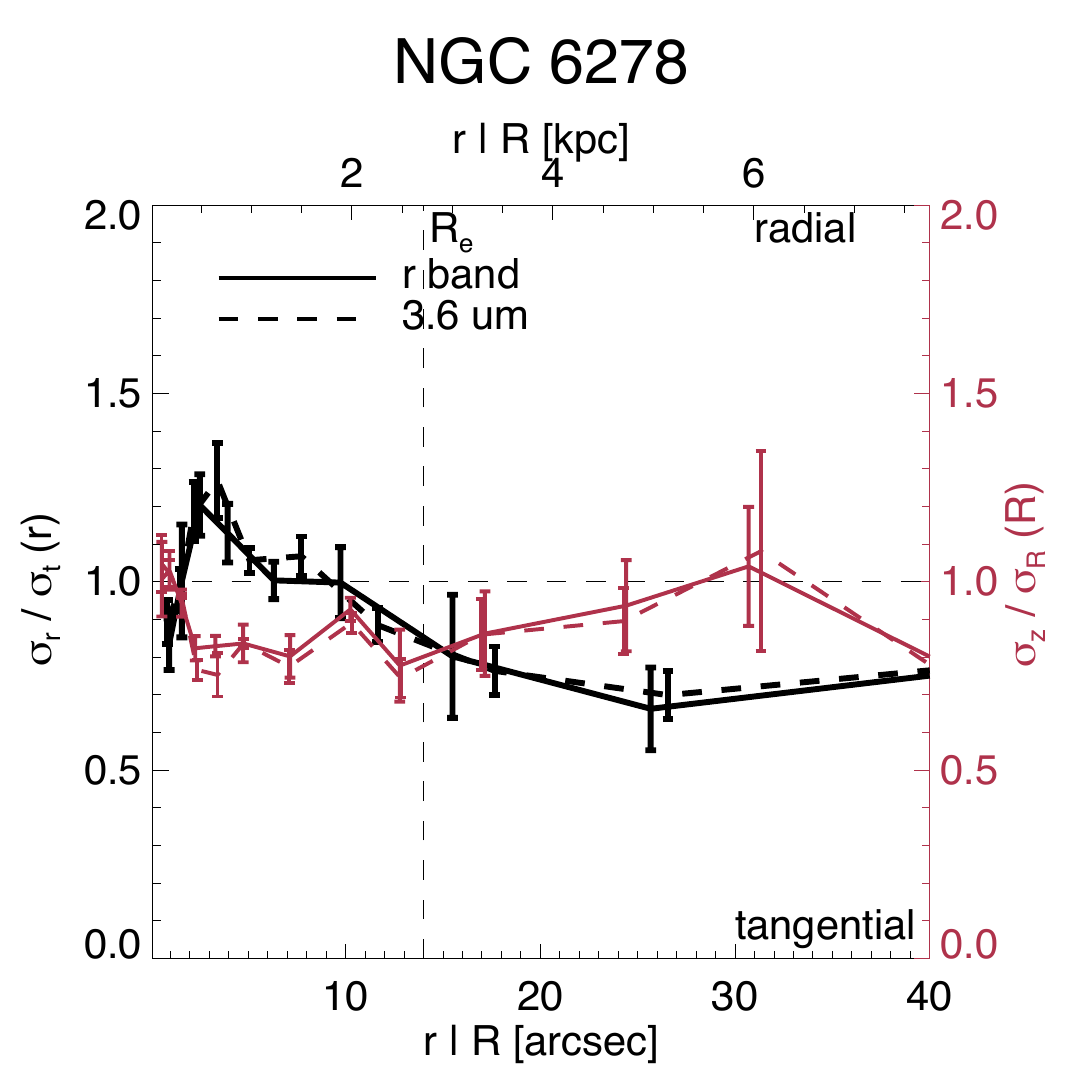}
\caption{The velocity anisotropy profiles $\sigma_r/\sigma_t$ as a function of the intrinsic radius $r$ in black, and $\sigma_z/\sigma_R$ as a function of $R$ on the disk plane in red. The solid curves represent velocity anisotropy profiles obtained by $r$-band models. The dashed curves represent those obtained by 3.6-$\mu$m models. The error bars indicate the scatters among models within $1\,\sigma$ confidence. The vertical dashed line represents the position of $1\,R_e$.}
\label{fig:anisotropy}
\end{figure*}

The internal dynamical properties of the galaxy can be investigated via the best-fitting models.
We show both the velocity anisotropy profiles in a spherical coordinate and that in a cylindrical coordinate in Figure~\ref{fig:anisotropy}.

$\sigma_r/\sigma_t$ in black is plotted along the intrinsic radius $r$, where $\sigma_t = \sqrt{ (\sigma_{\phi}^2 + \sigma_{\theta}^2)/2 }$; $\sigma_r$, $\sigma_{\phi}$ and $\sigma_{\theta}$ are the radial, azimuthal angular and polar angular velocity dispersion in a spherical coordinates. 
$\sigma_z/\sigma_R$ in red is plotted along the radius $R$ on the disk plane; $\sigma_z$ and $\sigma_R$ are the vetical and radial velocity dispersions in a cylindrical coordinates. 
We discuss  $\sigma_r/\sigma_t$ and $\sigma_z/\sigma_R$ separately.

The solid and dashed curves represent that obtained by $r$-band and 3.6-$\mu$m models, respectively. The error bars indicate the scatters among models within $1\,\sigma$ confidence intervals. The velocity anisotropy profiles from the two sets of models are consistent with each other.

$\sigma_r/\sigma_t$ is the velocity anisotropy widely used for early type galaxies, a value of unit indicates isotropic, a value larger (smaller) than unit indicates radially (tangentially) anisotropic.
NGC 4210 is close to isotropic with $\sigma_r/\sigma_t \sim 0.9$ in the inner $\sim 0.5\,R_e$, and becomes strongly tangentially anisotropic with $\sigma_r/\sigma_{t} \sim 0.5$ in the outer regions.  
NGC 6278 is radially anisotropic within $1\,R_e$, and also gets to be tangentially anisotropic in the outer regions. 
$\sigma_r/\sigma_t$ is a good indicator of the underlying orbital distribution, the more radial anisotropic in the inner regions indicates the existence of dynamical hot orbits as we show in Section~\ref{SS:bd}.

$\sigma_z/\sigma_R$ has been used as an indicator of different heating processes in the spiral galaxies. 
NGC 4210 is a typical Sb galaxy,  its $\sigma_z/\sigma_R$ decreases with radius and reaches $\sim 0.5$ in the outer regions, which is similar to $\sigma_z/\sigma_r = 0.5$ of the Milky Way (\citealt{Dehnen1998}; \citealt{Smith2012}).
For NGC 6278 as an S0 galaxy,  $\sigma_z/\sigma_R$ is nearly constant  with values close to unit along radius $R$.
NGC 6278 is near-spheroidal, thus binning the data along $R$ on the disk plane is not efficient to show the difference from the inner to outer regions.
The $\sigma_z/\sigma_R$ of these two galaxies we obtained are consistent with the variation of $\sigma_z/\sigma_R$ across Hubble sequence \citep{Gerssen2012}.

\subsection{Orbit distributions} 
\label{SS:orbital}
\subsubsection{The orbit distributions}
\label{SS:orbits}

\begin{figure*}
\centering\includegraphics[width=8cm]{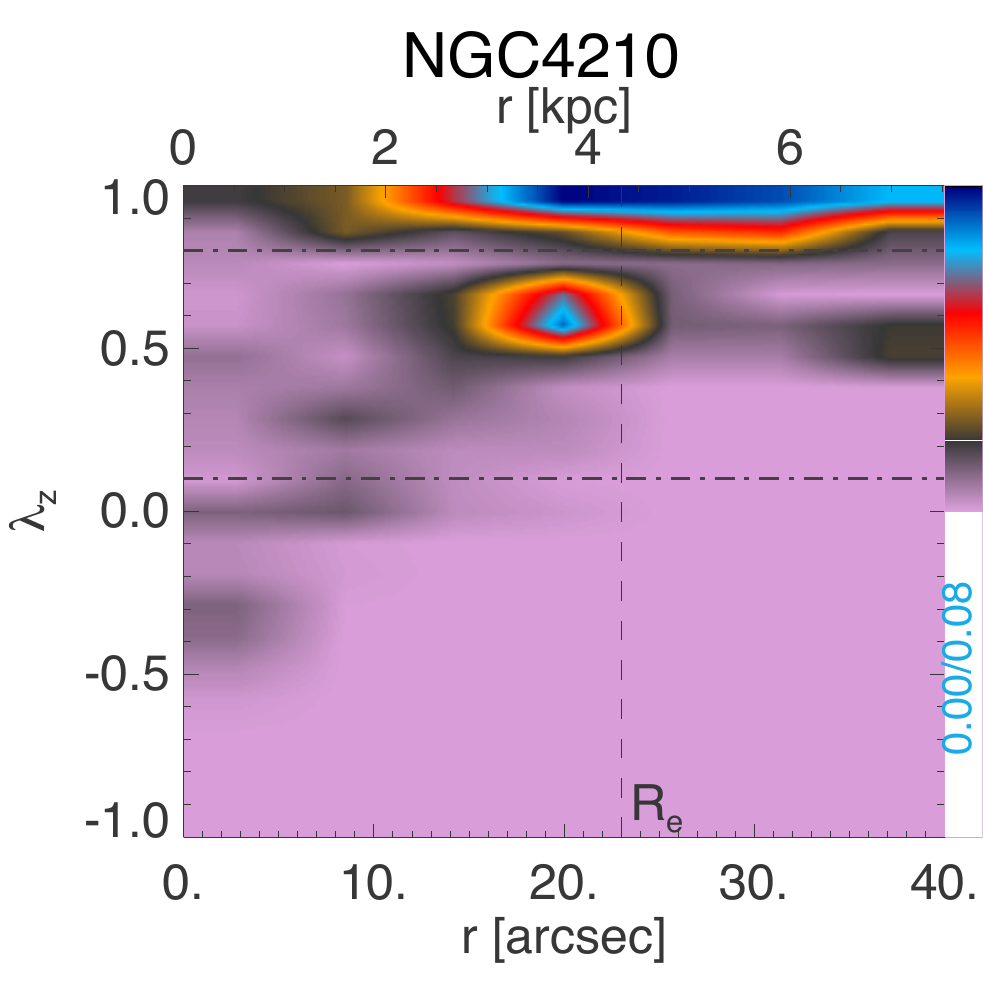}\includegraphics[width=8cm]{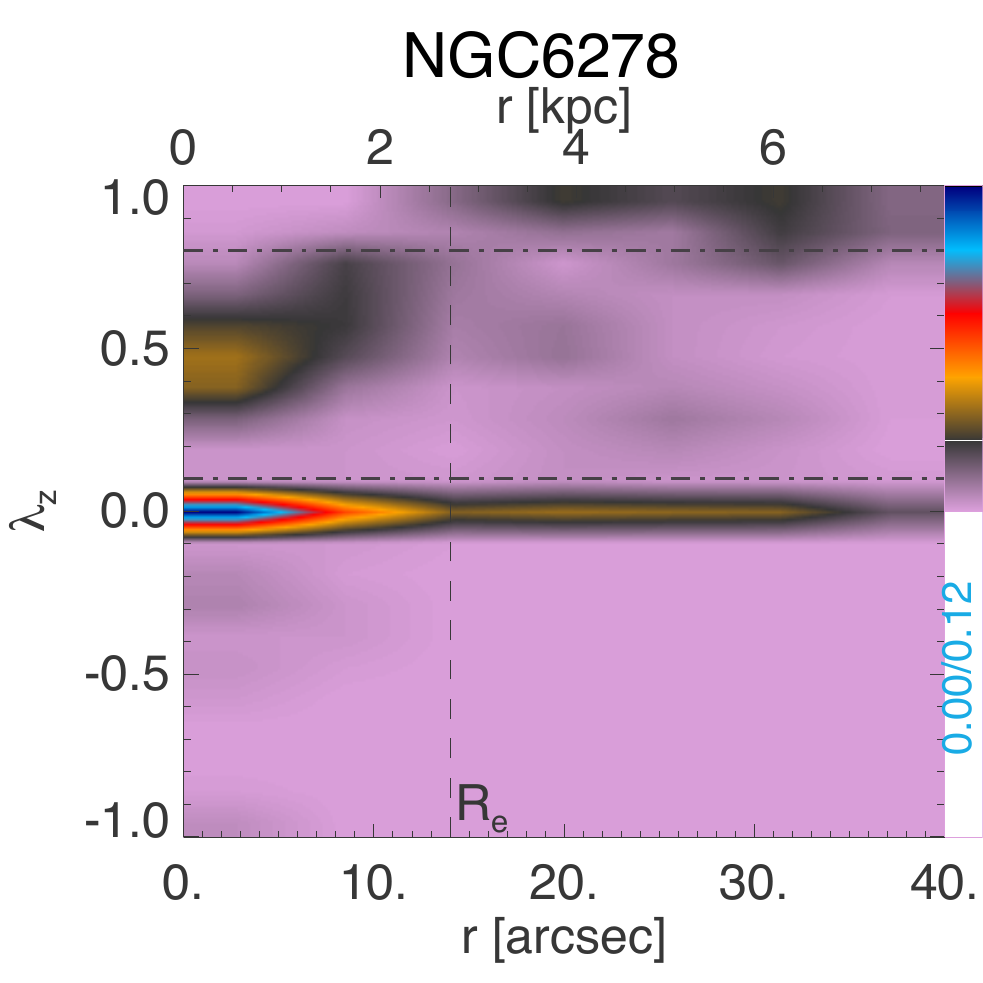}
\caption{The orbit distribution on the phase space of  circularity $\lambda_z$ versus intrinsic radius $r$ of the best-fitting $r$-band models for NGC 4210 (left) and NGC 6278 (right). 
The colour indicates the density of the orbits on the phase space, the two black dot-dashed lines indicate $\lambda_z = 0.8$ and $\lambda_z = 0.1$. The vertical dashed line represent the position of $1\,R_e$. }  
\label{fig:orbit}
\end{figure*}

We use the circularity $\lambda_z$ to indicate different orbit types:
\begin{equation}
\label{eqn:lz}
  \lambda_z = \overline{L_z} / (r \times \overline{V_{c}}),
\end{equation}
where $\overline{L_z} = \overline{xv_y - yv_x}$, $r = \overline{\sqrt{x^2 + y^2 + z^2}}$, and $\overline{V_{c}} = \sqrt{\overline{v_x^2 + v_y^2 + v_z^2 + 2v_x v_y + 2v_x v_z + 2 v_y v_z}}$, taken the average of the points $(x, y, z, v_x, v_y, v_z)$ saved with equal time step for that orbit.
Notice that $V_c$ here is defined as the summation of all the elements of the second moments matrix, representing $V_{\rm rms}$ with the cross elements included.

$|\lambda_z| =1$ indicates a circular orbit, while $\lambda_z = 0$ indicates a box or radial orbit.

Taken the radius $r$ and circularity $\lambda_z$ of each orbit, and considering their weights given by the solution from the best-fitting model, we get the orbit distribution on the phase space from the best-fitting $r$-band models as shown in Figure~\ref{fig:orbit}.  
The phase space have been divided into $7 \times 21$ bins, the colors indicate the total weights of orbit in each bin. The orbit distributions in the best-fitting 3.6-$\mu$m models (not shown) are similar. 

The orbit distributions on the phase space have clear structures, which suggests different formation history of the stars in different regions. In the inner regions, hot orbits dominate. There are also some counter-rotating orbits in the inner regions of NGC 4210, which may contribute to a bulge or a bar. 
A bar could be counter-rotating according to the disk (e.g., \citealt{Jung2016}), although we neglect the non-axisymmetry of bar in the image of NGC 4210, our models are still trying to fit all the kinematic features of the galaxy, including that induced by the bar. However, the counter-rotating orbits in our models are just regular orbits that produce similar kinematic features, not the real orbital structures of the bar because we do not include figure rotations as those dynamical models focusing on the bar (\citealt{Portail2016};\citealt{Vasiliev2015}; \citealt{Long2013}; \citealt{Wang2013}).

In the outer regions, the contribution of dynamical cold orbits increase and they are the dominant components in the outermost regions for these two galaxies. 

we make cuts at the two dips to separate the orbits into three components. We identify a cold component with the circular orbits ($\lambda_z>0.8$) and a hot component with the radial orbits ($\lambda_z<0.1$), while the orbits in between construct a warm component. This choice is consistent with the dynamical decomposition for some of simulated spiral galaxies \citep{Abadi2003}.

\subsubsection{The morphology and kinematics of the three components}
\label{SS:bd}

\begin{figure*}
\centering\includegraphics[width=8.6cm]{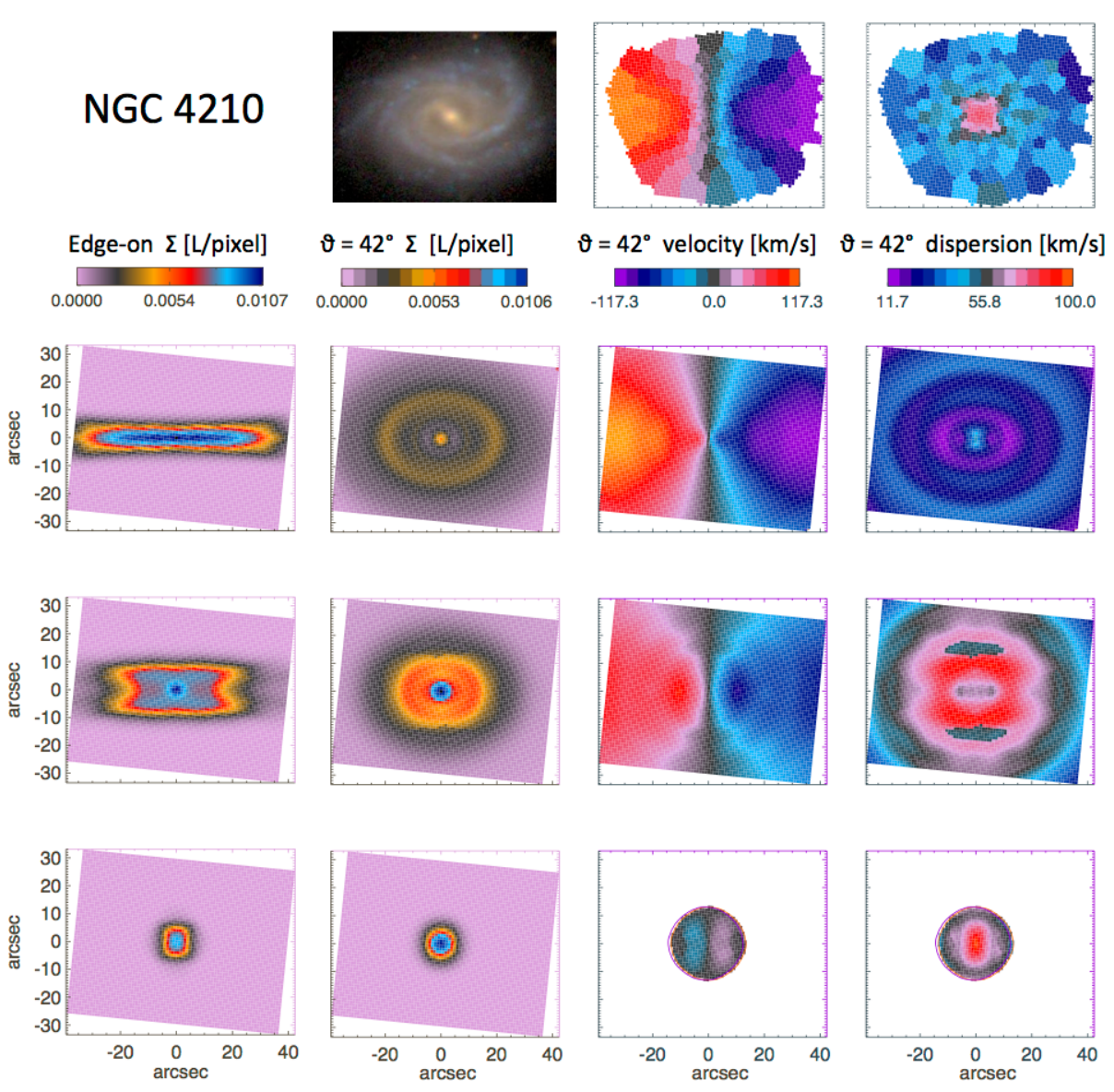}\includegraphics[width=8.3cm]{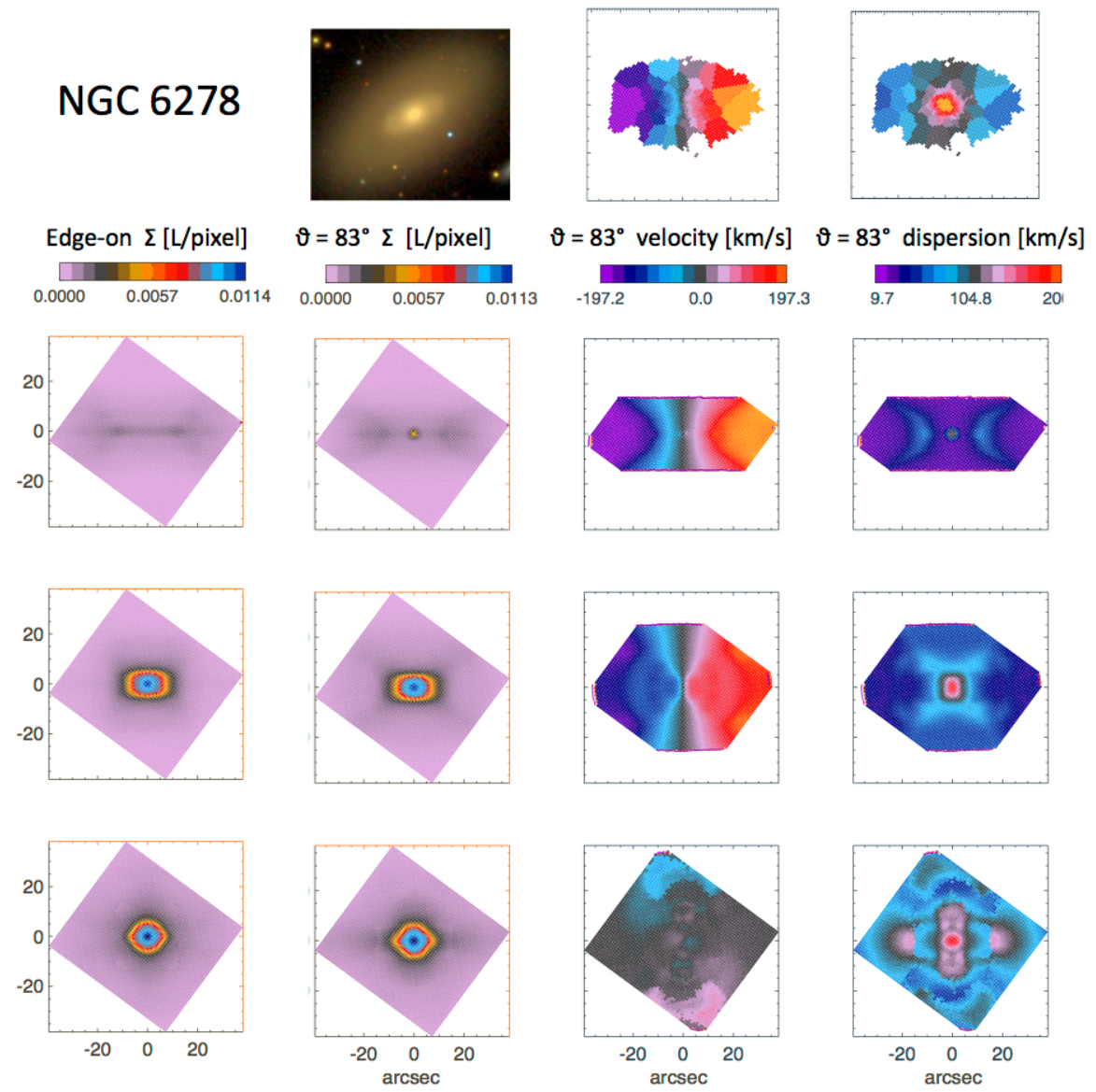}
\caption{The morphology and kinematics of the three components for NGC 4210 (left) and NGC 6278 (right). The first row shows a SDSS image and the CALIFA data of each galaxy, the following three rows represent the cold, warm and hot components rebuilt with each part of the orbits, with decreasing circularity from top to bottom. For each component, the four columns are the flux projected edge-on, and then flux, mean velocity and velocity dispersion projected with inclination angle $\vartheta$, from left to right. The total luminosity of the galaxy (three components together) is normalized to be unit. The mean velocity and velocity dispersion of one component is made to be white where the contribution of that component is neglectable.}  
\label{fig:hmzii}
\end{figure*}

We rebuilt the three components with the corresponding orbits as shown in Figure~\ref{fig:hmzii}. 
The first row shows a SDSS image and the CALIFA data of the galaxy.  The following three rows represent the cold, warm and hot component from top to bottom.
From left to right, the four columns are the flux projected edge-on, and then the flux, mean velocity and velocity dispersion projected with inclination angle $\vartheta$.

Figure~\ref{fig:hmzii} shows that the three dynamical components have well-defined and distinct morphologies and kinematic properties. 
For NGC 4210, the cold component is geometrically thin and dynamically cold with $V/\sigma = 4.98$. The warm component is thicker and with $V/\sigma = 1.58$. The hot component is round and concentrated, it has $V/\sigma = -0.44$ and contributes to the high dispersion in the central regions. Note that negative $V/\sigma$ of the hot component could be just caused by the hard cut in $\lambda_z$, it does not necessary indicate a counter-rotating bulge. 
NGC 6278 has similar three components in morphology and kinematics with $V/\sigma$ of 4.36, 1.07 and 0.01, respectively. 

By perturbing the kinematic data, we find that the uncertainty of luminosity fraction of each component in a single model (with fixed potential and orientation) is $< 5\%$. 
The variation of luminosity fraction of each component among models within $1\sigma$ confidence interval is $\sim 20\%$. 
Considering the $1\sigma$ variation, the luminosity fractions of the three components are $0.49\pm 0.10$, $0.45\pm0.10$ and $0.06\pm0.03$ for NGC 4210 and $0.12\pm0.06$, $0.39\pm0.10$ and $0.49\pm0.10$ for NGC 6278, respectively.

The fine structures, e.g., the X-shape in the flux map of the warm component of NGC 4210, or the ring-like structures in the velocity dispersion maps, are likely to be caused by hard cuts on $\lambda_z$.

\section{Comparison orbital versus photometric decomposition}
\label{S:discussion}
The cold, warm and hot components are separated purely dynamically, and they are not necessary to match any morphological structures. However, they do show either a disk-like or bulge-like morphology. 
In Figure~\ref{fig:phot}, we compare the projected surface brightness of the three components with the results from photometric decomposition. In this section, disk and bulge represent particularly the photometrically-identified disk-like and bulge-like structures.

The black and red dotted lines represent an exponential disk and a sersic bulge from the photometric decomposition by \cite{Mendez2017}, they also include an elongated bar which only contribute a small fraction of the luminosity and we do not show it in the figure. 
The surface brightness of the orbital decomposed cold, warm and hot components are shown with blue, orange and red solid lines, while the black solid line represents the combination of cold and warm components. The three dynamical components are only constrained within the coverage of the kinematic maps, so we only compare the surface brightness profiles within 40 arcsec for NGC 4210 and within 30 arcsec for NGC 6278. 
The luminosity fraction of different components are shown in the figure. 

The combination of our dynamical cold and warm components generally matches, although higher in the inner region, the photometrically-identified exponential disks, the warm component may also partly contribute to the bar and/or bulge in the inner regions. 
For both galaxies, the disks are dominated by the dynamical cold components beyond $\sim 1\,R_e$, while they are constructed mainly by the warm components inside.

The hot components are concentrated in the inner regions and have mass fractions generally consistent with the combination of bulges and bars.  The surface brightness profiles of the hot components are similar to the sersic profiles of the bulges.
 
The contribution of the cold and warm components to the disk is consistent with what was found from simulations, that a morphological disk-like structure could also contain significant non-circular orbits (\citealt{Obreja2016}; \citealt{Teklu2015}). The warmer disk-like component in the inner regions is consistent with the inside-out scenario in which stars born earlier have lower angular momentum \citep{Lagos2016}.
 
\begin{figure*}
\centering\includegraphics[width=8cm]{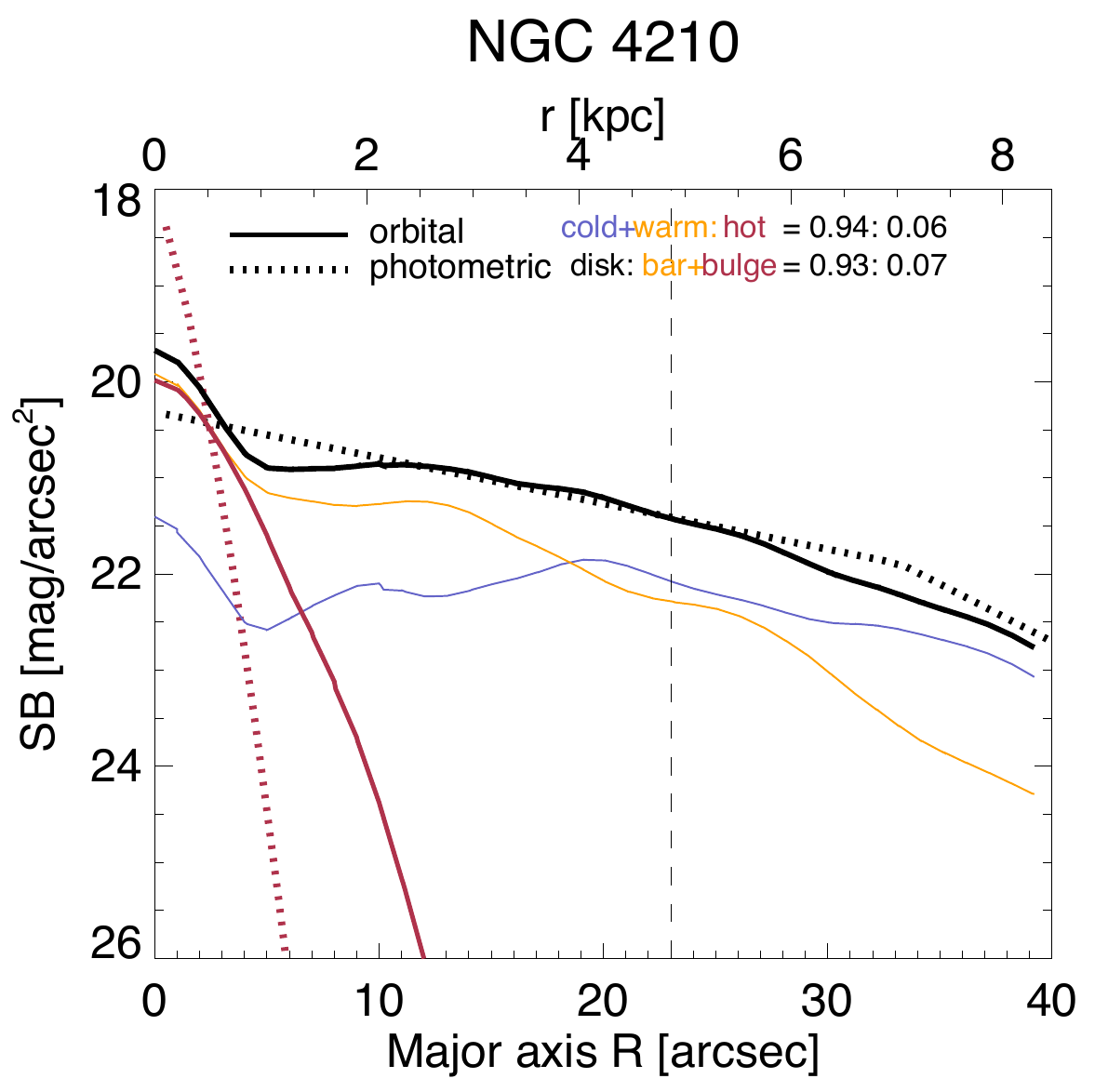}\includegraphics[width=8cm]{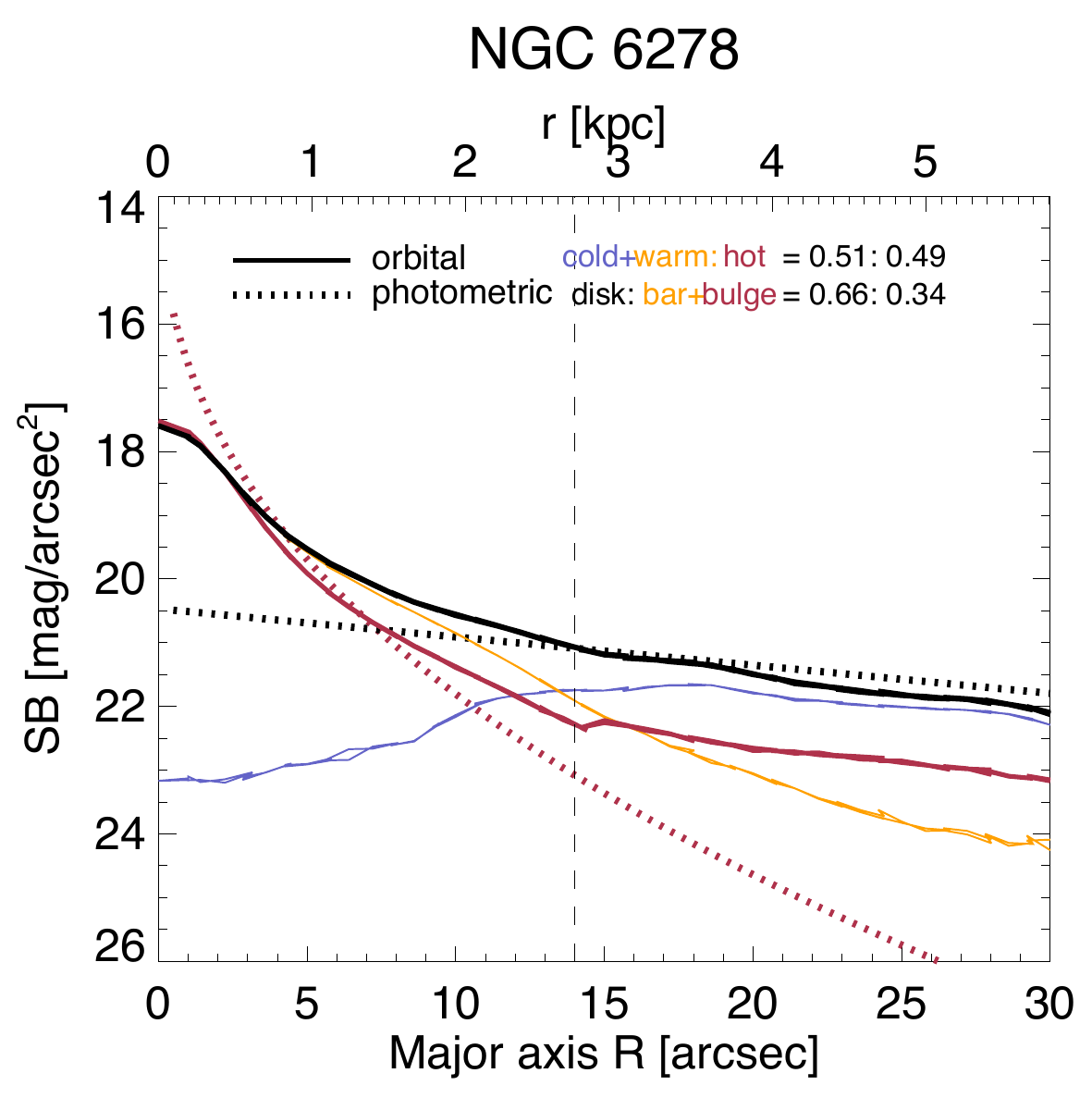}
\caption{The surface brightness profiles of different components. The blue, orange and red solid lines are the projected surface brightness of dynamically cold, warm and hot components from our best-fitting models, the thick black solid line is the combination of cold and warm components.  The black and red dotted lines are the exponential disk and the sersic bulge from the photometric decomposition of the $r$-band image in M$\acute{e}$ndez-Abreu (2016). The vertical black dashed line indicates the position of $1\, R_e$. 
}
\label{fig:phot}
\end{figure*}

\section{Summary}
\label{S:summary}

We create \swd models for two \emph{CALIFA} galaxies: an SBb galaxy NGC 4210 and an S0 galaxy NGC 6278. We have two sets of independent models, using two images at different wavelengths to construct the stellar mass, for each galaxy. The main results are:
\begin{itemize}
\item[-] With the \emph{CALIFA} kinematic data extending to $\sim 2 R_e$ of the galaxies,  the total mass profiles are well constrained with $1\sigma$ statistic uncertainties of $\sim 10\%$, within the data coverage.
Due to the degeneracy between the stellar mass and DM mass, the enclosed mass of stars and DM mass are constrained separately with uncertainties of $\sim 20\%$ and $\sim 50\%$.

\item[-] The assumption of constant stellar mass-to-light ratio affects the estimates of DM mass and luminous mass separately due to the degeneracy. However it does not affect our estimates of the total mass, and thus does not affect the internal dynamics of the galaxies obtained by our models. 

\item[-] The velocity anisotropy profiles of both $\sigma_r/\sigma_t$  and $\sigma_z/\sigma_R$ are well constrained.  $\sigma_r/\sigma_t$  profile is a good indication of the underlying orbital structures.  $\sigma_z/\sigma_R$ profiles we obtained for these two galaxies are consistent with the variation of $\sigma_z/\sigma_R$ across Hubble sequence. 
 
\item[-] We obtain the orbital distributions of the galaxies and dynamically decompose the galaxies into cold, warm and hot components based on the orbits' circularity. 
NGC 4210 is dominated by the cold and warm components with mass fractions of $0.49\pm 0.10$ and $0.45\pm0.10$ compared to $0.06\pm0.03$ for the hot component. 
NGC 6278 has a less massive but still well-defined cold component. The mass fractions of cold, warm and hot components are $0.12\pm0.06$, $0.39\pm0.10$ and $0.49\pm0.10$, respectively.

\item[-] The photometrically-identified exponential disks are dominated by the dynamical cold components beyond $\sim 1\,R_e$, while they are constructed mainly by the warm components in the inner regions. Our dynamical hot components are concentrated in the inner regions, similar to the photometrically-identified bulges. 

\end{itemize}

This is the first paper showing how the technique works. 
In the next paper we present and exploit  \swd models of a statistically representative sample of 300 CALIFA galaxies across the Hubble sequence.

\bibliographystyle{mn2e} 
\bibliography{spirals}

\section*{Acknowledgment}
We thank Miguel Querejeta for sending us the 3.6 $\mu$m stellar mass map of the two galaxies, and Jairo M{\'e}ndez-Abreu for sharing us the photometric results. GvdV acknowledge partial support from Sonderforschungsbereich SFB 881 "The Milky Way System" (subproject A7 and A8) funded by the German Research Foundation as well as funding from the European Research Council (ERC) under the European Union's Horizon 2020 research and innovation programme under grant agreement No 724857 (Consolidator Grant ‘ArcheoDyn'). J.~F-B. acknowledges support from grant AYA2016-77237-C3-1-P from the Spanish Ministry of Economy and Competitiveness (MINECO).
J.S. acknowledges support from an {\it Newton Advanced Fellowship} awarded by the Royal Society and the Newton Fund, and from the CAS/SAFEA International Partnership Program for Creative Research Teams.

\appendix

\section{Reason and solution of problems in modelling of fast-rotating galaxies}
\label{S:orbit-cut}

\subsection{GH expansion}
\label{SS:math}

\begin{figure}
\centering\includegraphics[width=\hsize]{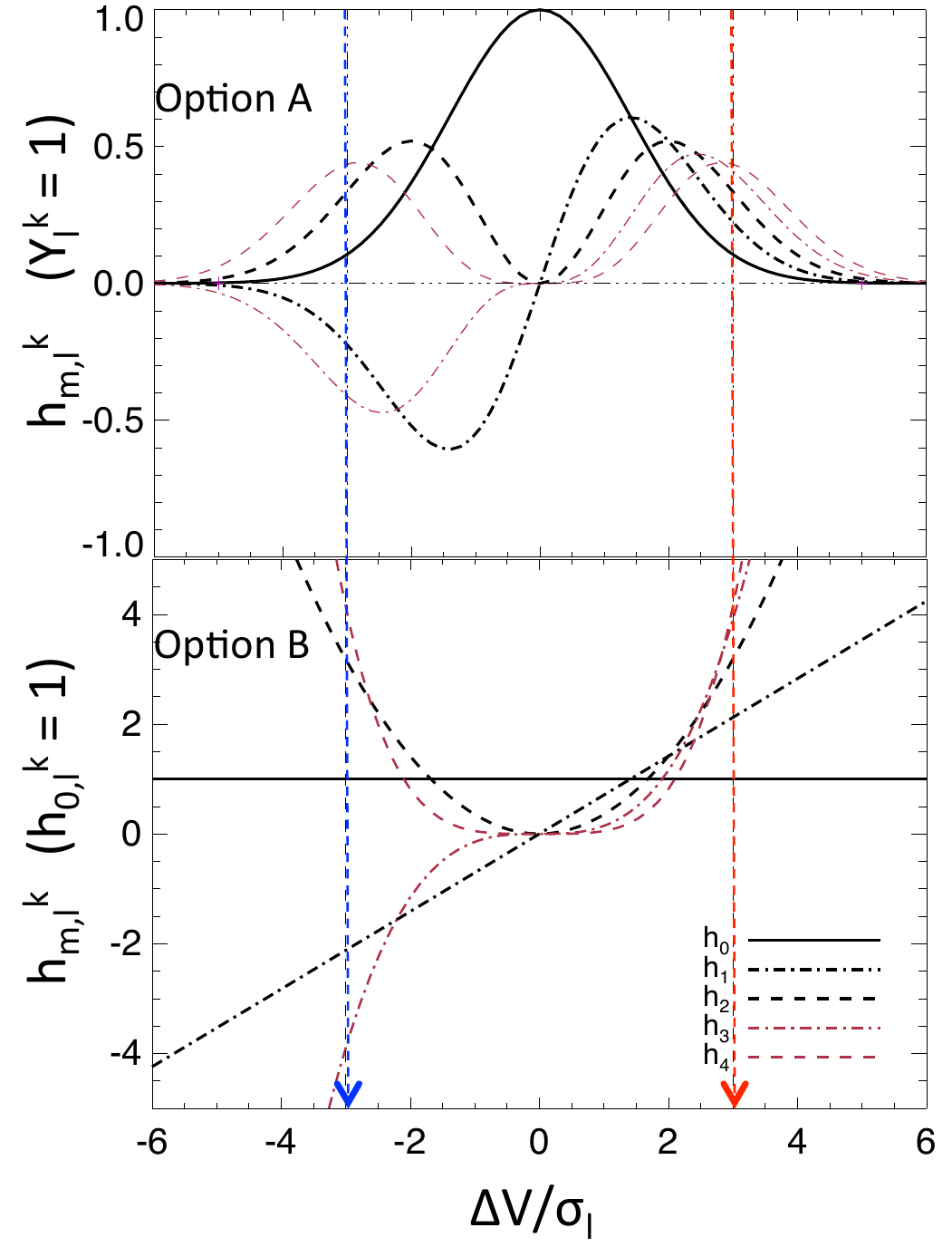}
\caption{{\bf Top:} the GH coefficients $h_{m,l}^k$ of a Gaussian profile defined by $(\Delta V + V_l, \sigma_l)$, described by a GH expansion around $(V_l, \sigma_l)$, as a function of their separation $\Delta V$, by adopting $\gamma_l^k =1$. {\bf Bottom:} Similar as the top panel but adopting $h_{0,l}^k =1$.}
\label{fig:GH}
\end{figure}
The GH coefficient of $f^k_l$ expanding as $\mathcal{GH}(v; \gamma_l^k, V_l, \sigma_l, h_{m,l}^k)$ can be obtained by:
\begin{multline}
\hat{h}_{m,l}^k = \gamma^k_l h_{m,l}^k = 
\\
\sqrt{2} \int_{-\infty}^{\infty} f^k_l(v) \exp \Bigg [ -\frac{1}{2} (\frac{v-V_l}{\sigma_l})^2 \Bigg] H_m(\frac{v-V_l}{\sigma_l}) dv,
\end{multline}
for m = 0,1,2,3,4.

Now we consider a specific case:
\begin{equation}
f^k_l(v; \Delta V) = \frac{1}{(\sqrt{2\pi} \sigma_l)} \exp \Bigg [-\frac{1}{2} (\frac{v-(\Delta V + V_l)}{\sigma_l})^2  \Bigg],
\end{equation}
which is a Gaussian profile with $(V_l, \sigma_l)$ being the observed values of that aperture. 
For this specific form of $f^k_l$, $h_{m,l}^k$ varies as a function of $\Delta V$.

Notice that we have a $\gamma^k_l$ degenerated with $h_{m,l}^k$. In order to obtain $h_{m,l}^k$ explicitly, we have two options: {\bf option A} assuming $\gamma^k_l =1$, then $h_{m,l}^k = \hat{h}_{m,l}^k$; and {\bf option B} assuming $h_{0,l}^k =1$, then we have $\gamma^k_l = \hat{h}_{0,l}^k$, and $h_{m,l}^k = \hat{h}_{m,l}^k / \hat{h}_{0,l}^k$. 
The top and bottom panel of Figure \ref{fig:GH} shows how $h_{m,l}^k\,(m=0,1,2,3,4)$ varies as a function of $\Delta V$ for the two options, respectively. 

For option A (Top panel), when $\Delta V = 0$, $f_l^k$ is identical with the center of the GH expansion, thus $h_{0,l}^k = 1$ and $h_{m,l}^k = 0$ (m=1,2,3,4). With $|\Delta V|$ increasing from zero, $h_{0,l}^k$ gradually decreases from $1$ to $0$, while $|h_{m,l}^k| \, (m=1,2,3,4)$ increase and get their maximum at $|\Delta V/\sigma| \sim 2$, then $|h_{m,l}^k| \, (m=1,2,3,4)$ decrease and become zeros again when $|\Delta V/\sigma| > 5$.

For option B (Bottom panel), when $\Delta V = 0$, still $h_{0,l}^k = 1$ and $h_{m,l}^k = 0$ (m=1,2,3,4). With $|\Delta V|$ increasing from zero, $h_{0,l}^k$ keeps unit, while $|h_{m,l}^k| \, (m=1,2,3,4)$ monotonously and sharply increase (note the large scale of y axis in the bottom panel comparing to the top panel). 

We describe these two options in detail in the following sections and option A will be modified for modelling fast rotating galaxies. 
\subsection{Option A}
\label{SS:O1}
For whatever reason, $\gamma^k_l =1$ was taken in the \swd models when describing the distribution of each orbit bundle $f_l^k$ by GH expansion around the observational aperture $(V_l, \sigma_l)$
(\citealt{Rix1997}; \citealt{Cretton&Bosch1999}; \citealt{vdB2008})). 

$h_{0,l}$ can not be fitted in a consistent way for this option. Because $h_{0,l}$, as a normalization parameter degenerated with $\gamma_l$, was fixed ($h_{0,l} = 1$) when extracting observational data. With $h_{0,l}^k$ varying from 1 to 0 in option A as shown in top panel of Figure~\ref{fig:GH}, the following two equations can not be satisfied simultaneously:
$\chi^2_{\mathrm{S}} = \sum_{l} \Bigg[ \frac{S_l - \sum_k w_k S_l^k}{0.01S_l} \Bigg]^2$ and
$\chi^2_{h_0} = \sum_{l} \Bigg[ \frac{S_l h_{0,l} - \sum_k w_k S_l^k h_{0,l}^k }{S_l dh_{0,l}} \Bigg]^2$.
Because the fitting of surface brightness ($\chi^2_{\mathrm{S}}$) is the one obviously we have to satisfy for the model normalization, thus the fitting of $h_{0,l}$ is skipped. Only $h_{m,l} \,(m=1,2,3,4)$ are considered in the fitting of kinematics as shown in equation~\ref{eqn:kin}.

Now consider a specific orbit $k$ contributing to the aperture $l$ with a VP of $f_l^k(v; \Delta V = 0)$, 
The GH expansion of $f_l^k$ around $(V_{l}, \sigma_{l})$ results in $h_{0,l}^k = 1$ and $h_{m,l}^k = 0$ (m = 1,2,3,4). 

Its counter-rotating partner orbit $k'$, thus has a Gaussian VP of $f_l^{k'}$ with $(V_k', \sigma_k') = (-V_{l}, \sigma_{l})$, which has $\Delta V = 2V_{l}$ comparing to $(V_{l}, \sigma_{l})$. 
When expanding $f_l^{k'}$ around $(V_{l}, \sigma_{l})$, the values of $h_{m,l}^{k'}$ (m=0,1,2,3,4) depend on $\Delta V/\sigma$ ($=2V_{l} /\sigma_{l}$) (see Figure~\ref{fig:GH}, top panel). 

When $|V_{l}/\sigma_{l}|>2.5$, thus $|\Delta V/\sigma|>5$ for orbit $k'$ comparing to $(V_{l}, \sigma_{l})$, we get $h_{1,l}^{k'} \approx h_{2,l}^{k'} \approx 0$, which is almost identical with its partner orbit $k$ having $h_{1,l}^k = h_{2,l}^k = 0$. 
Limited to reliable estimate for only $h_{m,l}$ (m=1,2) for use as model constrains, these two orbits become indistinguishable in the model. 
When $1.5<|V_{l}/\sigma_{l} |<2.5$, thus $3<|\Delta V/\sigma|<5$, we get $h_{m,l}^{k'}$ (m=1,2) not zero but still small values, by which the couple of orbits $k$ and $k'$ are still hard to be distinguished from each other.  

We notice that higher order GH coefficients $h_{1,l}^{k'}$ (m=3,4) have larger values when $3<|\Delta V/\sigma|<5$, while $h_{1,l}^{k} = 0$ (m=3,4) keeps for orbit $k$. Thus including higher order of GH coefficient $h_{3,l}, h_{4,l}$ helps to distinguish these two orbits. However, the quality of \emph{CALIFA} data can only provide realiable $h_{m,l}$ (m=1,2) \citep{Falcon-Barroso2016}.

In practice, we may not have such orbits $k$ and $k'$ with shape of VPs exactly the same as Gaussian profiles defined by $(\pm V_l, \sigma_l)$. The above analysis just illustrates the condition that a pair of counter-rotating orbits with high $V/\sigma$ are hard to be distinguished in the model.

If one of the orbits $k$ is highly weighted to the observations, as for fast rotating galaxies, the model will mistakenly take in significant contributions of its counter-rotating partner, thus causing problems in the modelling. 
The problem can be solved by just cutting all the counter-rotating orbits in the regions where $|V_{l}/\sigma_{l}|$ are high. \citet{Cretton1999} excluded the counter-rotating orbits whose circular radius are larger than 
a limiting radius. Within the limiting radius, all observations have $|V_{l}/\sigma_{l}|<1.5$. This procedure works in their modelling. 

We choose a different solution to avoid determining the orbit cut limiting radius from galaxy to galaxy, and to allow the contribution of counter-rotating orbits. 
If a counter-rotating orbits $k'$ passing aperture $l$ with $|V_k' -V_{l}| / \sigma_{l} > 3$ and $V_k'  V_l < 0$, this orbit should contribute little to the velocity distribution of this aperture $l$. When the orbit $k'$ meets this criterion in $\geq 2$ observational apertures, we exclude it from the model. The orbit is still included when it only meets the criterion in one aperture to avoid the case that a bad bin in the data would exclude orbits incorrectly. 
The orbits with $|V_k'-V_{l}| / \sigma_{l} < 3$, even counter-rotating (with $V_k'  V_l < 0$), are included in the model, such orbits have $h_{1,l}^{k'},  h_{2,l}^{k'} $ values distinguishable from those of their counter-rotating partners $k$.
We will test how it works in Appendix B, and this option will be mentioned as option A (taking $\gamma^k_l =1$ and being modified by cutting counter-rotating orbits as described).  
 
As a result, this orbit exclusion suppresses the counter-rotating components in strongly rotating disk, such structures, if any, could be found with data of high quality $h_3$ and $h_4$. We do not expect such structures to be common, and our procedure should not bias the orbit distribution for most of the galaxies. 

\subsection{Option B}
\label{SS:O2}
Taking $h_{0,l}^k = 1$ is consistent with the way we extracting the observational data with $h_{0,l} =1$ fixed. 
$h_{0,l}$ can be included in the fitting in a consistent way as $h_{m,l} \, (m = 1,2,3,4)$. Following \citet{Magorian&Binney1994}, we calculate the measure error $dh_{0,l} = d\sigma_l/(2\sigma_l)$. 

In this case,
$\chi^2_{h_0} = \sum_{l} \Bigg[ \frac{S_l h_{0,l} - \sum_k w_k S_l^k h_{0,l}^k }{S_l dh_{0,l}} \Bigg]^2$ with $h_{0,l}=1$ and $h_{0,l}^k =1$, 
is equivalent to 
$\chi^2_{\mathrm{S}} = \sum_{l} \Bigg[ \frac{S_l - \sum_k w_k S_l^k}{0.01S_l} \Bigg]^2$ but with different errors. 
Thus the fitting of $h_{0,l}$ is actually included in the fitting of surface brightness.

As shown in the bottom panel of Figure~\ref{fig:GH}, when keeping $h_{0,l}^k=1$, $|h_{m,l}^k| \, (m=1,2,3,4)$ increase monotonously with $|\Delta V / \sigma|$. Thus the counter-rotating orbits issue should disappear straightforwardly. 

This option (taking $h_{0,l}^k = 1$) will also be tested in appendix~\ref{S:simulation}, and it will be mentioned as option B.
\section{Application to simulated galaxies}
\label{S:simulation}

To test our model's ability of recovering the underlying mass profile and internal orbit distribution with \emph{CALIFA}-like kinematic data for spiral galaxies,  we apply our model to mock data created from a simulated spiral galaxy. The two options of extracting GH coefficients for orbit bundles described in appendix~\ref{S:orbit-cut} will be applied separately, we call the code optimized with option A as Model A, as the latter Model B.  In the following sections, we will show that the two models work comparable well on recovering the mass profiles. Model A works reasonably well on recovering the orbit distribution, while model B, unexpectedly, does not. 

\subsection{The mock data}
\label{SS:mockdata}

\begin{figure}
\centering\includegraphics[width=6cm]{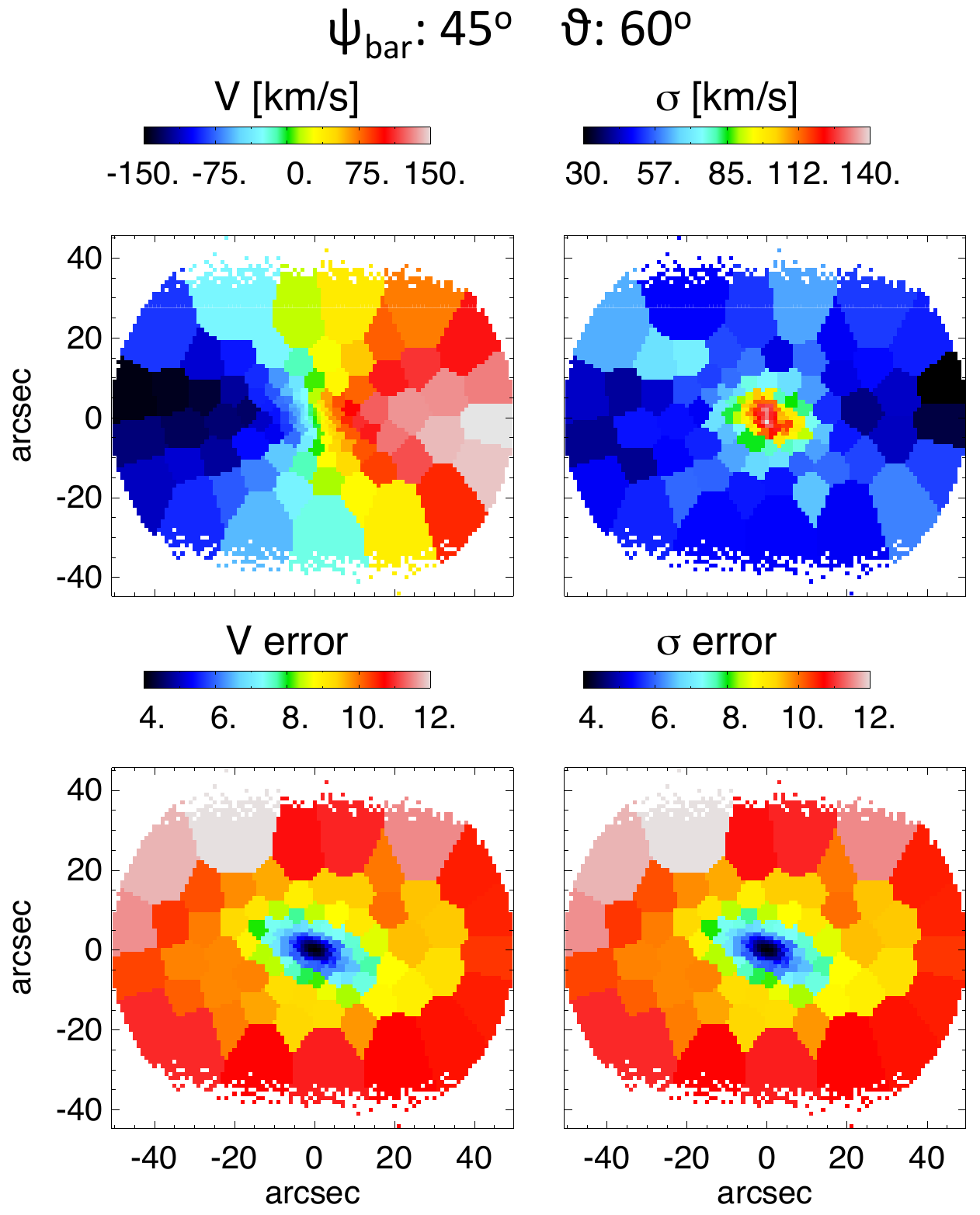}
\caption{A kinematic map created from a simulated galaxy, projected with the position angle of the bar $\psi_{bar} = 45^{\circ}$ and the inclination angle of the disk $\vartheta_{true} = 60^{\circ}$. The top panels are the mean velocity $V$ and velocity dispersion $\sigma$.  The $V$ and $\sigma$ maps have been perturbed with the error maps shown in the bottom panels. }
\label{fig:kin40}
\end{figure}

We use the N-body simulation from \citet{Shen2010} of a Milky-Way like galaxy.  
The simulation contains $10^6$ equal-mass particles, with the total stellar mass of $4.25\times10^{10}\,M_{\odot}$.  A rigid logrithmic DM halo is included; the potential of the DM halo $\phi(r) = \frac{1}{2} V_c^2 \ln(1 + r^2/R_c^2)$ with the scale radius $R_c = 15 \, \mathrm{kpc}$ and scale velocity $V_c = 250 \, \mathrm{km}\, \mathrm{s^{-1}}$. 
We take a snapshot from the simulation at $t = 2.4$ Gyr, which corresponds to a well-developed barred spiral galaxy.

We place the galaxy at a distance of 41 Mpc ($ 5\, \mathrm{arcsec} = 1\, \mathrm{kpc}$), then project it to the observational plane. 
The projected mock data will be affected by the viewing angles (position angles of the bar $\psi_{bar}$ and inclination angles of the disk plane $\vartheta$) of the galaxy, thus affect the internal properties that could be constrained from our models. 
21 sets of mock data are created with their viewing angles shown in Table~\ref{tab:mock}, we call them $S_1$ to $S_{21}$ as listed.
We omit the very face-on cases ($\vartheta < 30^o$), for which the uncertainty caused by de-projection becoming large.   

We take each mock data set as an independent galaxy and observe each galaxy with spatial resolution of 1 arcsec per pixel, to get the surface mass density. For the kinematic data,  we first divide the particles into each pixel with the size of 1 arcsec, then process Voronoi binning \citep{Cappellari2003} to get a signal-to-noise threshold of $S/N= 40$, then calculate the mean velocity and velocity dispersion with the particles in each bin. 
We use a simple logarithmic function inferred from the \emph{CALIFA} data to construct the errors of the mean velocity and velocity dispersion \citep{Sassa2015}.  

We then perturb the kinematic data by adding random values inferred from the error maps. A typical case of the final kinematic maps and the error maps ($S_{11}$) is shown in Figure~\ref{fig:kin40}. 

The surface mass density is taken as an image of the galaxy with constant stellar mass-to-light ratio $\Upsilon = 1$.
We perform a 2D axisymmetric MGE fit to the surface mass density, which is used as the tracer density as well as the stellar mass distribution in our model. 

\begin{table}
\caption{21 mock data sets. $\psi_{bar}$ is the positional angle of the bar, $\psi_{bar} = 0^o$ denotes the long axis of bar aligned with long axis of the galaxy. $\vartheta$ is the inclination angle of the disk, $\vartheta = 0^o$ is face-on and $\vartheta = 90^o$ is edge-on. }
\label{tab:mock}
\begin{tabular}{*{3}{l}}
\hline
Name   &  $\psi_{\mathrm{bar}}$ & $\vartheta$  \\
\hline

S$_1$ & 0 & 30 \\
S$_2$ & 0 & 40 \\
S$_3$ & 0 & 50 \\
S$_4$ & 0 & 60 \\
S$_5$ & 0 & 70 \\
S$_6$ & 0 & 80 \\
S$_7$ & 0 & 90 \\
\hline
S$_8$ & 45 & 30 \\
S$_9$ & 45 & 40 \\
S$_{10}$ & 45 & 50 \\
S$_{11}$ & 45 & 60 \\
S$_{12}$ & 45 & 70 \\
S$_{13}$ & 45 & 80 \\
S$_{14}$ & 45 & 90 \\
  \hline
S$_{15}$ & 90 & 30 \\
S$_{16}$ & 90 & 40 \\
S$_{17}$ & 90 & 50 \\
S$_{18}$ & 90 & 60 \\
S$_{19}$ & 90 & 70 \\
S$_{20}$ & 90 & 80 \\
S$_{21}$ & 90 & 90 \\
\hline
 \end{tabular}
\end{table}

\subsection{Best-fitting models}
\label{SS:mock_fit}
We take each of those 21 mock data sets as an independent galaxy, and applying the same modelling process to each of them. 

We still take $S_{11}$ as an example to show the best-fitting models in Figure~\ref{fig:kinbest}. The upper panels are the mock data; mean velocity on the left and velocity dispersion on the right, with contours of the surface mass density overplotted. The middle and bottom panels are our best-fitting kinematic maps from model A and model B; with contours of the MGE fits to the surface mass density overplotted. The shape of bars are clear in the contours of the original surface mass density, while the non-axisymmetry of mass density caused by the bars is not included in the MGEs. 
The kinematics are generally matched well by our models, with $\min (\chi_{\mathrm{kin}}^2) /N_{\mathrm{kin}} = 1.1$ obtained for the best-fitting model, $N_{\mathrm{kin}}$ is the total bins of kinematic data (number of apertures times number of GH moments). Unlike to the two CALIFA galaxies in Section~\ref{SS:califa_fit}, the kinematic data here are not symmetrized, thus the data points are independent from each other. $\chi^2_{\mathrm{kin}} /N_{\mathrm{kin}} \sim 1$ is what we expected for a good model. 

\begin{figure}
\centering\includegraphics[width=6.5cm]{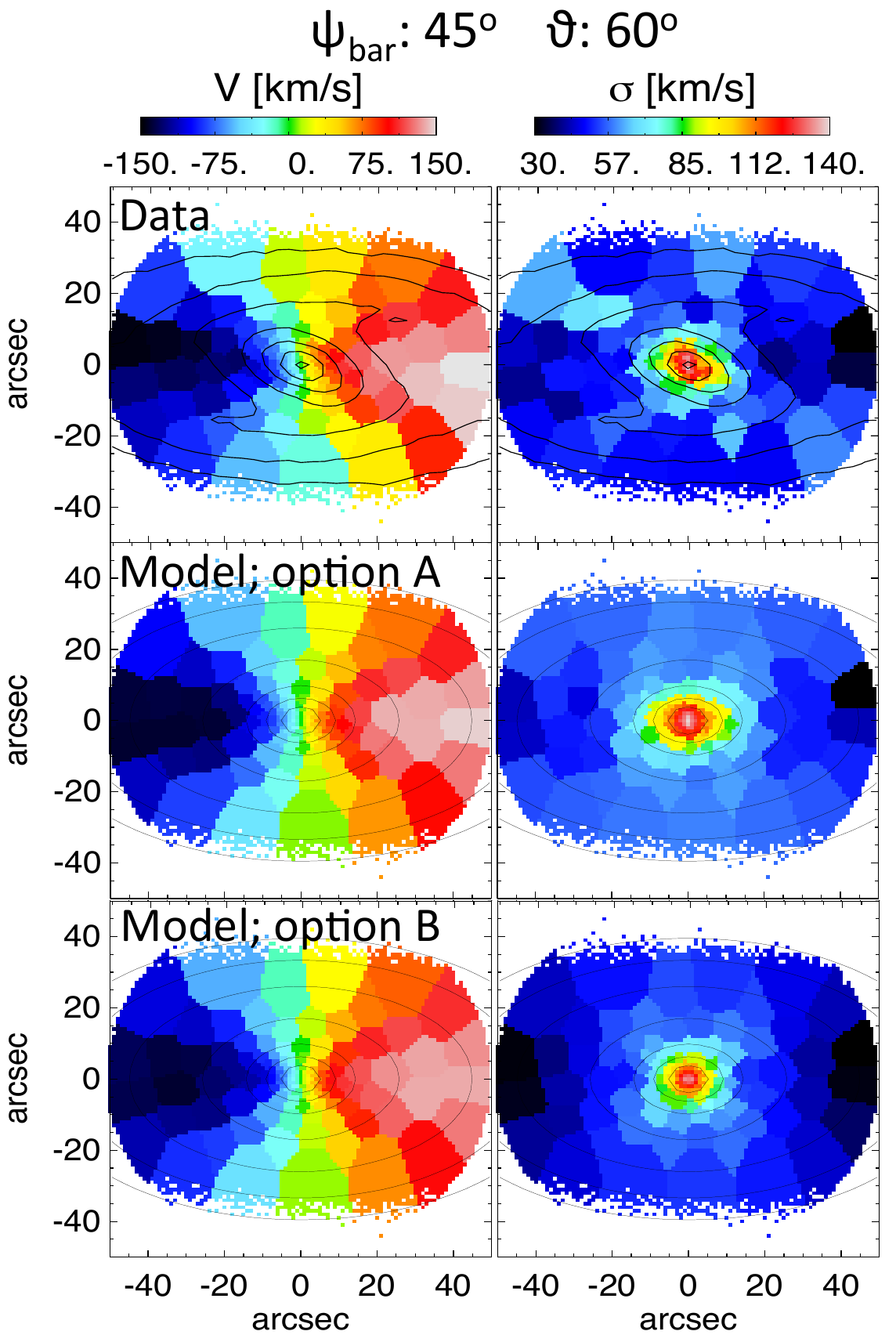}
\caption{The best-fitting models of $S_{11}$. The upper panels are the mock kinematic data; mean velocity (left) and velocity dispersion (right), with contours of the surface mass density overplotted. The middle and bottom panels are our best-fitting models of model A and model B, respectively; with contours of the axisymmetric MGE fits to the surface mass density overplotted. }
\label{fig:kinbest}
\end{figure}

\subsection{Recovery of mass profiles}
\label{SS:mock_mass}

\begin{figure*}
\centering\includegraphics[width=5.3cm]{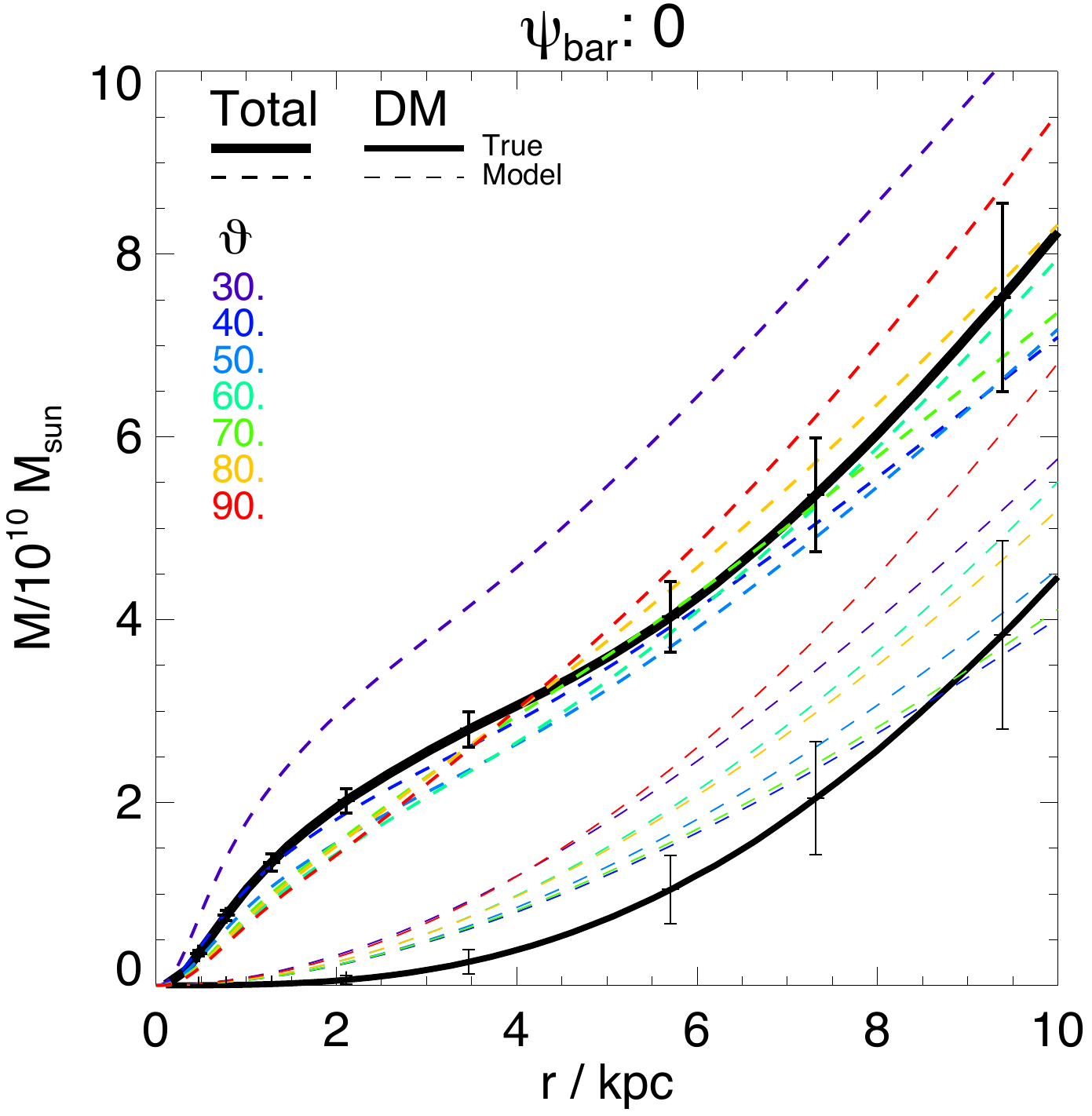}\includegraphics[width=5.3cm]{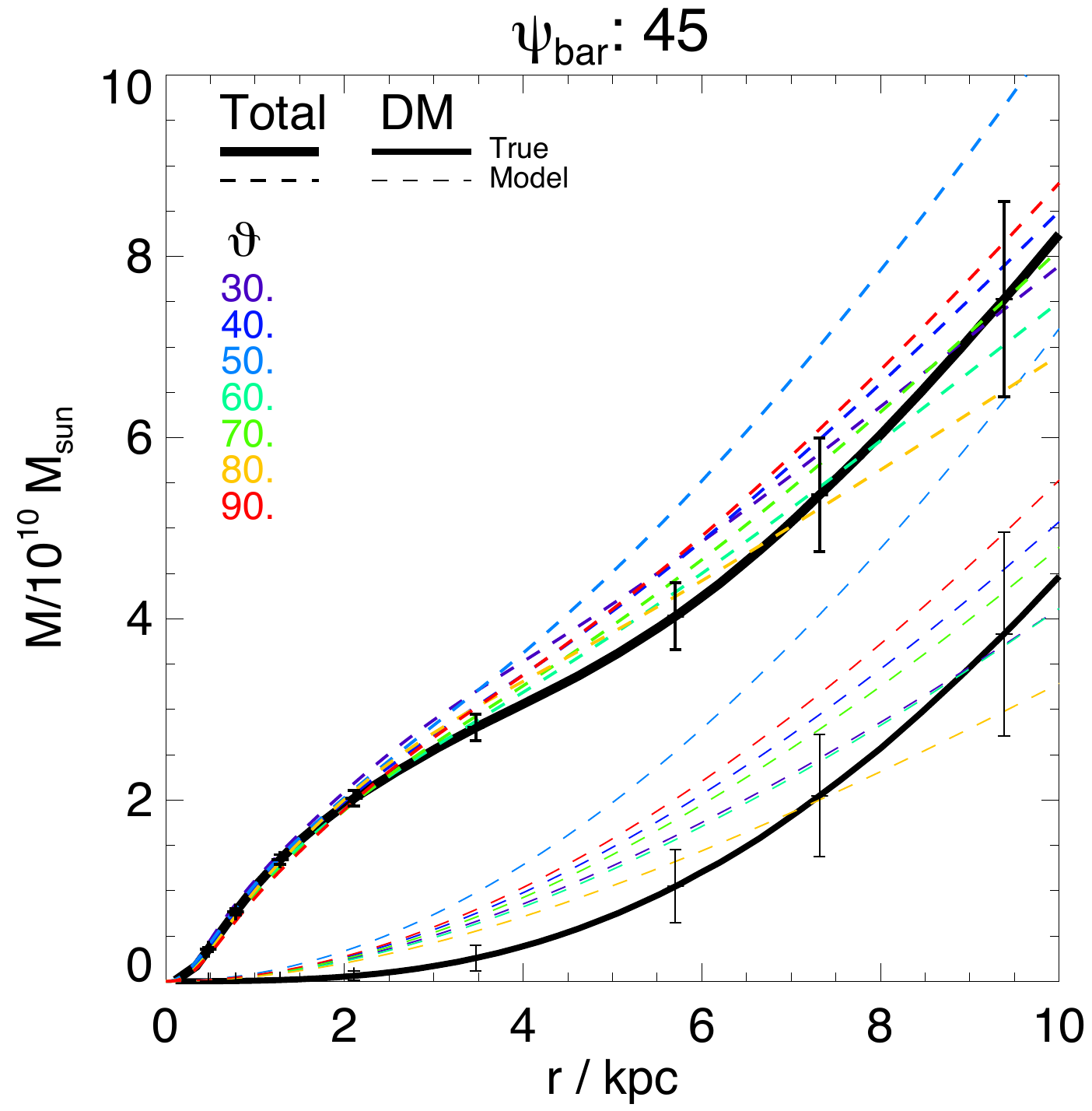}\includegraphics[width=5.3cm]{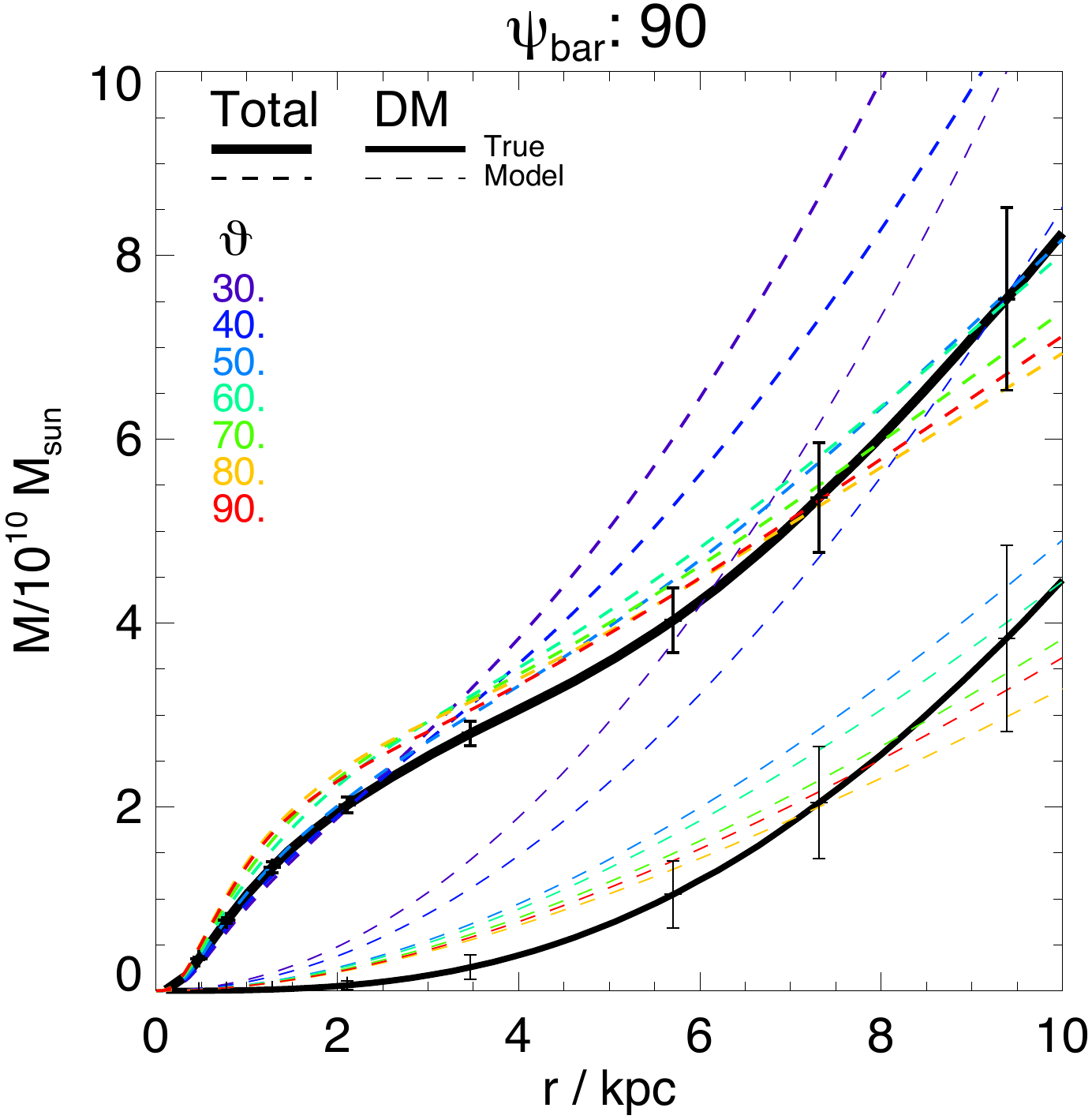}
\caption{The enclosed mass profiles obtained from model option A. The true enclosed mass profiles of the simulated galaxy are plotted with back solid lines, the thicker one represents the total mass profile and the thinner one represents the dark matter mass profile. The left, middle and right panels show the mass profiles from the best-fitting models of $S_1-S_7$ ($\psi_{\mathrm{bar}} = 0^o$), $S_8-S_{14}$ ($\psi_{\mathrm{bar}} = 45^o$) and $S_{15}-S_{21}$ ($\psi_{\mathrm{bar}} = 90^o$), respectively. In each panel, dashed lines with colors from blue to red indicate inclination angle $\vartheta$ from $30^o$ to $90^o$. Each dashed line represents the mean mass profiles of the models with $\chi^2_{\mathrm{kin}} - \min{\chi^2_{\mathrm{kin}}} < \sqrt{2N_{kin}}$, the error bars show the typical values of the $1\sigma$ scatter of mass profiles among those models for each single data set.  
Note that the error bars are not errors of the true mass profiles although they are located with the black solid lines. }
\label{fig:mock_mass}
\end{figure*}

\begin{figure*}
\centering\includegraphics[width=5.3cm]{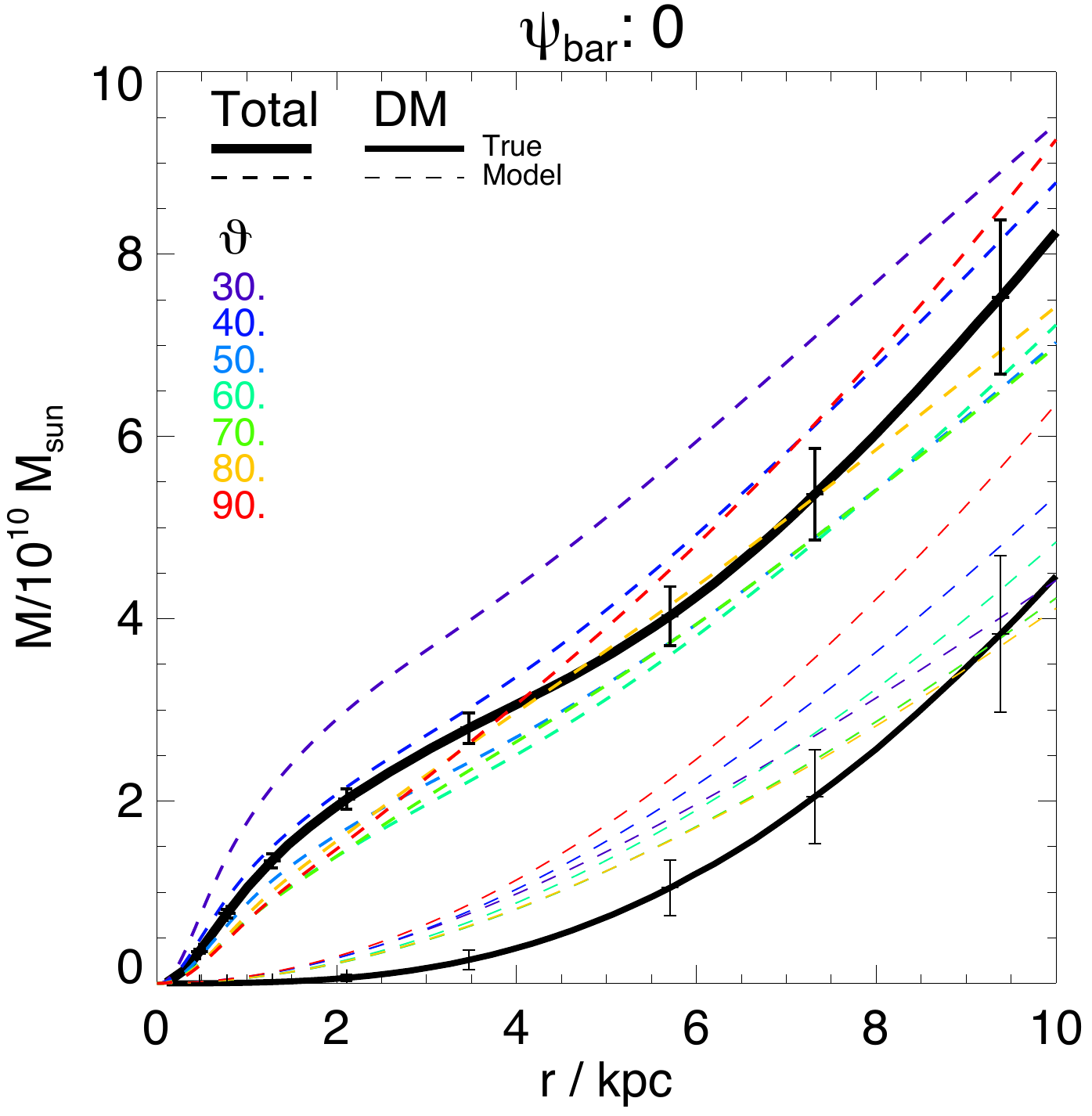}\includegraphics[width=5.3cm]{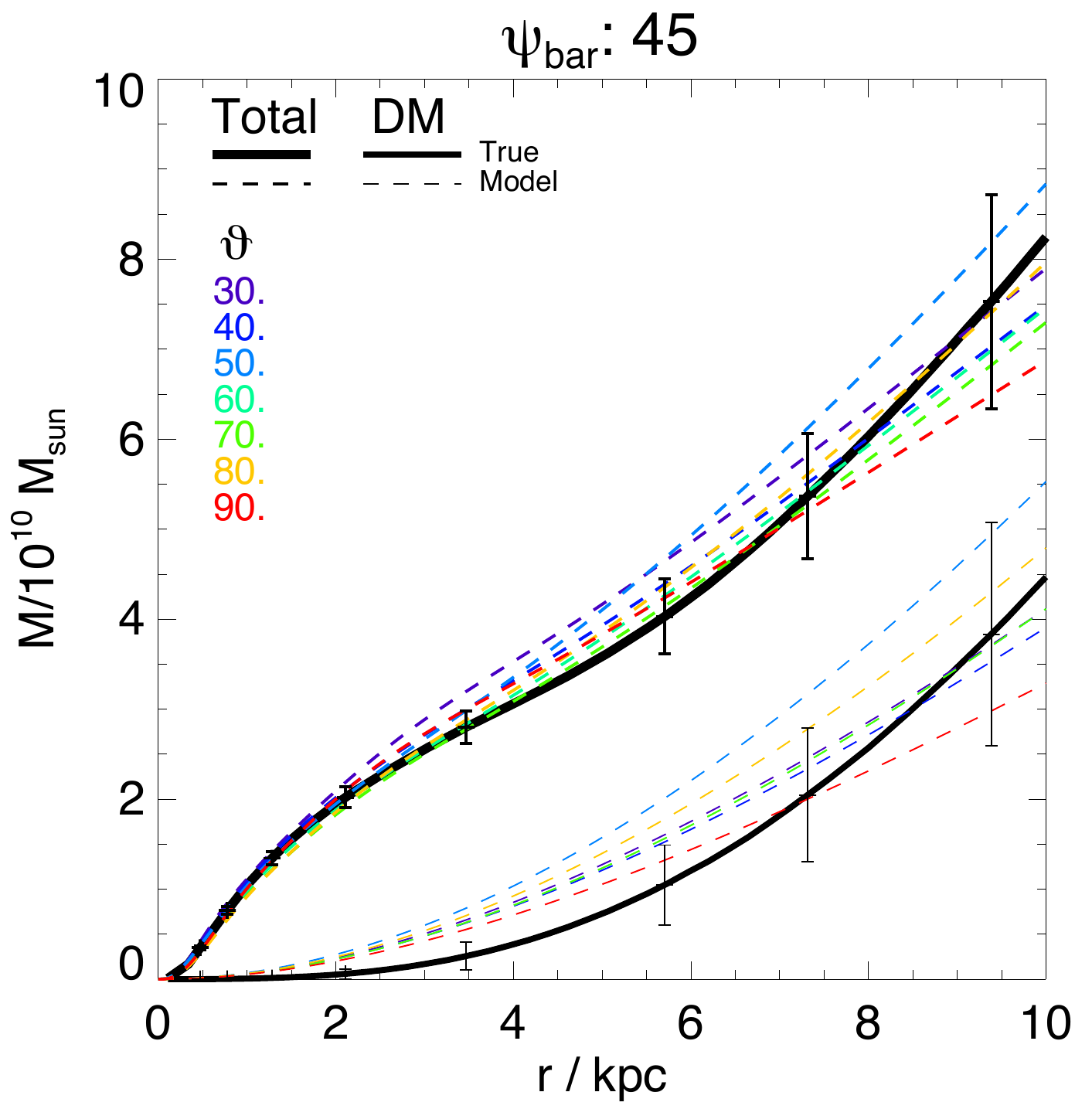}\includegraphics[width=5.3cm]{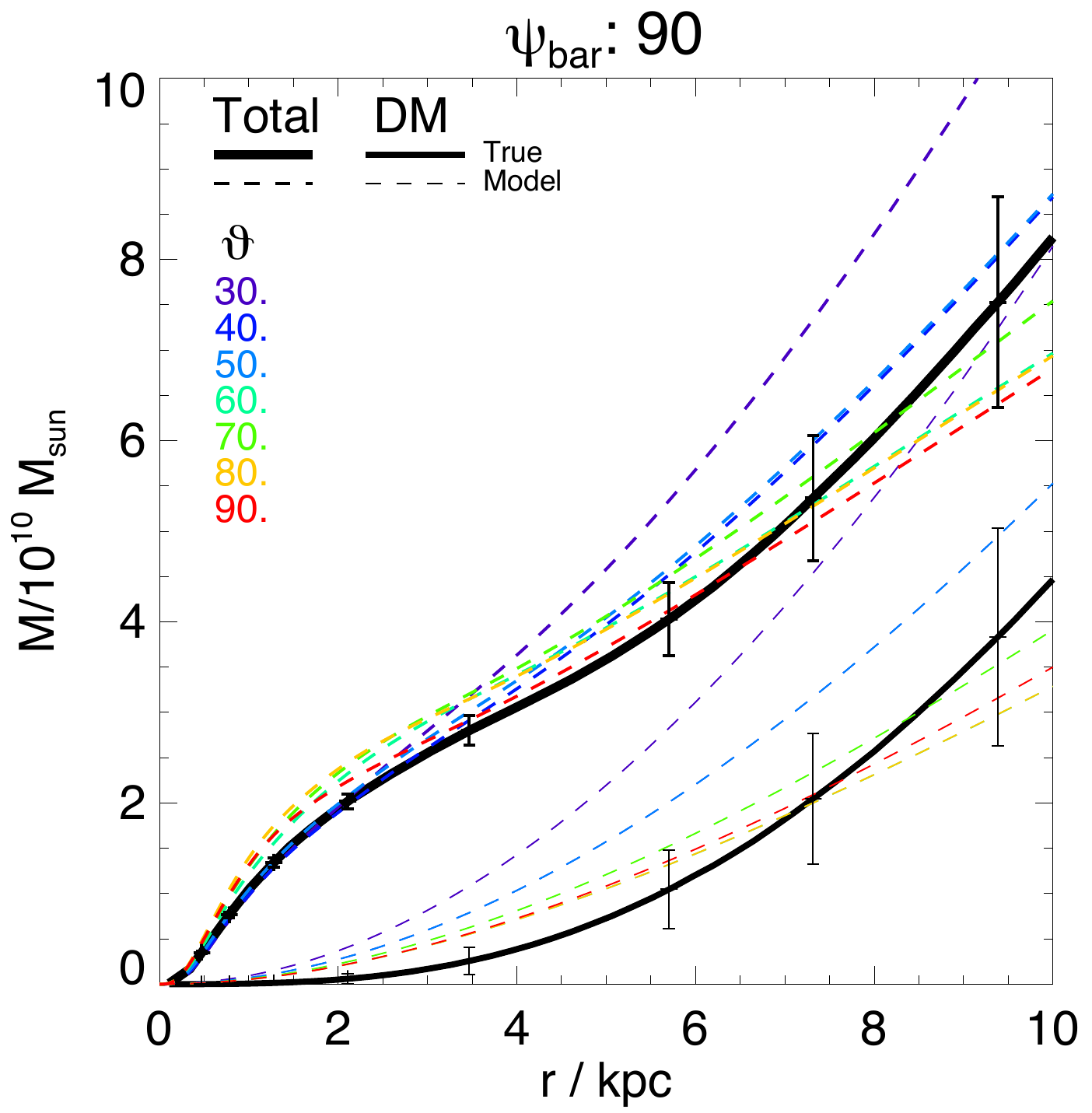}
\caption{Similar to Figure~\ref{fig:mock_mass}, but from model option B.}
\label{fig:mock_massO2}
\end{figure*}

To illustrate the recovery of the underlying mass profiles, we calculate the enclosed total mass and dark matter mass profiles as shown in Figure~\ref{fig:mock_mass} from model A and Figure~\ref{fig:mock_massO2} from model B. 
The true enclosed mass profiles of the simulated galaxy are plotted with back solid lines, the thick one represents the total mass profile and the thinner one represents the dark matter mass profile. The mass profiles from the best-fitting models of $S_1-S_7$ ($\psi_{\mathrm{bar}} = 0^o$), $S_8-S_{14}$ ($\psi_{\mathrm{bar}} = 45^o$) and $S_{15}-S_{21}$ ($\psi_{\mathrm{bar}} = 90^o$) are shown in the left, middle and right panels, respectively. 

The $\chi^2_{\mathrm{kin}}$ has significant fluctuation with standard deviation of $\sim \sqrt{2 N_{\mathrm{kin}}}$ in \swd models. 
\footnote{In a single model with fixed potential and orbit library, by perturbing the kinematic data, we find that $\chi^2_{\mathrm{kin}}$ of the best solutions vary. And the $1\sigma$ standard deviation is between $\sim \sqrt{2N_{kin}}$ to $\sim 2 \sqrt{2N_{kin}}$. Here we take $\sqrt{2N_{kin}}$ as $1\sigma$ of $\chi^2_{\mathrm{kin}}$ fluctuation. }
Each dashed line in Figure~\ref{fig:mock_mass} is the mean mass profiles of the models with 
$\chi^2_{\mathrm{kin}} - \min{\chi^2_{\mathrm{kin}}} < \sqrt{2N_{kin}}$ for each mock data set.
The error bars show the typical $1\sigma$ scatter of mass profiles among those models. For the enclosed total mass at $r = 10$ kpc, masses represented by dashed lines are consistent with the true values within the error bar for 16/21 cases for model option A and 17/21 for model option B. 
Thus, the $1\sigma$ $\chi^2_{\mathrm{kin}}$ fluctuation of $\sqrt{2 N_{kin}}$ works as the $1\sigma$ confidence level, in respective to total enclosed mass within $10$ kpc.  
The total enclosed mass at this large radius is least affected by the axisymmetric assumption of our model, thus with errors could be dominated by statistical error. 
We also note that the $1\sigma$ error is $\sim 10\%$ of the total mass with this \emph{CALIFA}-like mock data. 

From now on, we take $\Delta \chi^2_{\mathrm{kin}} = \sqrt{2N_{kin}}$ as the $1\sigma$ confidence level of \swd models with \emph{CALIFA}-like kinematic data. And qualify how well our model recovers another properties within $1\sigma$ errors defined by this confidence level. 

In the inner regions, the total mass obtained by our models is less accurate due to the ignorance of bar. And we notice that the deviation is larger for the projections with  $\psi_{\mathrm{bar}} = 0^o$. 

The $1\sigma$ error of dark matter mass only is larger, with relative uncertainty of $\sim 20\%$. However, we systemically over-estimated the dark matter mass by $1-2\sigma$ errors in the inner regions.  
This is caused by the NFW DM model we are using, it concentrates more DM in the inner regions than the real logrithmic DM halo of this simulated galaxy. This accounts for the possible systemic errors caused by non-perfect DM models. We expect the DM to be recovered better for real galaxies. From Cosmological Hydrodynamic simulations, DM halos are found to be closer to NFW halos \citep{Xu2016}, rather than logrithmic halos.

\subsection{Recovery of orbit distribution}
\label{SS:mock_orbit}

\begin{figure}
\centering\includegraphics[width=7cm]{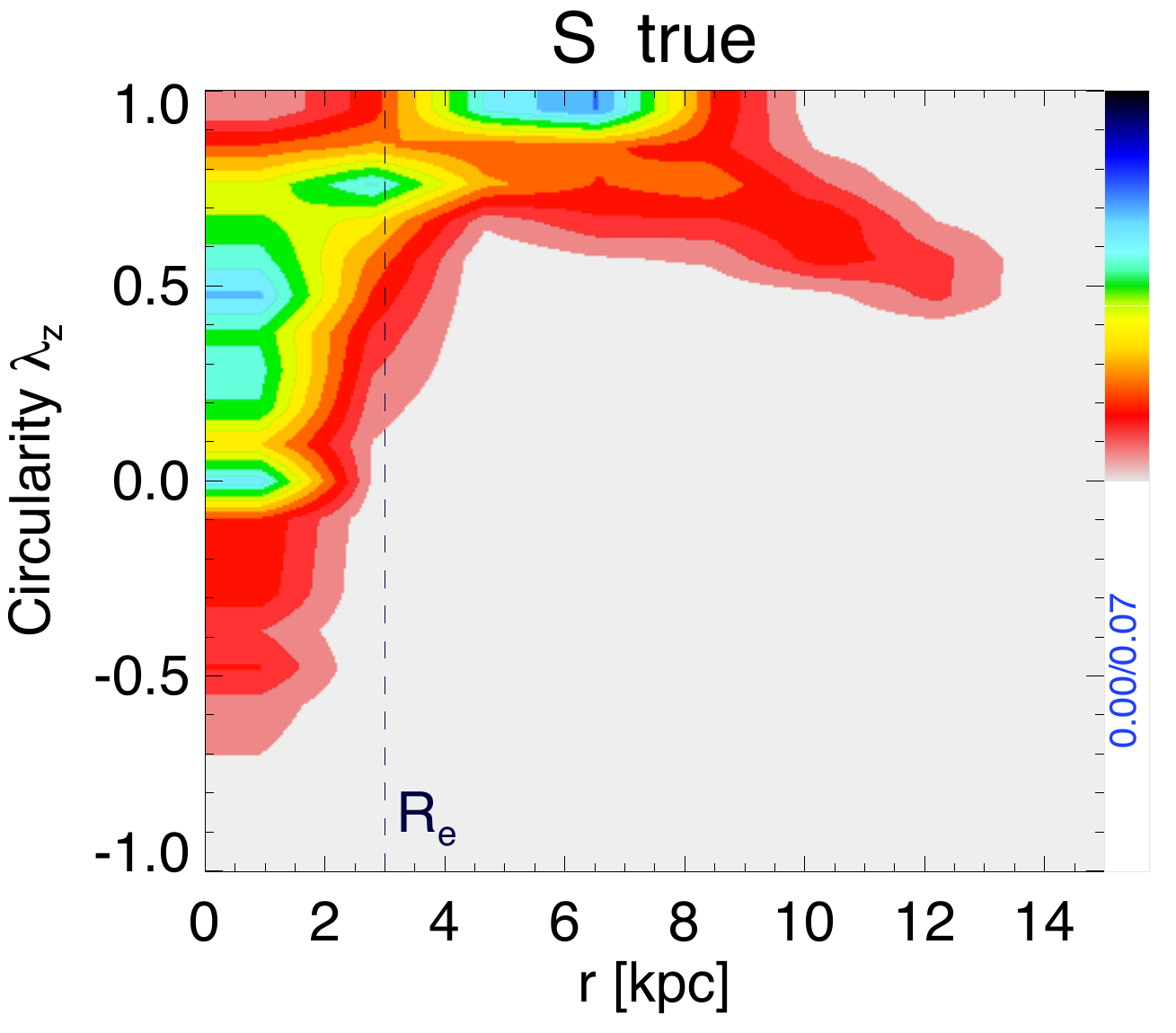}
\caption{The true orbit distribution of this simulated galaxy shown as the orbits' probability distribution on the phase space $r$ vs. $\lambda_z$. }
\label{fig:true_orbits}
\end{figure}

\begin{figure*}
\includegraphics[width=5.2cm]{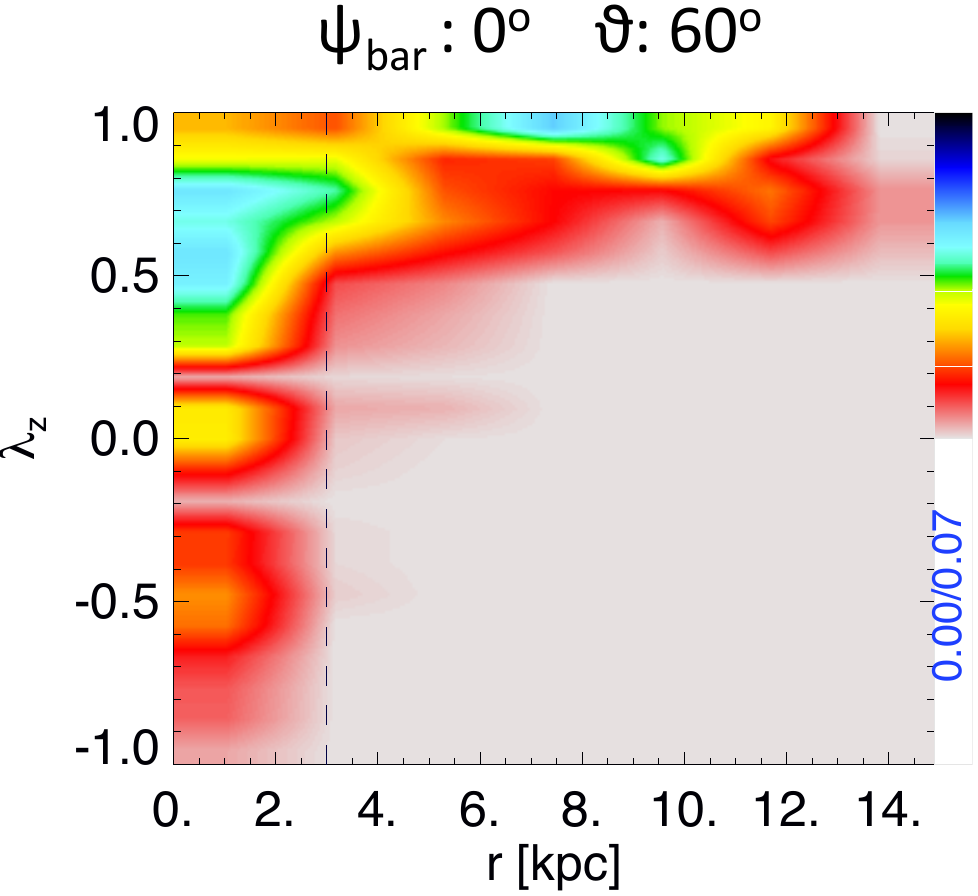}\includegraphics[width=5.2cm]{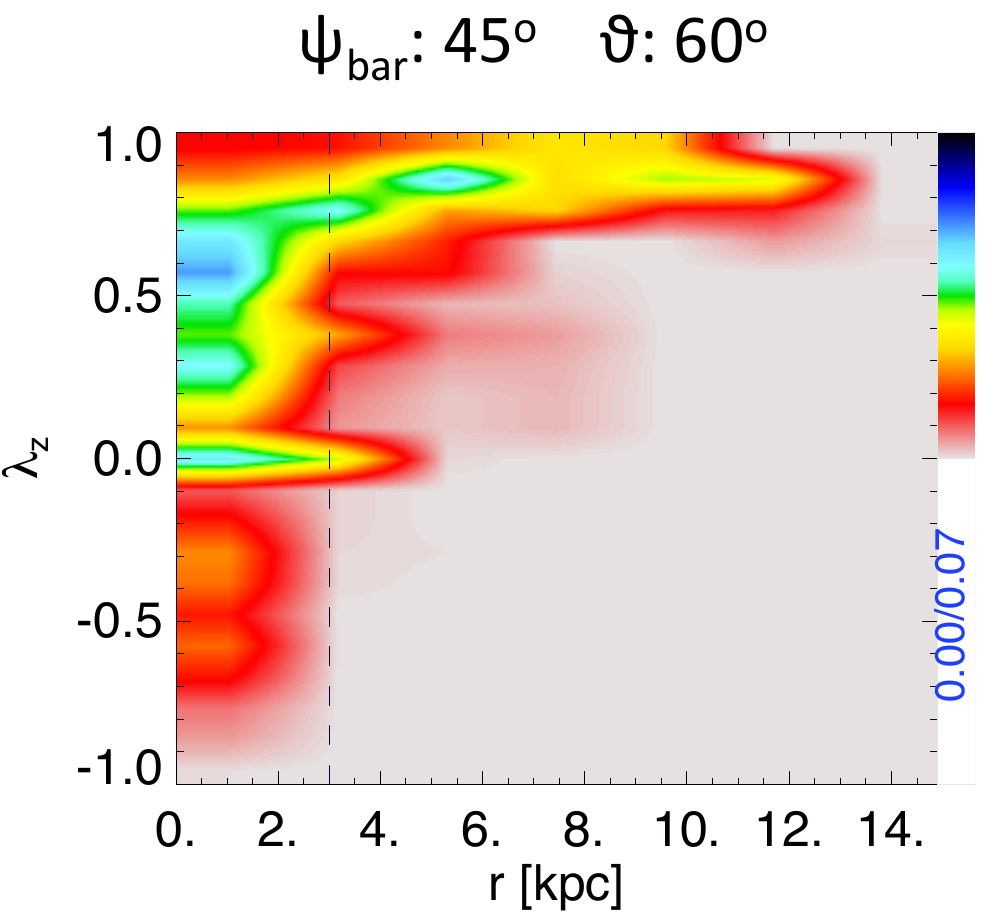}\includegraphics[width=5.2cm]{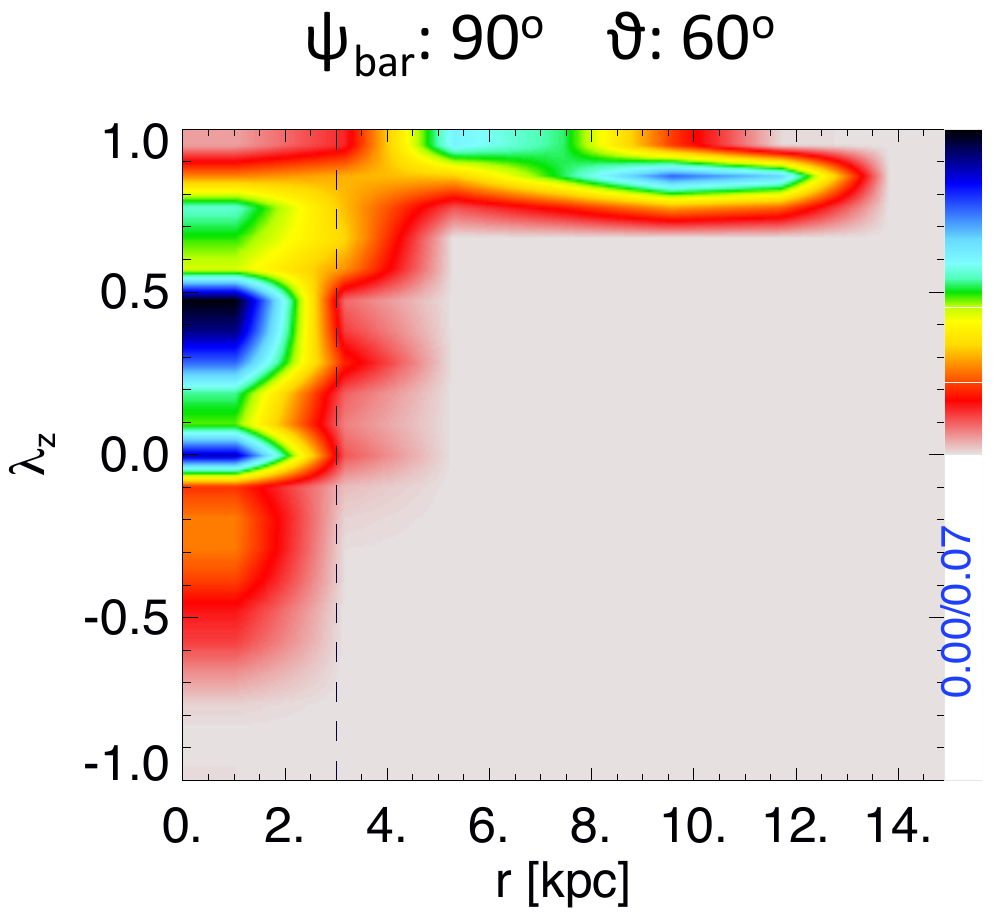}
\centering\includegraphics[width=5.2cm]{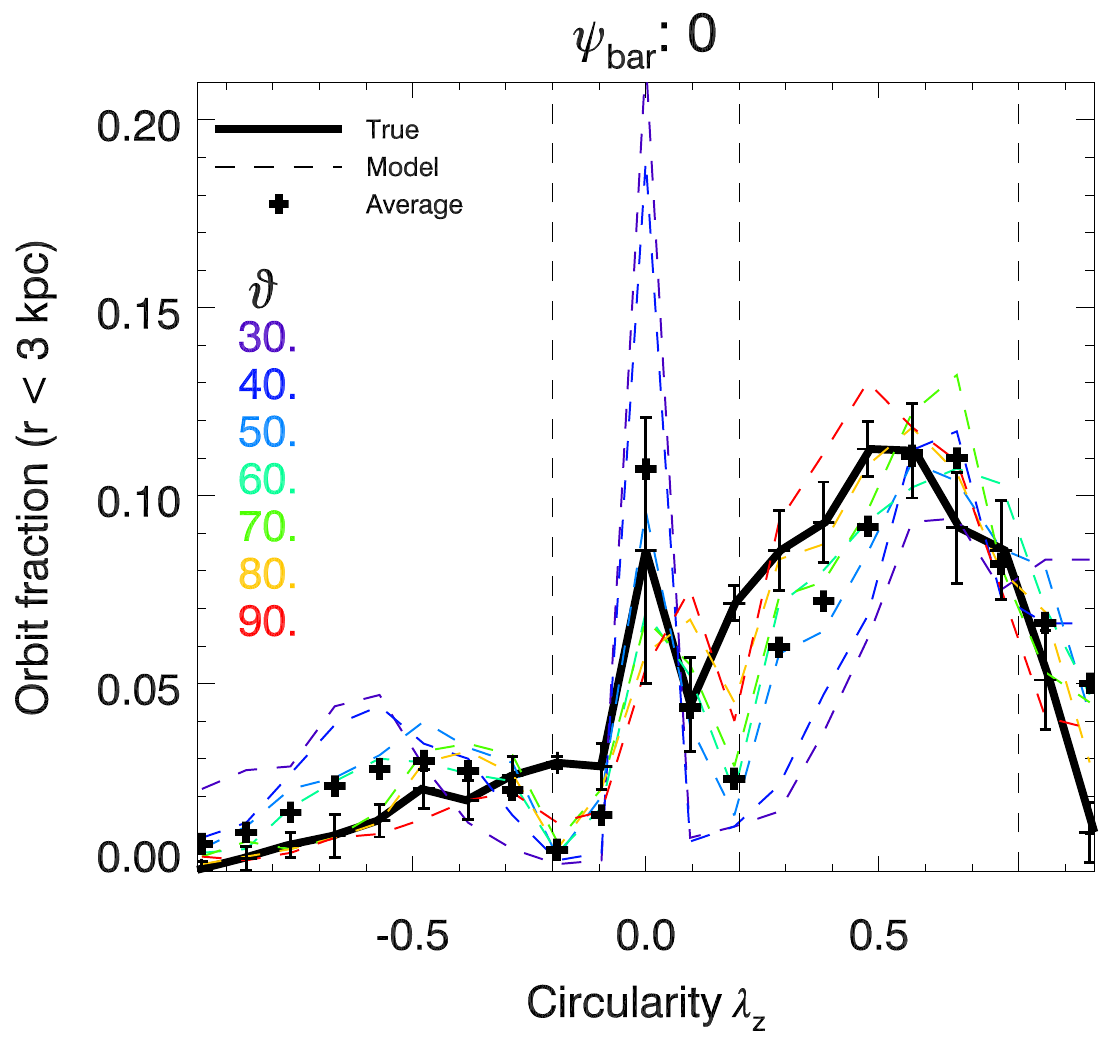}\includegraphics[width=5.2cm]{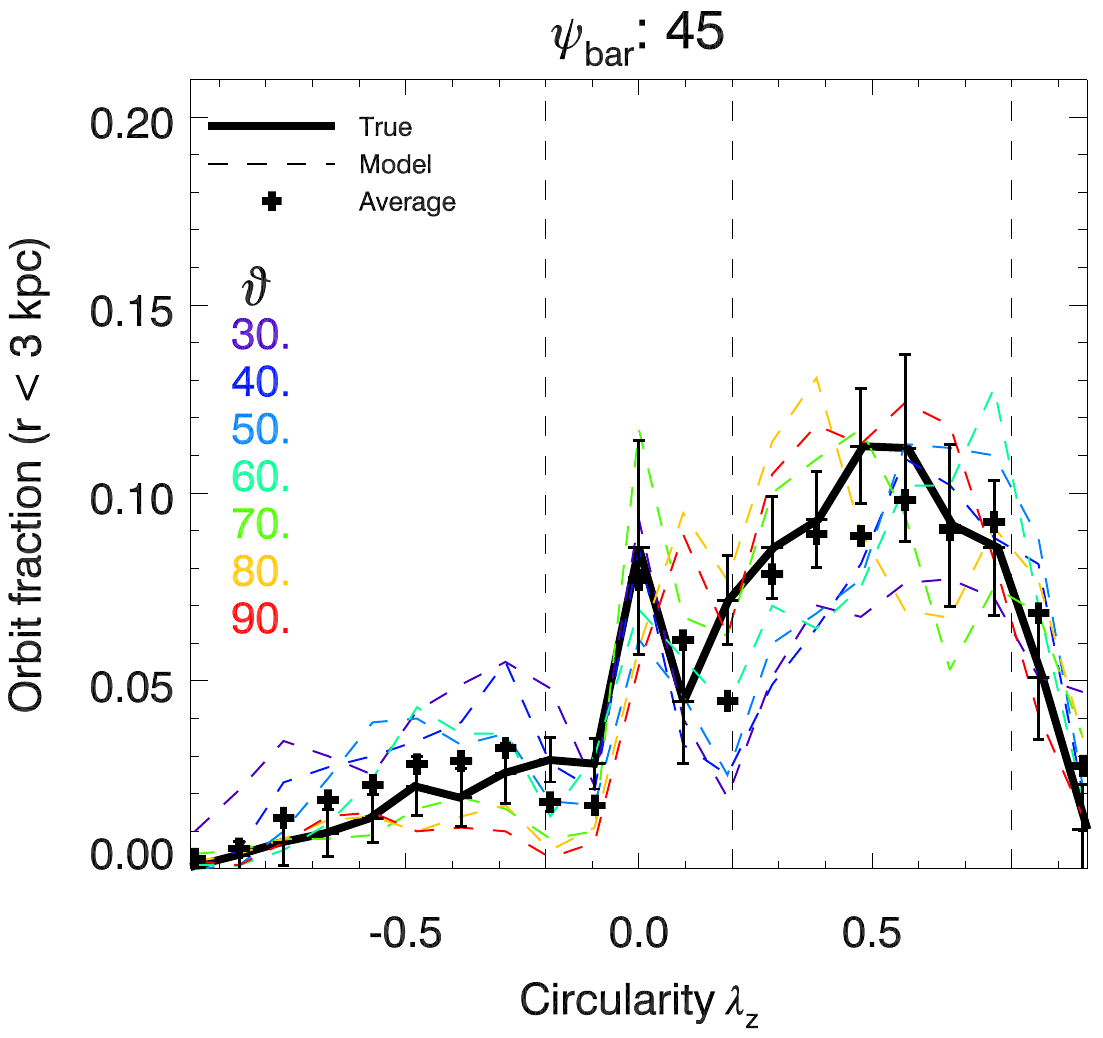}\includegraphics[width=5.2cm]{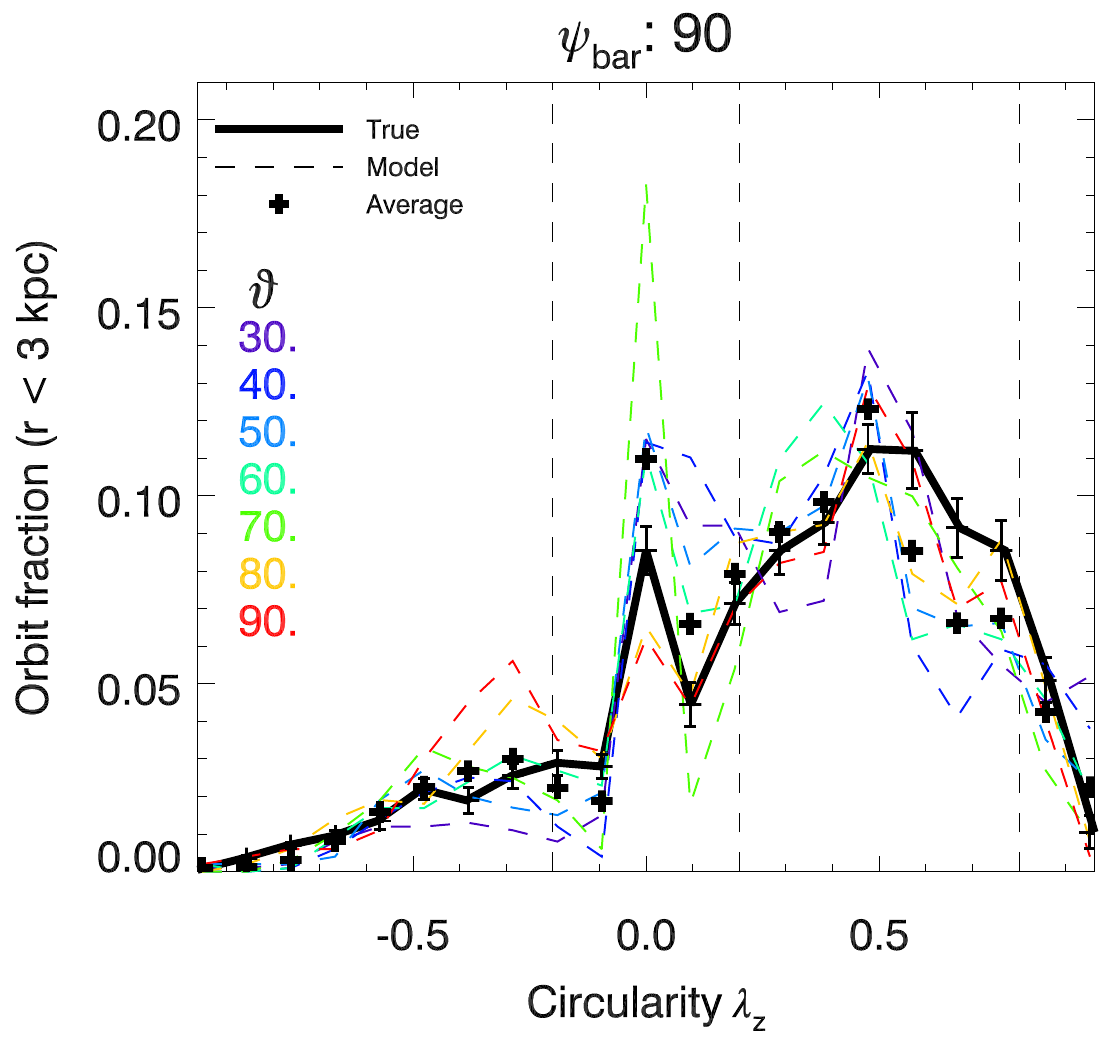}
\includegraphics[width=5.2cm]{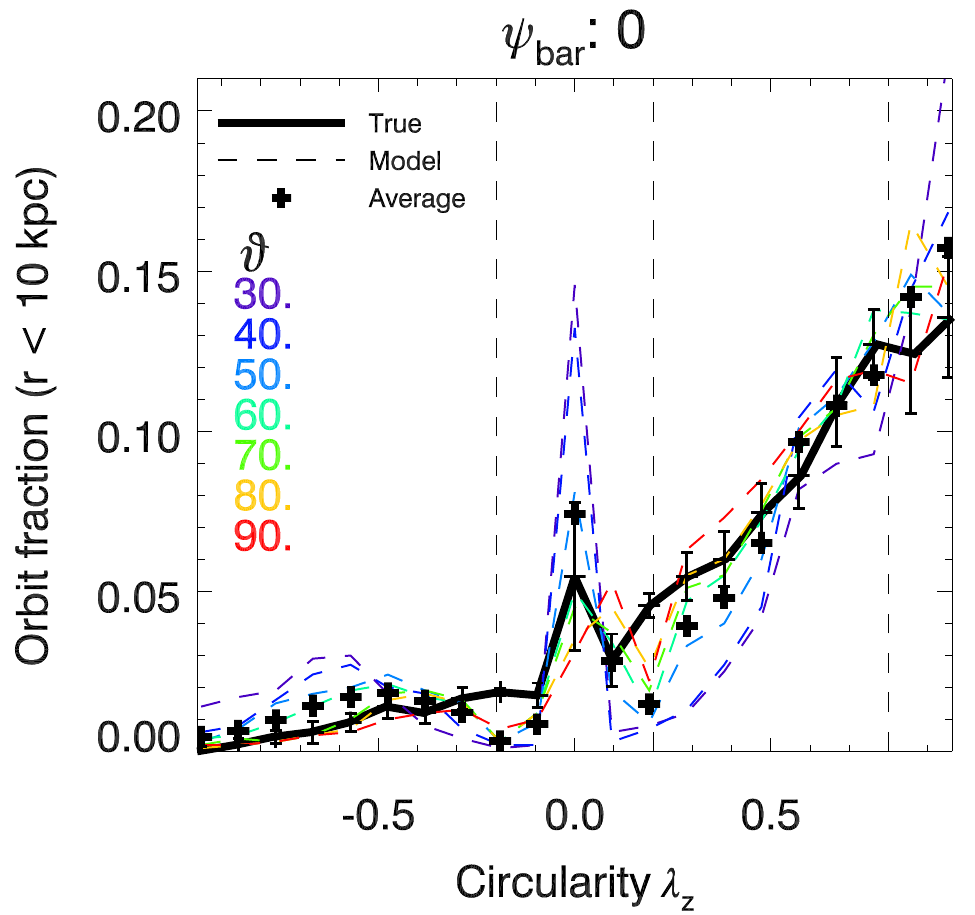}\includegraphics[width=5.2cm]{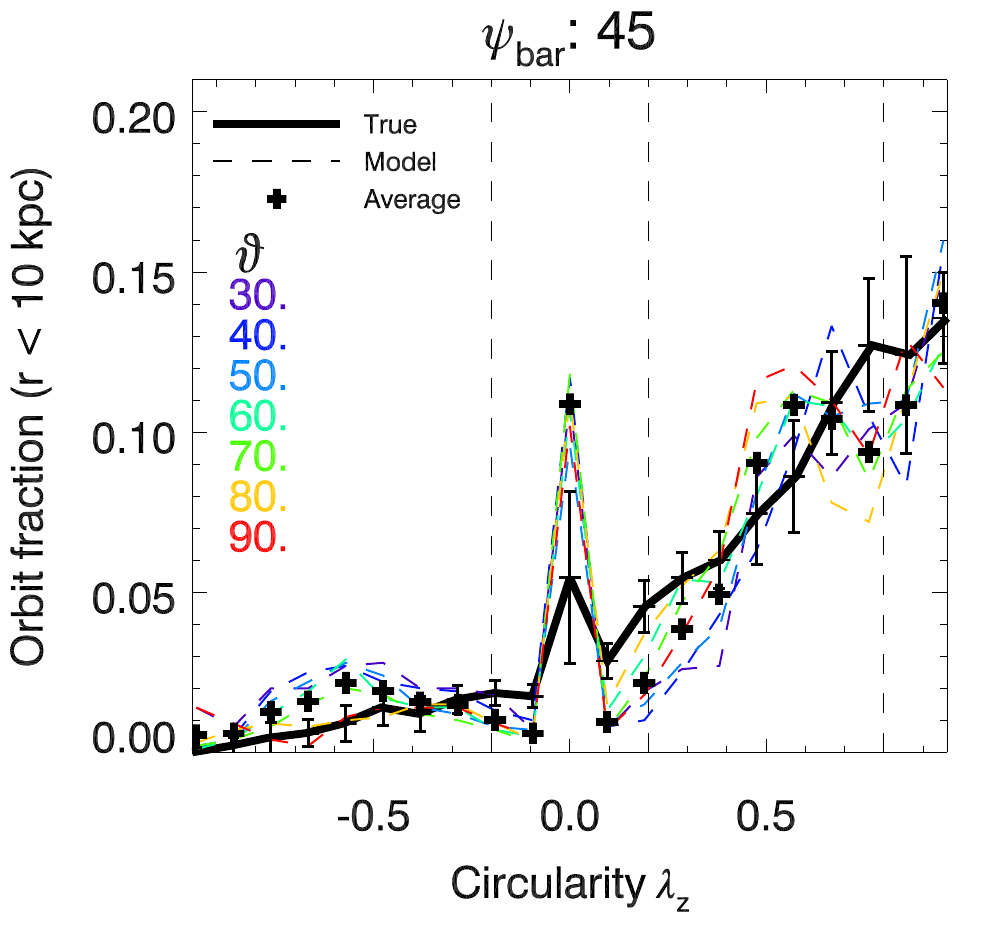}\includegraphics[width=5.2cm]{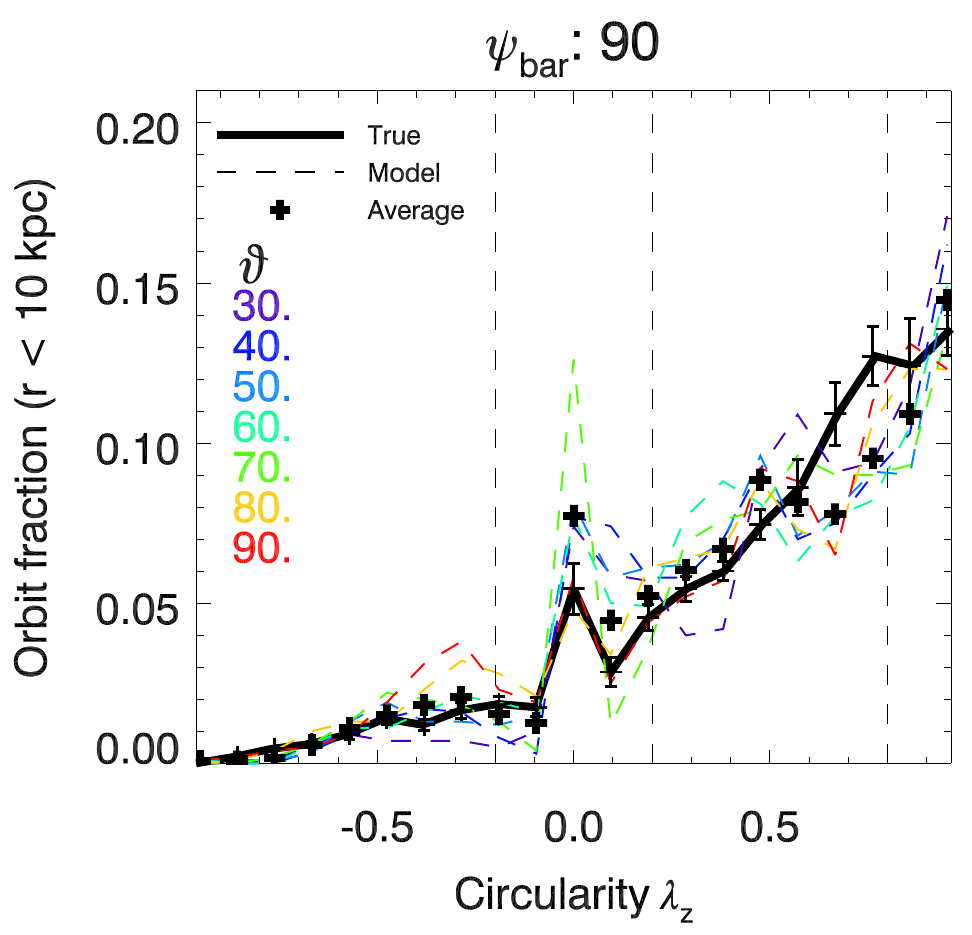}
\caption{Orbit distributions obtained by model option A. 
{\bf Top:} The probability distribution of orbit on $r$ vs. $\lambda_z$ of the best-fitting models of $S_{4}$ ($\psi_{\mathrm{bar}} = 0^o$, $\vartheta = 60^o$), $S_{11}$ ($\psi_{\mathrm{bar}} = 45^o$, $\vartheta = 60^o$) and $S_{18}$ ($\psi_{\mathrm{bar}} = 90^o$, $\vartheta = 60^o$).
{\bf Middle and bottom:} Orbit's $\lambda_z$ distribution, within 3 kpc (top panels) and within 10 kpc (bottom panels). The left, middle and right panels show the $\lambda_z$ distribution from the best-fitting models of $S_1-S_7$ ($\psi_{\mathrm{bar}} = 0^o$), $S_8-S_{14}$ ($\psi_{\mathrm{bar}} = 45^o$) and $S_{15}-S_{21}$ ($\psi_{\mathrm{bar}} = 90^o$), respectively. In each panel, black solid line is the true $\lambda_z$ distribution of the simulated galaxy, dashed with colors from blue to red indicate that obtained from models with mock data with inclination angle $\vartheta$ from $30^o$ to $90^o$. Each dashed line is the average orbit distribution of models among $1\sigma$ confidence level (defined as $\sqrt{2N_{\mathrm{kin}}}$) with each mock data set, the error bars show typical values of the $1\sigma$ error of the colorful lines (note that the error bars are not errors of the true orbit distribution although they are located with the black solid line.). Crosses denote the average of the seven projections with different $\vartheta$ in each panel. }
\label{fig:o1_orbits}
\end{figure*}

As described in Section~\ref{SS:orbits}, we characterize each orbit by its average radius $r$, and 
$\lambda_z$, indicating the circularity of the orbit, as defined in Equation~\ref{eqn:lz}.
The internal orbit distribution of a galaxy is described by the probability density of orbits on the phase-space $r$ vs. $\lambda_z$. 

To obtain the true orbit distribution of the simulated galaxy at the time of this snapshot taken, one has to freeze the potential, and let the particles run more than a few times of orbital period. Thus we can get the time averaged $r$ and $\lambda_z$ for each orbit that occupied by particles.  
We take an approximately-equal but simpler approach by taking space average instead of time average, using only one snapshot of the simulation here. We consider the particles with close values of energy $E$, angular momentum $L_z$ and the total angular momentum amplitude $L$ are on the same (or similar) orbits even the galaxy is not a perfect axisymmetric system. Then we calculate averaged $r$ and $\lambda_z$ with these particles, and this orbit is weighted with the number of particles. 
The true orbit distribution of this galaxy constructed in this way is shown in Figure~\ref{fig:true_orbits}. 

In figure~\ref{fig:o1_orbits}, we show the orbit distribution obtained by taken model option A. 
The top panels show the orbit distribution on $r$ vs. $\lambda_z$ from the best-fitting models of $S_{4}$ ($\psi_{\mathrm{bar}} = 0^o$, $\vartheta = 60^o$), $S_{11}$ ($\psi_{\mathrm{bar}} = 45^o$, $\vartheta = 60^o$) and $S_{18}$ ($\psi_{\mathrm{bar}} = 90^o$, $\vartheta = 60^o$).
To quantitatively compare the $\lambda_z$ distributions, we integrate over $r$ and show, in the middle and bottom panels, the $\lambda_z$ distribution for orbits within $r < 3$ kpc and $r<10$ kpc. The left, middle and right panels show the orbits' $\lambda_z$ distribution from models of $S_1-S_7$ ($\psi_{\mathrm{bar}} = 0^o$), $S_8-S_{14}$ ($\psi_{\mathrm{bar}} = 45^o$) and $S_{15}-S_{21}$ ($\psi_{\mathrm{bar}} = 90^o$), respectively.  

With model option A, we generally recovered the main substructures of the orbit distribution. Counter-rotating orbits at large radius $r > 3$ kpc are cleaned as expected. At $r<3$ kpc, orbits span a large range of $\lambda_z$ constructing the bulge/bar-like kinematics, while at $r>3$ kpc, high $\lambda_z$ orbits dominate in constructing the disk. 

Figure~\ref{fig:o2_orbits} is similar to Figure~\ref{fig:o1_orbits}, but showing the orbit distribution obtained by model option B. First of all, it also worked to clean the counter-rotating orbits at large radius and get some substructures of orbit distribution on the phase space $r$ vs. $\lambda_z$ by looking the top panel. However, with quantitatively comparison in the middle and bottom panels, the orbit distribution obtained by this model option B is discretized and includes too much hot and cold orbits, which deviates from the true orbit distribution significantly. 

With reasonable well recovery of orbit distribution by model option A, we see that projections of the galaxy matter.
Bar orientation $\psi_{\mathrm{bar}}$ of the mock data affects the models' orbit distribution in the inner regions. The models' orbit distribution could easier be biased from kinematic data with a bar aligning with major axis of the disk ($\psi_{\mathrm{bar}} = 0^o$) (also the mass profiles shown in Figure~\ref{fig:mock_mass}).   

Inclination angles of the projection is also important for us to recover the real orbit distribution. In general $ 50^o \le \vartheta \le 80^o$ is the favored area. Projections with $\vartheta > 80^o$ and $30^o \le \vartheta <50^o$ still work but could have larger deviations. We include all projections with $\vartheta \ge 30^o$ to evaluate the quality of our models' orbit recovery. The models' internal orbit distribution could be biased significantly constrained by mock data projected with $\vartheta < 30^o$, which takes $\sim 5\%$ of \emph{CALIFA} galaxies.

We divide the orbits into cold, warm, hot and counter-rotating (CR) components at $\lambda_z = 0.8, 0.2$ similarly to that shown in Section~\ref{SS:orbits}, and summarize the quality of orbit recovery in Figure~\ref{fig:st400}. 
with model A, the orbit fractions of the four components are generally recovered well with deviation $\overline{d}<0.05 $, $|\overline{d}| = |\overline{f_{\mathrm{model}} - f_{\mathrm{true}}}| = |\overline{f_{\mathrm{model}}} - \overline{f_{\mathrm{true}}}|$, averaging from seven sets of models with kinematic maps inclined from $30^o < \vartheta <90^o$ but with the same $\psi_{\mathrm{bar}}$.
After accumulating to just four components, the recovery with model B looks better than it shown in Figure~\ref{fig:o2_orbits}, although we still see significant over-estimate of the cold component and under-estimates of warm components. We adopt model option A for modelling of \emph{CALIFA} galaxies.

\begin{figure*}
\centering\includegraphics[width=5.2cm]{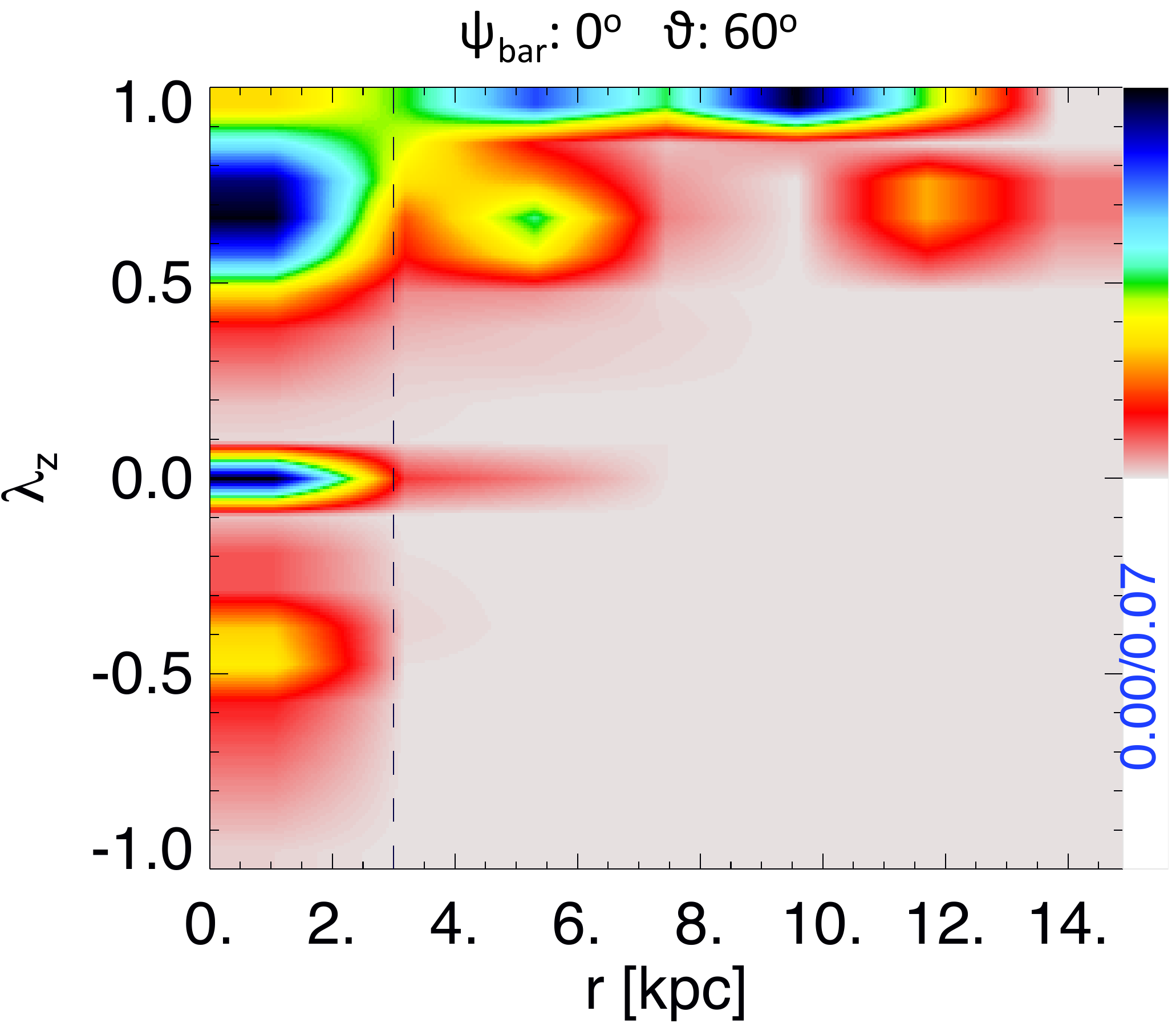}\includegraphics[width=5.2cm]{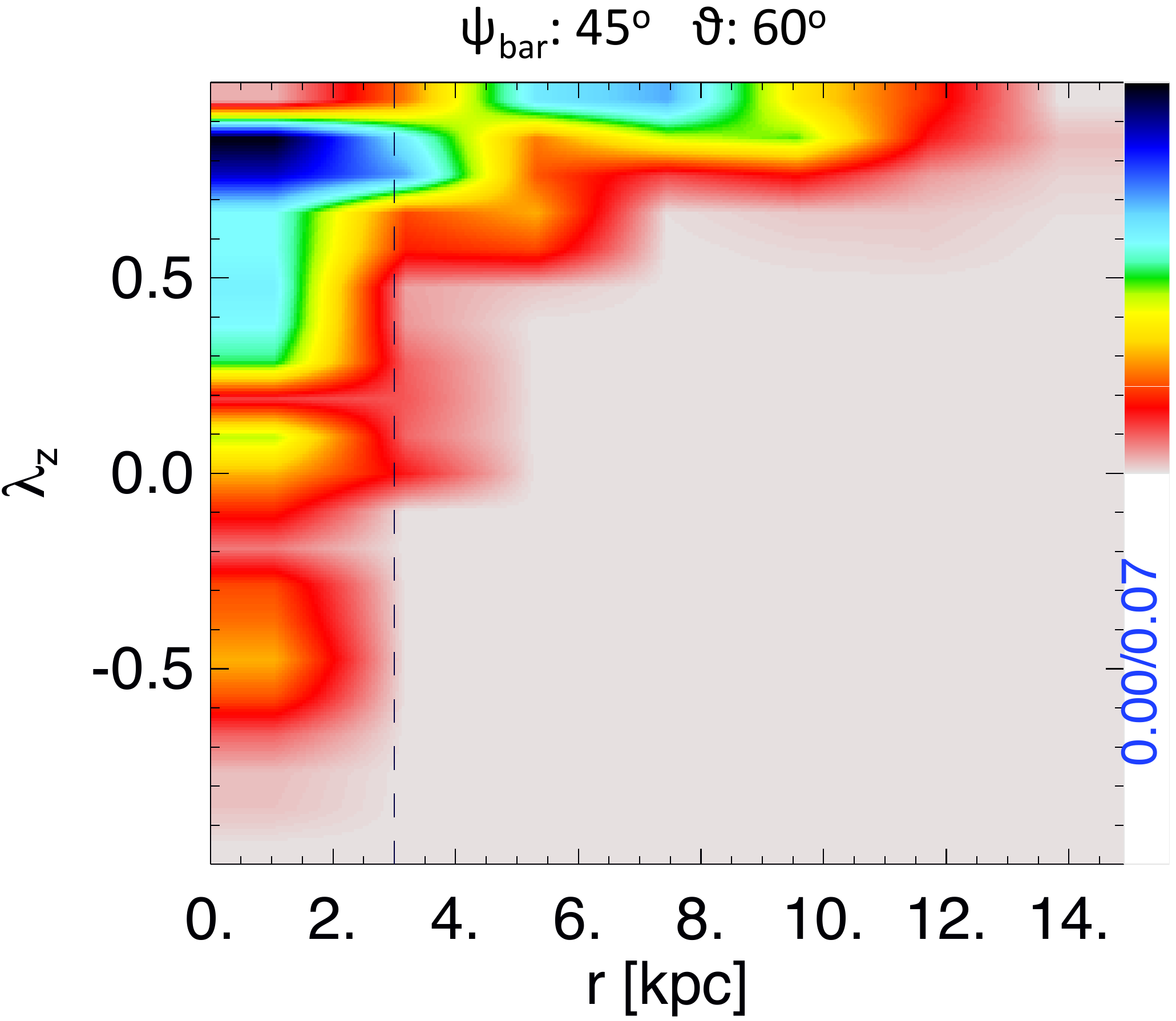}\includegraphics[width=5.2cm]{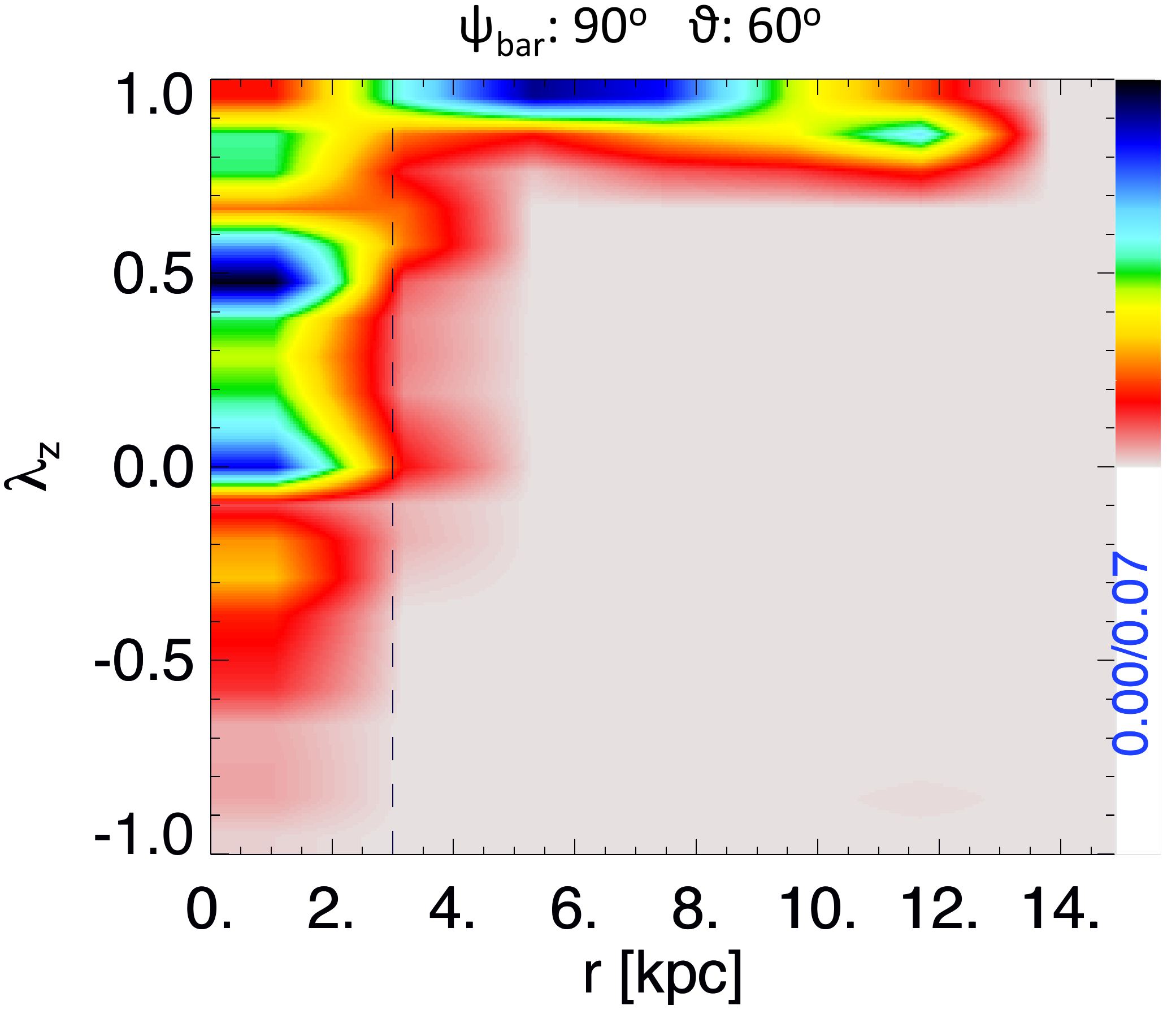}
\centering\includegraphics[width=5.2cm]{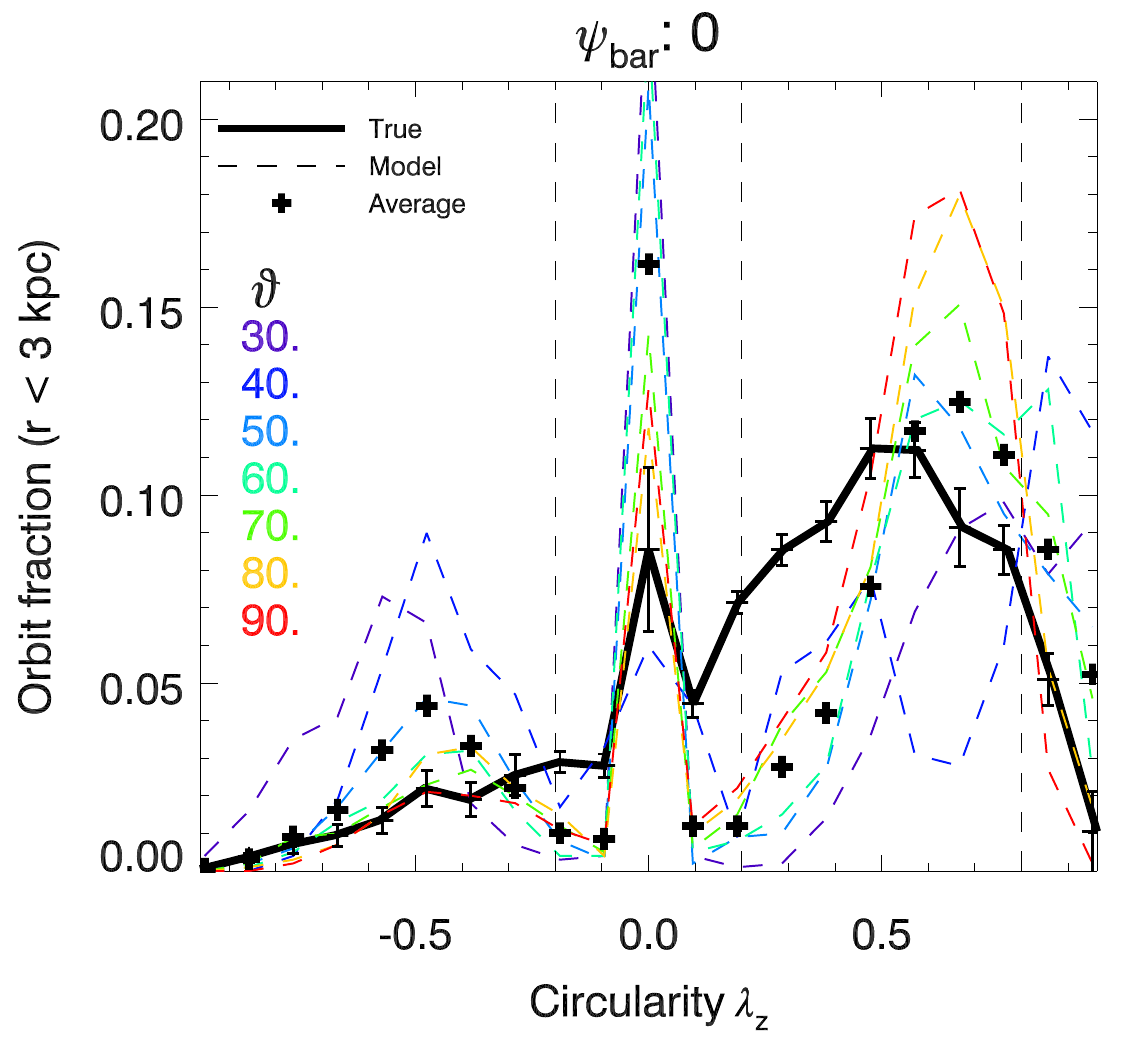}\includegraphics[width=5.2cm]{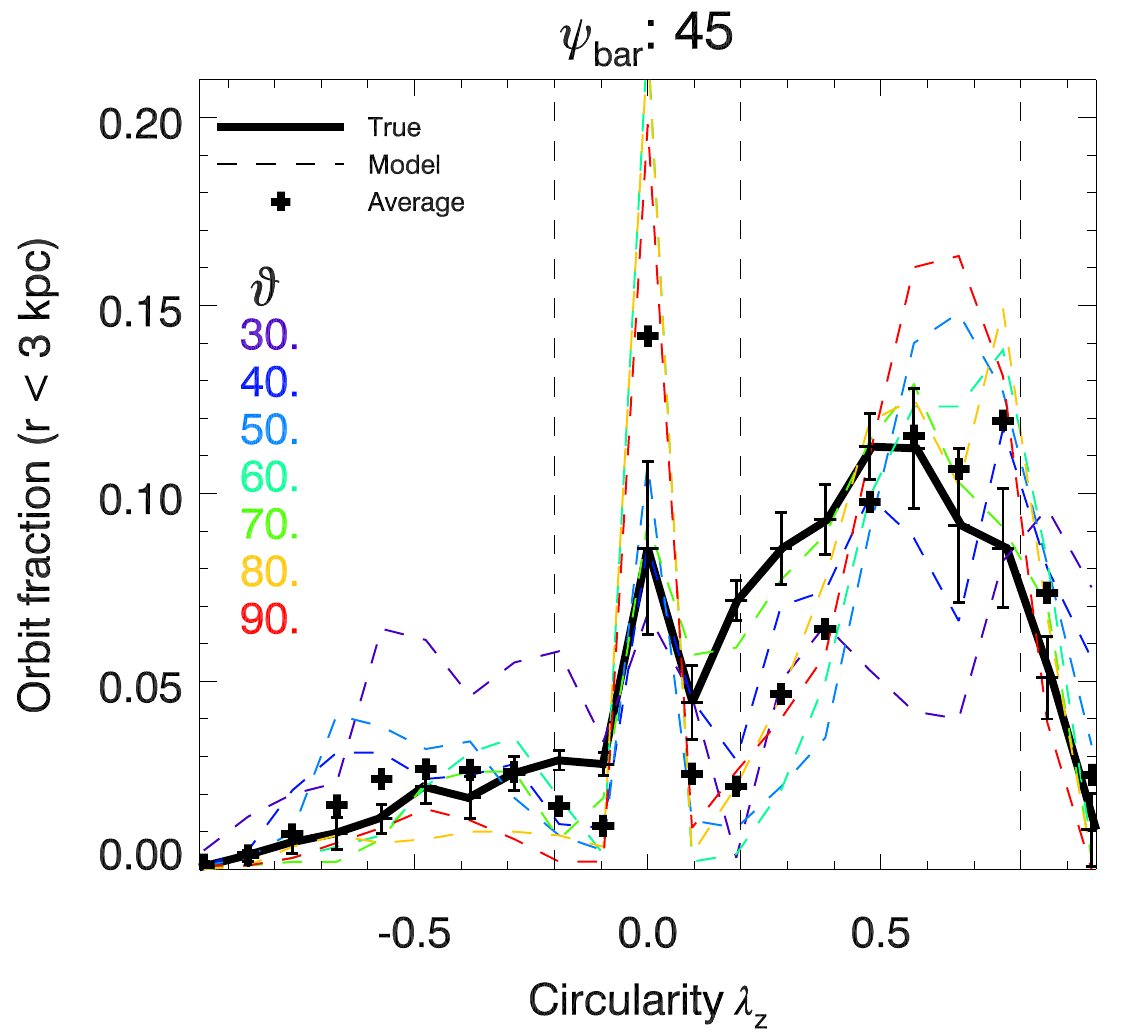}\includegraphics[width=5.2cm]{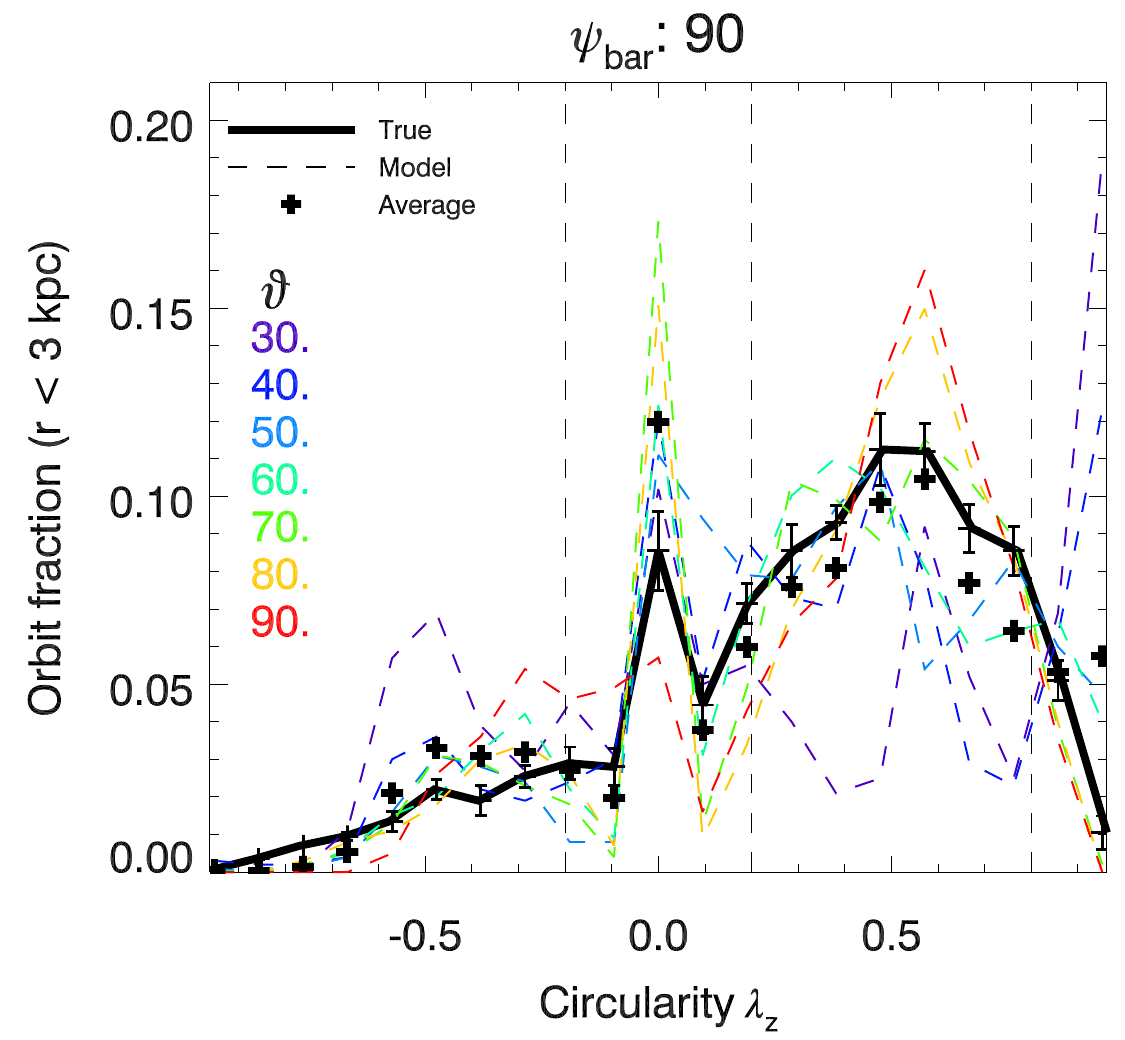}
\centering\includegraphics[width=5.2cm]{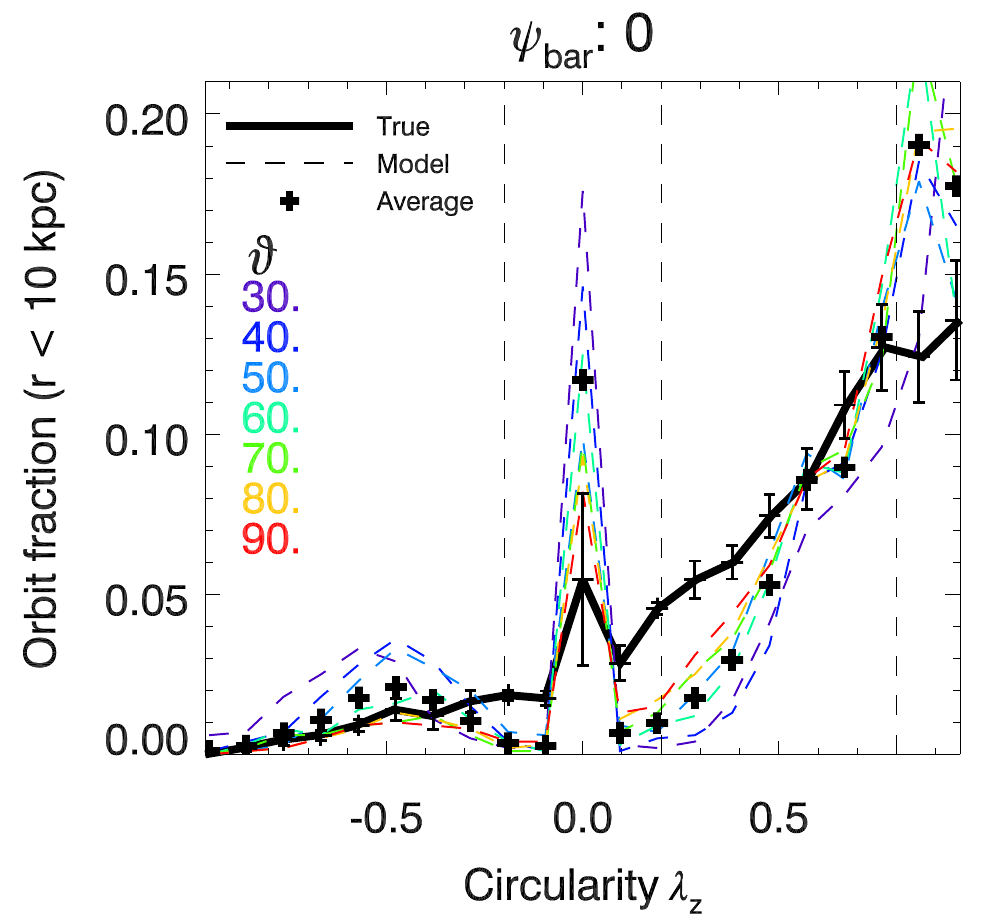}\includegraphics[width=5.2cm]{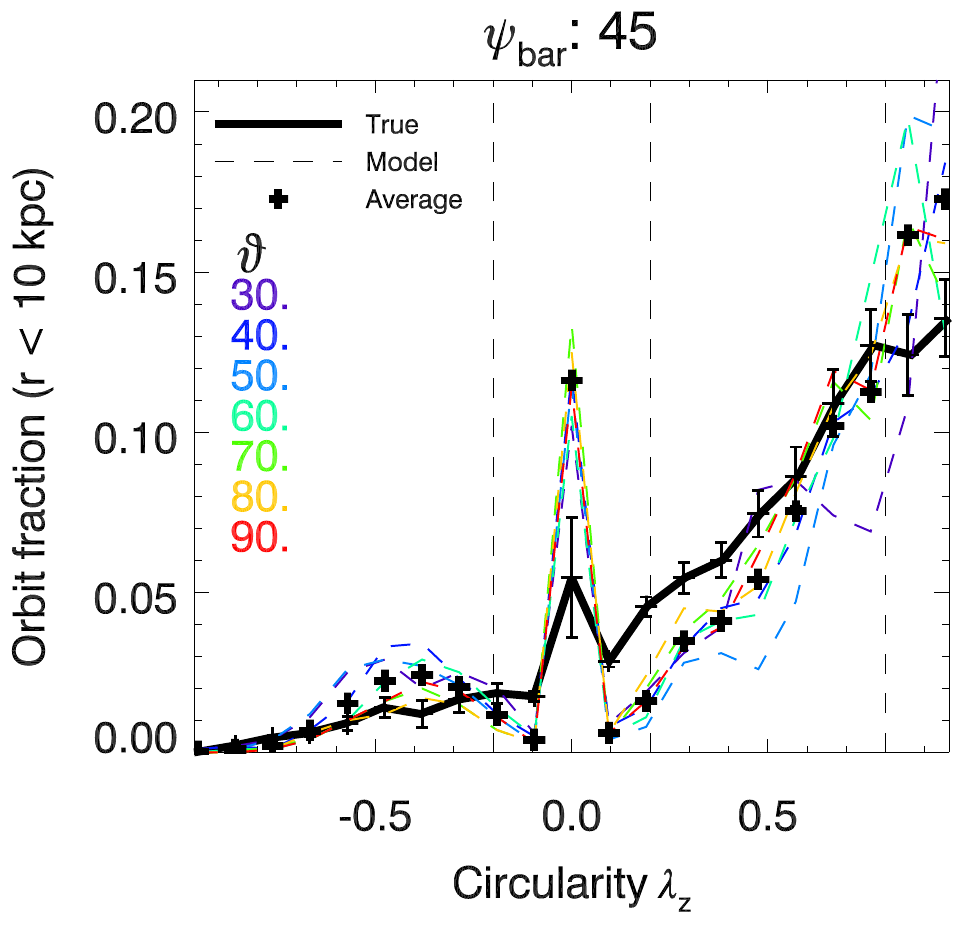}\includegraphics[width=5.2cm]{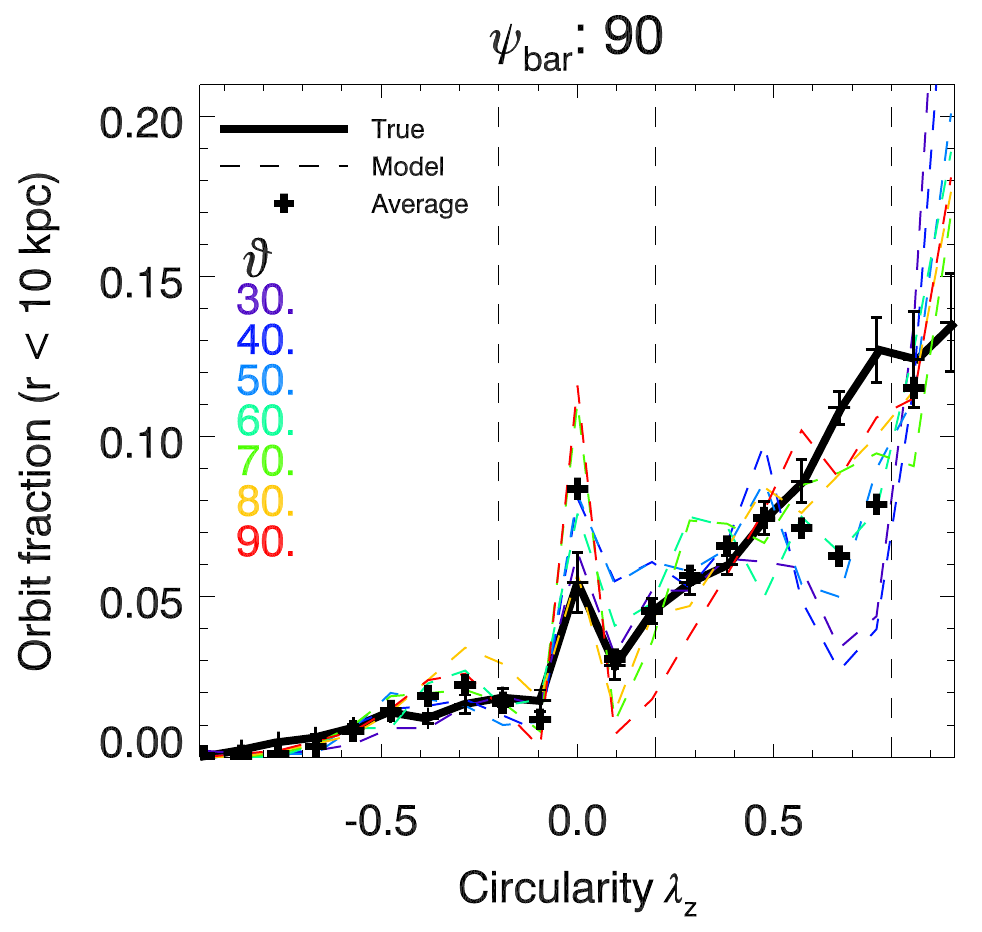}
\caption{Similar to Figure~\ref{fig:o2_orbits}, but with model option B. The orbits' $\lambda_z$ distributions obtained by this option B are more discretized, and they deviates from the true orbit distribution significantly, the galaxy's orbit distributions on $\lambda_z$ are not recovered well.}
\label{fig:o2_orbits}
\end{figure*}

\begin{figure}
\centering\includegraphics[width=8cm]{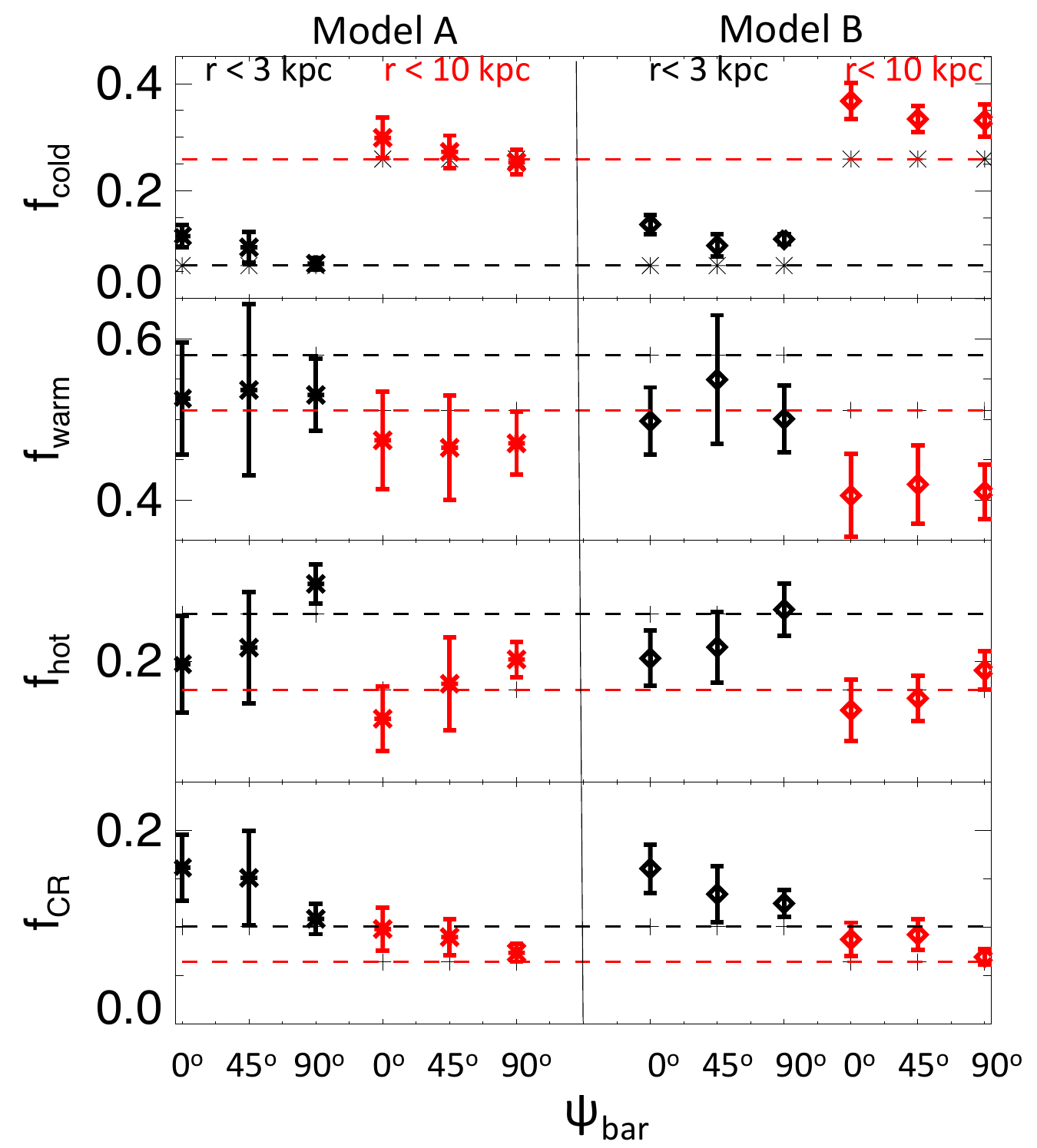}
\caption{The recovery of cold, warm, hot and CR orbit fractions. The black symbols represent orbit fractions within $r <3$ kpc and red symbols within $r<10$ kpc. The dashed lines are the true values of the simulated galaxy, while the thick symbols are that obtained by our models, each point represents the average of the 7 sets of models with kinematic data inclined from $30^o< \vartheta <90^o$, but with the same $\psi_{\mathrm{bar}}$ as labeled on the x-axis. The error bars are the typical $1\sigma$ scatter of each single set of model. The left six lanes are obtained from model option A and the right six lanes are from model option B. }
\label{fig:st400}
\end{figure}

\section{The parameters of the MGEs}
\label{S:MGE}

\begin{table}
\caption{MGE fits of the $r$-band image and the $3.6-\mu$m image for NGC 4210 and NGC 6278. $L_j$ is the central flux in unit of $L_{\mathrm{sun}}/pc^2$, $\sigma_j$ is the size in unit of arcsec, and ${q'}_j$ is the flattening of each Gaussian component. The position angles of all Gaussian components are fixed to be 0. }
\label{tab:mge}
\begin{tabular}{*{8}{l}}
\hline
\multicolumn{4}{l}{NGC 4210}  \\
\hline
\multicolumn{4}{l}{$r$-band}  \\
 j &  $L_j$  &  $\sigma_j$ &  ${q'}_j$   \\
 1 &     1804.6  &   0.493  &   0.872  \\   
 2 &    362.3  &   1.514  &  0.745  \\
 3 &    522.4   &   2.014   &  0.999    \\ 
 4 &    177.7   &   19.49   &  0.760   \\  
\hline
\multicolumn{4}{l}{$3.6-\mu$m}  \\
 1 &   4575.9   &      0.406 &   0.899 \\   
 2 &   1588.9   &      1.665 &   0.802 \\  
 3 &    694.1   &      2.292 &   0.999 \\  
 4 &     98.3   &      6.318 &   0.999 \\ 
 5 &    417.2   &     19.584 &   0.808 \\  
 6 &      1.4   &     45.00 &   0.999 \\
\hline
\hline
\multicolumn{4}{l}{NGC 6278}  \\
\hline
\multicolumn{4}{l}{$r$-band}  \\
 j &  $L_j$  &  $\sigma_j$ &  ${q'}_j$   \\
 1 &     4141.36  &   0.792 &    0.998 \\   
 2 &     9300.93  &   0.792 &     0.922 \\  
 3 &     1793.64  &    2.454 &     0.846 \\     
 4 &     610.450  &    6.409 &     0.555 \\   
 5 &     137.000  &    22.086 &     0.502 \\  
 6 &     11.5600  &    24.500 &    0.998 \\     
\hline
\multicolumn{4}{l}{$3.6-\mu$m}  \\
 1 &  55768.3    &     0.79   &  0.993  \\
 2 &  15000.3    &     1.500  &  0.813   \\
 3 &   2920.0    &     2.672  &  0.999    \\
 4 &   3708.2    &     5.809  &  0.623   \\
 5 &    669.1    &    16.360  &  0.623    \\
 6 &    177.2    &    28.00  &  0.623   \\
 7 &     40.6    &    28.00  &  0.999   \\
  \hline
  \hline
 \end{tabular}
\end{table}

\bsp 

\label{lastpage}

\end{document}